\documentclass[aps,prb,twocolumn,floats,epsfig]{revtex4}
\usepackage{amssymb}
\usepackage{amsbsy}
\usepackage{amsmath}
\usepackage{epsfig}
\usepackage{color}
\usepackage[colorlinks,linktocpage,bookmarks=false,citecolor=blue,linkcolor=red,urlcolor=blue]{hyperref}

\newcommand{\ing}{\includegraphics}
\newcommand{\bib}{\bibitem}
\newcommand{\beq}{\begin{equation}}
\newcommand{\eeq}{\end{equation}}
\newcommand{\bea}{\begin{eqnarray}}
\newcommand{\eea}{\end{eqnarray}}

\newcommand{\ep}{\epsilon}

\newcommand{\lamb}{\lambda}

\newcommand{\calD}{{\cal D}}
\newcommand{\calG}{{\cal G}}
\newcommand{\w}{\omega}
\newcommand{\bra}[1]{\langle#1|}
\newcommand{\ket}[1]{|#1\rangle}
\newcommand{\braket}[2]{\langle #1|#2\rangle}
\newcommand{\non}{\nonumber}

\newcommand{\Tr}{{\rm Tr}}
\newcommand{\mean}[1]{\langle #1 \rangle}
\newcommand{\half}{\frac{1}{2}}

\begin{document}

\title{Entanglement measures and non-equilibrium dynamics of quantum many-body systems: a path integral approach}

\author{Roopayan Ghosh$^{(1)}$, Nicolas Dupuis$^{(2)}$, Arnab Sen$^{(1)}$, and K. Sengupta$^{(1)}$}

\affiliation{$^{(1)}$ School of Physical Sciences, Indian
Association for the Cultivation of Science, 2A and 2B Raja S. C.
Mullick Road, Jadavpur 700032, India \\ $^{(2)}$ Sorbonne Universit\'e, CNRS,
Laboratoire de Physique Th\'eorique de la Mati\`ere Condens\'ee, LPTMC, F-75005 Paris, France}

\date{December 11, 2019}

\begin{abstract}

We present a path integral formalism for expressing matrix elements
of the density matrix of a quantum many-body system between any two
coherent states in terms of standard Matsubara action with
periodic(anti-periodic) boundary conditions on bosonic(fermionic)
fields. We show that this enables us to express several entanglement
measures for bosonic/fermionic many-body systems described by a
Gaussian action in terms of the Matsubara Green function. We apply
this formalism to compute various entanglement measures for the
two-dimensional Bose-Hubbard model in the strong-coupling regime,
both in the presence and absence of Abelian and non-Abelian
synthetic gauge fields, within a strong coupling mean-field theory.
In addition, our method provides an alternative formalism for
addressing time evolution of quantum-many body systems, with
Gaussian actions, driven out of equilibrium without the use of
Keldysh technique. We demonstrate this by deriving analytical
expressions of the return probability and the counting statistics of
several operators for a class of integrable models represented by
free Dirac fermions subjected to a periodic drive in terms of the
elements of their Floquet Hamiltonians. We provide a detailed
comparison of our method with the earlier, related, techniques used
for similar computations, discuss the significance of our results,
and chart out other systems where our formalism can be used.

\end{abstract}

\maketitle

\section{Introduction}
\label{intro}

The density matrix $\hat \rho$ is of central importance for
describing properties of any quantum statistical mechanical system.
For a many-body quantum system with a Hamiltonian $\hat H$, it is
given by
\begin{equation}
\hat \rho = \exp[-\beta \hat H]/Z, \quad  Z = {\rm Tr}
\exp[-\beta \hat H] , \label{dmexp1}
\end{equation}
where ${\rm Tr}$ denotes the sum over all possible field configurations
of the system, $\beta=1/(k_B T_0)$, $k_B$ is the Boltzmann constant,
and $T_0$ is the temperature. Its importance stems from the fact
that any physical observable pertaining to such a system, described
by an operator $\hat O$, satisfies $\langle \hat O \rangle = {\rm
Tr}[\hat \rho \hat O]$.\cite{huangref} It is well known that such
statistical mechanical many-body systems may be described using path
integrals. Indeed, the partition function $Z$ of a quantum many-body
system or expectation of any of its operator, $\langle \hat O
\rangle$, is conveniently described as a coherent state path
integral over of bosonic or fermionic fields describing its
constituent particles \cite{norref1}. It is therefore useful to have
an expression for the matrix element of the density matrix between
two arbitrary coherent states: $\rho_{fi} =\langle \phi_f |\hat \rho
|\phi_i\rangle$, where $|\phi_{f,i}\rangle$ are bosonic or fermionic
coherent states. However, computation of $\rho_{fi}$ cannot be
trivially carried out even for the simplest Gaussian systems. The
key problem that one faces in trying to evaluate such a quantity is
that it cannot be directly expressed in terms of a path integral
with standard periodic or anti-periodic boundary conditions on
bosonic or fermionic fields. This difficulty stems from the fact
that unlike expression of $Z$, the presence of fixed initial and
final coherent states ($\phi_f$ and $\phi_i$) does not allow for a
straightforward implementation of these boundary conditions
\cite{bcdiss}. Consequently, it is not clear if one can use
Matsubara formalism for such computations.

There are at least two key quantities where the knowledge of
$\rho_{fi}$ proves to be useful. The first is the entanglement
entropy of a quantum system which can be directly expressed in terms
of the reduced density matrix ${\hat \rho}_{\rm red}$ which can be
obtained from $\hat \rho$ by tracing out appropriate degrees of
freedom \cite{pleniorev,huertorev,eisertref1,eisertref2,ryu1}. There
are several measures for a system's entanglement; the most
frequently used ones are the von Neumann entropy $S$ and the R\'enyi
entropies $S_n$. These are most easily computed by separating the
closed quantum system into a subsystem and a bath and integrating
out the bath degrees of freedom. In terms of the reduced density
matrix ${\hat \rho}_{\rm red}$, obtained from $\hat\rho$ by
integrating out the bath degrees of freedom, $S$ and $S_n$ are
expressed as
\begin{align}
S &= -{\rm Tr} [ \hat\rho_{\rm red} \ln \hat\rho_{\rm red}] , \label{sedef} \\
S_n &= \frac{1}{1-n} \ln [{\rm Tr} (\hat\rho_{\rm red}^n)  ] , \label{sendef}
\end{align}
where one has the relation $S= \lim_{n \to 1} S_n$. A scheme for
computation of $S_n$ has been pointed out in the context of
one-dimensional (1D) conformal quantum field theories in Refs.\
\onlinecite{cardyref1,calabreseref} by introducing replica fields.
Here we note that the knowledge of the matrix element $\rho_{fi}$
enables us to compute $\rho^{\rm red}_{fi}$ and thus allows for a
direct computation of $S_n$. The key difficulty in carrying this out
in the context of general bosonic or fermionic theories stems, once
again, from one's inability to express $\rho_{fi}$ in terms of
correlation functions that can be computed using Matsubara
formalism. In contrast, the Hamiltonian formulation for computing
matrix elements of ${\hat \rho}_{\rm red}$ for systems with Gaussian
action indicates that such matrix elements can be expressed in terms
of Gaussian correlators computed using standard Green function
methods \cite{peschelref,henleyref}. However, such a formulation
works only if the system Hamiltonian is quadratic. To the best of
our knowledge, the results of Refs.\
\onlinecite{peschelref,henleyref} have not been derived using path
integral methods. Such a derivation would be extremely useful since
there are several systems, such as the Bose-Hubbard model treated
with strong coupling mean-field approximation \cite{dupuis1} or the
non-linear sigma model in the large-$N$ limit \cite{subir1}, where
the effective action of the system can be brought to a Gaussian
form. In contrast, due to complicated frequency dependence of the
effective action \cite{dupuis1} or presence of constraint conditions
\cite{subir1}, there are no simple Gaussian Hamiltonians that
describe such systems. The knowledge of $\rho_{fi}$ (and hence
$\rho_{fi}^{\rm red}$) for such systems may throw light on their
entanglement measures.

The second instance where the knowledge of $\rho_{fi}$ would prove
useful is in the field of non-equilibrium quantum dynamics of
many-body systems. The study of out-of-equilibrium dynamics of
quantum systems has been a subject of active theoretical and
experimental research in recent years
\cite{rev1,rev2,rev3,rev4,rev5,rev6}. The key quantity which
controls such dynamics is the evolution operator $\hat U$ for a
quantum system given by
\begin{eqnarray}
\hat U(t,0) = {\mathcal T} e^{- i \int_0^t dt' \hat H(t')/\hbar}
\label{unievol1}
\end{eqnarray}
where $t$ is the final time up to which we track the evolution,
$\hbar$ is the Planck's constant, ${\mathcal T}$ is the time
ordering operator, and the initial time is set to zero without loss
of generality. From Eqs.\ \eqref{unievol1} and \eqref{dmexp1}, one finds
that $U_{fi}(t) = \langle \phi_f |\hat U(t,0)|\phi_i\rangle$ can be obtained
from $\rho_{fi}$ (with $\hat H \to \hat H(\tau')$ in Eq.\
\eqref{dmexp1} in which case $\hat \rho = {\mathcal
T}_{\tau}\exp[-\int_0^{\beta \hbar} \hat H(\tau') d\tau']/Z$) via a
Wick rotation $t \to -i \beta \hbar$.  We note any correlation
function of a non-equilibrium system (and hence any physical
observable) can be computed from the knowledge of $\hat U$. This
relation is usually not utilized due to the practical difficulty
associated with such Wick rotation; however for cases where
$\rho_{fi}$ is analytically known, this method can provide a
different technique for computing the evolution operator without
having to resort to Keldysh formulation \cite{kelrev,kelref}. This
is particularly useful for Gaussian actions subjected to periodic
drives where they are expected to provide analytic expression for
the system's Floquet Hamiltonian \cite{rev4,asen1}. However, the
relation between $\rho_{fi}$ and $U_{fi}$, to the best of our
knowledge, has not been utilized so far for computing physical
quantities in driven many-body systems.

In this work, we present a formalism for computing matrix elements
of the density matrix of a many-body quantum system between two
arbitrary coherent states. The main results that we obtain from our
analysis are as follows. First, we provide a general framework for
computing $\rho_{fi}$ for arbitrary fermionic and bosonic systems in
terms of their correlation functions which can be computed using
standard Matsubara formalism. Our method shows that this is possible
in principle for arbitrary quantum systems; however, practical
computation are most easily done for systems with Gaussian actions.
We note that such actions need not correspond to non-interacting
systems with quadratic Hamiltonian but can arise out of mean-field
or large-$N$ approximations of strongly interacting quantum systems.
Second, we use this formalism to compute several measures of
entanglement such as von Neumann and R\'enyi entropies of strongly
interacting bosons in an optical lattice, both without and in the
presence of Abelian and non-Abelian artificial gauge fields, within
a strong coupling mean-field theory \cite{dupuis1,sinha1,kush1}. For
bosons without gauge fields, our results regarding $S_n$ and $S$
qualitatively agree with those of Refs.\ \onlinecite{bentang1,
entro1}. Moreover, we also compute the entanglement negativity of
such bosons both in the presence and absence of artificial gauge
fields using our formalism. To the best of our knowledge,
entanglement measures for boson systems with artificial gauge fields
and entanglement negativity for strongly interacting bosons with or
without gauge fields have not been computed so far. Since $S_2$ has
been experimentally measured for superfluid bosons, we expect these
computations to have direct relevance to future experiments
\cite{rislam1}. Third, we use our formalism to compute the Floquet
Hamiltonian for a class of integrable many-body models represented
by free Dirac fermions subjected to periodic drive according to a
square pulse protocol and show that it agrees with the known result
in the literature \cite{rev6,asen1}. We then use it to obtain exact
analytical expressions for the return probability and the counting
statistics for the fermionic density and order parameter for such
driven Dirac fermions in terms of the elements of the Floquet
Hamiltonian. We note that such expressions have been obtained
earlier for quench protocols; our results constitute a
generalization of these results to periodically driven systems. We
use them to discuss the behavior of these quantities for Ising model
in a transverse field, represented by 1D Dirac fermions, near its
critical point. Finally, we summarize our main results, point out
other physical systems where our formalism can be applied, and
discuss the relation of our results with those in the existing
literature.

The plan of the rest of the paper is as follows. In Sec.\
\ref{genform} we detail the general formalism for expressing the
matrix element of the density matrix of quantum many-body systems
between two arbitrary coherent states and chart out how the results
obtained can be used to compute several measures of entanglement
entropies and the entanglement negativity for these systems. We
shall also discuss the application of this formalism to address
non-equilibrium dynamics of driven quantum systems. This is followed
by Sec.~\ref{bhmodel}, where we use this formalism to compute the
R\'enyi and von Neumann entropies and the entanglement negativity of
the Bose-Hubbard model both in the presence and absence of Abelian
and non-Abelian gauge fields. Next, in Sec.\ \ref{rpcs}, we obtain
analytical expressions of the return probability and the counting
statistics of fermionic number and order parameter operators for a
class of integrable quantum many-body systems driven out of
equilibrium by a periodic drive. Finally, we summarize our results,
discuss their applicability to other systems, and conclude in
Sec.~\ref{diss}. Further details of some aspects of our calculations
and the relation of our results to those in the existing literature
are discussed in the appendices.

\section{General formalism}

\label{genform}

In this section, we provide a path-integral based formalism for
computing the matrix elements of a many-body density matrix between
two arbitrary coherent states. This is done in Sec.\ \ref{genform1}
and is followed by its application for computing von Neumann and
R\'enyi entanglement entropies and entanglement negativity in Sec.\
\ref{genform2}. The use of this formalism for out-of-equilibrium
systems is charted out in Sec.\ \ref{noneq}.

\subsection{Matrix elements of a density matrix}
\label{genform1}

In this subsection, we shall compute $\rho_{fi}$ where the coherent
states $|\phi_f\rangle$ and $|\phi_i\rangle$ can either be bosonic
or fermionic (represented by Grassmann functions). To compute this
matrix element, we express the density matrix, given by Eq.\
\eqref{dmexp1}, as a path integral over coherent states using the
standard Suzuki-Trotter decomposition,\cite{norref1}
\begin{multline}
\rho_{fi} = \frac{1}{Z} \int \prod_{k=1}^{N-1} d\phi_k^* d\phi_k \exp\biggl\{ - \sum_{k=1}^{N-1} \sum_{\alpha} |\phi_{k\alpha}|^2 \\  + \sum_{k=2}^{N-1} \Bigl[ \sum_\alpha \phi_{k\alpha}^* \phi_{k-1\alpha} - \ep H(\phi_k^*,\phi_{k-1}) \Bigr]
+ \sum_\alpha (\phi_{f\alpha}^* \phi_{N-1\alpha} \\ + \phi_{1\alpha}^* \phi_{i\alpha} ) - \ep H(\phi_f^*,\phi_{N-1}) - \ep H(\phi_1^*,\phi_i) \biggr\}  ,
\label{dmexp2}
\end{multline}
where $\epsilon= \beta/N$ and $N\to\infty$, $k$ denotes index for time slices,
$\alpha$ stands for spatial and spin indices of the fields, and we use the short-hand notation $d\phi_k^* d\phi_k = \prod_\alpha d\phi_{k,\alpha}^* d\phi_{k,\alpha}$. The $\phi_k$'s are c-numbers for bosons and Grassmann numbers for fermions. We now introduce additional variables, $\phi_{N\alpha}=\eta\phi_{i\alpha}$ and $\phi^*_{N\alpha}=\phi^*_{f\alpha}$ and use
\begin{align}
1 ={}& \int d\phi^*_N d\phi_N \prod_\alpha [ \delta (\phi^*_{N\alpha}-\phi^*_{f\alpha})  \delta (\phi_{N\alpha}-\eta\phi_{i\alpha}) ] \nonumber \\
={}& e^{\eta\phi^*_f\phi_i}
\int d\phi^*_N d\phi_N \int d\lambda^* d\lambda \exp \Bigl\{ -\sum_\alpha \Bigl[ |\phi_{N\alpha}|^2 \nonumber \\ & + (\phi^*_{N\alpha}-\phi^*_{f\alpha}) \lambda_\alpha - \lambda^*_\alpha (\phi_{N\alpha}-\eta \phi_{i\alpha}) \Bigr] \Bigr\}
\label{id1}
\end{align}
(with $\phi^*_f\phi_i\equiv \sum_\alpha \phi^*_{f\alpha}\phi_{i\alpha}$)
to rewrite Eq.~(\ref{dmexp2}) as
\begin{multline}
\rho_{fi} = \frac{1}{Z} e^{\eta\phi^*_f\phi_i} \int d\lambda^* d\lambda \int \prod_{k=1}^{N} d\phi_k^* d\phi_k   \\\times
\exp\biggl\{ - \sum_{k=1}^{N} \Bigl[ \sum_\alpha \phi_{k\alpha}^* (\phi_{k\alpha} - \phi_{k-1\alpha} ) +  \ep H(\phi_k^*,\phi_{k-1}) \Bigr] \\
-\sum_\alpha \Bigl[ (\phi^*_{N\alpha}-\phi^*_{f\alpha}) \lambda_\alpha - \lambda^*_\alpha (\phi_{N\alpha}-\eta \phi_{i\alpha}) \Bigr]
\biggr\}
\label{dmexp2a}
\end{multline}
with the boundary conditions $\phi_0=\eta\phi_N$ and $\phi^*_0=\eta\phi^*_N$ ($\eta=\pm 1$ for bosons/fermions). Eq.~(\ref{id1}) can be
trivially verified by carrying out the Gaussian integration over $\phi_N$ and
$\lambda$. In the continuum time limit $N\to\infty$, this gives
\begin{multline}
\rho_{fi} = \frac{1}{Z} e^{\eta\phi^*_f\phi_i} \int \calD[\lamb^*,\lamb] \int \calD[\phi^*,\phi]   \exp \biggl\{ -S[\phi^*,\phi] \\ + \sum_\alpha \Bigl[ \eta \lamb^*_\alpha (\phi_\alpha(0) - \phi_{i\alpha}) - (\phi^*_\alpha(\beta) - \phi^*_{f\alpha}) \lamb_\alpha ) \Bigr]
\biggr\}
\label{dmexp3}
\end{multline}
(using the freedom to replace $\phi(\beta)$ by $\eta\phi(0)$), where
{\color {blue} $S[\phi^*,\phi]  = \int_0^{\beta} d\tau (\phi^{\ast}
\partial_{\tau} \phi + H[\phi^{\ast}, \phi])$ } is the usual Euclidean
action and the field $\phi$ satisfies (anti)periodic boundary
conditions: $\phi(\beta)=\eta\phi(0)$ and
$\phi^*(\beta)=\eta\phi^*(0)$.

Equation~(\ref{dmexp3}) is one of the key results of this work: It
shows that the matrix element $\rho_{fi}=\langle\phi_f|\hat
\rho|\phi_i\rangle$ of the density matrix can be expressed as a
standard Euclidean path integral with (anti)periodic conditions
provided that we introduce a Lagrange multiplier field $\lamb$ to
enforce the constraint $\phi^*_f=\phi(\beta)$ and $\phi_i=\phi(0)$.
In Appendix~\ref{app1a} we show that the path integral reproduces
the exact result for $\rho_{fi}$ in the case of a single degree of
freedom.

Next we integrate out the $\phi$ fields to obtain
\begin{align}
\rho_{fi} ={}& e^{\eta\phi^*_f\phi_i}  \int \calD[\lamb^*,\lamb]
\exp\Bigl\{ W[\lamb^*,\lamb] \non \\ & + \sum_\alpha ( - \eta \lamb^*_\alpha \phi_{i\alpha} +  \phi^*_{f\alpha} \lamb_\alpha ) \Bigr\} ,
\end{align}
where
\begin{align}
W[\lamb^*,\lamb] ={}& \sum_{n=1}^\infty \frac{1}{(n!)^2} \sum_{\{\alpha_i,\alpha'_i\}}\lamb^*_{\alpha_n} \cdots \lamb^*_{\alpha_1} \non \\ & \times W^{(2n)}_{\alpha_1\cdots\alpha_n,\alpha'_n\cdots\alpha'_1} \lamb_{\alpha'_1} \cdots  \lamb_{\alpha'_n}
\end{align}
is the generating functional of connected correlation functions
\begin{align}
W^{(2n)}_{\alpha_1\cdots\alpha_n,\alpha'_n\cdots\alpha'_1} ={}& (-\eta)^n \langle\phi_{\alpha_1}(0) \cdots\phi_{\alpha_n}(0) \non \\ & \times \phi^*_{\alpha'_n}(\beta) \cdots \phi^*_{\alpha'_1}(\beta)\rangle_{S[\phi^*,\phi],c}
\end{align}
of the $\phi$ field. One can then integrate out the $\lambda$ field to obtain the final form of $\rho_{fi}$ given by
\beq
\rho_{fi} = \exp\{ \eta\phi^*_f\phi_i + \tilde W[\phi^*_f,\phi_i] \} ,
\label{dmexp5}
\eeq
where
\begin{multline}
\tilde W[\phi^*_f,\phi_i] = \ln \int \calD[\lamb^*,\lamb]\, e^{ W[\lamb^*,\lamb] }   + \sum_{n=1}^\infty \frac{1}{(n!)^2} \\ \times \sum_{\{\alpha_i,\alpha'_i\}} \phi^*_{f\alpha_n} \cdots \phi^*_{f\alpha_1} \tilde W^{(2n)}_{\alpha_1\cdots\alpha_n,\alpha'_n\cdots\alpha'_1} \phi_{i\alpha'_1} \cdots  \phi_{i\alpha'_n} ,
\label{dmexp6}
\end{multline}
with
\begin{align}
\tilde W^{(2n)}_{\alpha_1\cdots\alpha_n,\alpha'_n\cdots\alpha'_1} ={}& (-\eta)^n
\langle \lamb_{\alpha_1} \cdots\lamb_{\alpha_n} \non \\ & \times \lamb^*_{\alpha'_n} \cdots \lamb^*_{\alpha'_1}\rangle_{S_{\rm eff}[\lamb^*,\lamb],c}
\label{dmexp7}
\end{align}
and $S_{\rm eff}[\lamb^*,\lamb]=-W[\lamb^*,\lamb]$. We note that Eqs.~(\ref{dmexp5}-\ref{dmexp7}) constitute an
expression of $\rho_{fi}$ for any generic fermionic and bosonic
systems in terms of their correlation functions which can in
principle be computed using Matsubara formulation. In practice,
however, these correlators get quite complicated with increasing order.

To make further analytical progress, we now concentrate on systems
that can be represented by Gaussian actions,
\begin{equation}
S[\phi^*,\phi] = \int_0^\beta d\tau d\tau' \sum_{\alpha,\alpha'} \phi^*_{\alpha}(\tau) G^{-1}_{\alpha\alpha'}(\tau-\tau') \phi_{\alpha'}(\tau') ,
\label{gaction}
\end{equation}
where $G_{\alpha\alpha'}(\tau-\tau')=\langle
\phi_\alpha(\tau)\phi^*_{\alpha'}(\tau')\rangle$. We note that this
does not necessarily restrict our analysis to non-interacting
systems since $G$ may include self-energy terms. Hence the Gaussian
action $S$ may represent several interacting and constrained systems
for which two-particle and higher-order Green functions can be neglected
compared to the single-particle one. This is typically possible when
the system concerned can be treated using a large-$N$ approximation
or within mean-field theory. Concrete examples
include the Bose-Hubbard model in its superfluid and Mott insulating
phases near the critical point treated within a strong coupling
mean-field approximation, the $O(N)$ non-linear sigma model in the
large-$N$ limit, and spin models such as the $d=1$ Ising model or
the $d=2$ Kitaev model which have exact free fermionic
representations.

For the Gaussian action~(\ref{gaction}) one easily finds
\beq
\begin{split}
e^{W[\lamb^*,\lamb]} &= e^{- \sum_{\alpha,\alpha'} \lamb^*_{\alpha} G_{\alpha\alpha'}(0^+) \lamb_{\alpha'}} , \\
e^{\tilde W[\phi^*_f,\phi_i]} &= [ \det G(0^+)]^{-\eta} \, e^{- \eta \sum_{\alpha,\alpha'} \phi^*_{f\alpha} G^{-1}_{\alpha\alpha'}(0^+) \phi_{i\alpha'}}
\end{split}
\eeq
and
\beq
\begin{split}
\rho_{fi} &= [ \det G(0^+)]^{-\eta}\, e^{\eta \sum_{\alpha,\alpha'} \phi^*_{f\alpha} L_{\alpha\alpha'} \phi_{i\alpha'} } ,  \\
L_{\alpha\alpha'} &= \delta_{\alpha\alpha'}-
G^{-1}_{\alpha\alpha'}(0^+) .
\end{split}
\label{quadres1}
\eeq
Here we have used $G(\beta)=\eta G(0)$ and interpreted $G(0)$ as $G(0^+)$ as can be inferred from a careful analysis of the discrete-time path integral (see, e.g., Appendix~\ref{app1a}). Note that $G^{-1}(0^+)$ should be understood as the inverse matrix of $G(0^+)$, i.e. $\sum_{\alpha''}G_{\alpha\alpha''}(0^+)G^{-1}_{\alpha''\alpha'}(0^+)=\delta_{\alpha\alpha'}$.
One can verify that the density operator has the correct normalization, since its trace
\begin{align}
& \frac{1}{Z} \int d\phi^* d\phi \, e^{-\sum_\alpha |\phi_\alpha|^2 } \langle \eta\phi| e^{-\beta \hat H} |\phi\rangle \non \\ &= [ \det G(0^+)]^{-\eta} \int d\phi^* d\phi \,
e^{- \sum_{\alpha,\alpha'} \phi^*_{\alpha} G^{-1}_{\alpha\alpha'}(0^+) \phi_{\alpha'} }
\end{align}
is equal to unity.

The matrix $L$ in~(\ref{quadres1}) can be more explicitly written using
\beq
\begin{split}
G_{\alpha\alpha'}(0^+) &= \delta_{\alpha\alpha'} + G_{\alpha\alpha'}(0^-) , \\
G_{\alpha\alpha'}(0^-) &= \frac{1}{\beta} \sum_{\omega_n} G_{\alpha\alpha'}(i\omega_n) e^{i\omega_n 0^+} ,
\end{split}
\label{matgreen1} \eeq where the first equality follows from
boson/fermion (anti)commutation relations and $\omega_n$ ($n$
integer) is a Matsubara frequency. Thus Eqs.~(\ref{quadres1}) and
(\ref{matgreen1}) provide us with an analytic expression for the
matrix elements of $\hat\rho$ for bosonic/fermionic systems with
Gaussian action in terms of their Matsubara one-particle Green
functions.

The above arguments may be easily generalized to cases where the
action of the system breaks $U(1)$ symmetry,
\begin{align}
S[\phi^*,\phi] ={}& \frac{1}{2} \int_0^\beta d\tau d\tau' \sum_{\alpha,\alpha'} \bigl(\phi^*_{\alpha}(\tau),\phi_{\alpha}(\tau)\bigr) \non \\ & \times \calG^{-1}_{\alpha\alpha'}(\tau-\tau')
\begin{pmatrix}  \phi_{\alpha'}(\tau') \\ \phi^*_{\alpha'}(\tau') \end{pmatrix},
\end{align}
where
\begin{align}
\calG_{\alpha\alpha'}(\tau-\tau') &= \begin{pmatrix}
\langle \phi_{\alpha}(\tau)\phi^*_{\alpha'}(\tau') \rangle &
\langle \phi_{\alpha}(\tau)\phi_{\alpha'}(\tau') \rangle \\
\langle \phi^*_{\alpha}(\tau)\phi^*_{\alpha'}(\tau') \rangle &
\langle \phi^*_{\alpha}(\tau)\phi_{\alpha'}(\tau') \rangle
\end{pmatrix} \non \\
&\equiv  \begin{pmatrix}
G_{\alpha\alpha'}(\tau-\tau') & F_{\alpha\alpha'}(\tau-\tau') \\
F^\dagger_{\alpha\alpha'}(\tau-\tau') & \eta G_{\alpha'\alpha}(\tau'-\tau)
\end{pmatrix} .
\end{align}
A straightforward calculation in the same lines as charted out above
leads to
\begin{equation}
\rho_{fi} = [ \det \calG(0^+)]^{-\eta/2}\, e^{\frac{\eta}{2}
\sum_{\alpha,\alpha'} \Phi^\dagger_{\alpha} L_{\alpha\alpha'}
\Phi_{\alpha'} } , \label{quadres2}
\end{equation}
where
\begin{align}
L_{\alpha\alpha'} &= I_0 \delta_{\alpha\alpha'} - I_0
\calG^{-1}_{\alpha\alpha'}(0^+) I_0 , \quad I_0 =
\begin{pmatrix} 1 & 0
\\ 0 & \eta \end{pmatrix} \label{quadres2a}
\end{align}
and $\Phi_{\alpha} = (\phi_{i\alpha},\phi^*_{f\alpha})^T$,
$\Phi^\dagger_{\alpha} = (\phi^*_{f\alpha},\phi_{i\alpha})$. The
$2\times 2$ matrix $\calG_{\alpha\alpha'}(0^+)$ can be written
as\cite{note1}
\begin{equation}
\calG_{\alpha\alpha'}(0^+) = \begin{pmatrix}
    G_{\alpha\alpha'}(0^+) & F_{\alpha\alpha'}(0^+) \\
    \eta F^*_{\alpha\alpha'}(0^+) & \eta G_{\alpha'\alpha}(0^+)
\end{pmatrix} .
\label{matgreen2}
\end{equation}

In the next section, we shall use Eqs.~(\ref{quadres1}) and
(\ref{quadres2}) to compute several entanglement measurements for
these systems. Moreover, these equations shall be used to compute
correlation functions in driven bosonic/fermionic systems in Sec.\
\ref{noneq}.

\subsection{Entanglement measures}
\label{genform2}

In this section, we are going to relate expressions for several
entanglement measures for ground states of many-body quantum systems
whose action is Gaussian and is given by Eq.\ \eqref{gaction}, to its
Matsubara Green function $G(0^+)$ at zero temperature. This can be
done directly from Eqs.\ \eqref{matgreen1} and \eqref{matgreen2} via the
introduction of replica fields \cite{cardyref1,peschelref}; this
procedure is charted out in App.\ \ref{app2}. Here we connect Eq.\
\eqref{matgreen1} and \eqref{matgreen2} to standard methods used in the
literature for such computation \cite{pleniorev,huertorev}.

In what follows, we consider a lattice model with Gaussian action
and broken $U(1)$ symmetry for which $F \ne 0$; results for systems
with no $U(1)$ symmetry breaking can be obtained from our analysis
by setting $F=0$. One can easily check that the density matrix
obtained in the preceding section yields the correct value of the
equal-time correlator
\begin{align}
\mean{\hat\phi_i \hat\phi^\dagger_j} &= \Tr(\hat\rho \hat\phi_i \hat\phi^\dagger_j) \nonumber \\
&= \int d\phi^*d\phi \, e^{-\sum_\ell|\phi_\ell|^2} \eta\phi_j^*
\bra{\eta\phi} \hat\rho \ket{\phi} \phi_i ,
\end{align}
where $i$ and $j$ denote sites of the lattice (we ignore the spin index).
Using~(\ref{quadres2}) for the matrix element of the density matrix
between the coherent states $\ket{\eta\phi}$ and $\ket{\phi}$, one
obtains \beq \mean{\hat\phi_i \hat\phi^\dagger_j} = K \int
d\phi^*d\phi \, \phi_i \phi_j^* \, e^{-\half \sum_{\ell,\ell'}
\Phi^\dagger_\ell \calG_{\ell\ell'}^{-1}(0^+) \Phi_{\ell'} } ,
\label{corr2} \eeq where $\Phi^\dagger_\ell=(\phi_\ell^*,\phi_\ell)$
and $K=[\det\, G(0^+)]^{-\eta/2}$. Performing the integral gives
$\mean{\hat\phi_i \hat\phi^\dagger_j}=G_{ij}(0^+)$ as expected. A
similar analysis gives
\begin{align}
\mean{\hat\phi_i \hat\phi_j} &= K \int d\phi^*d\phi \, \phi_i \phi_j \, e^{-\half \sum_{\ell,\ell'} \Phi^\dagger_\ell \calG_{\ell\ell'}^{-1}(0^+) \Phi_{\ell'} } \nonumber \\
&= F_{ij}(0^+) .
\label{corr3}
\end{align}
Given the structure~(\ref{matgreen2}) of the matrix $\calG_{ij}(0^+)$, the quadratic form appearing in Eqs.~(\ref{corr2}) and (\ref{corr3}) can be diagonalized by a Bogoliubov transformation.\cite{frerot15,blaizotbook} In Fourier space,
\begin{align}
\half \sum_{{\bf k}} \Phi^\dagger_{\bf k} \calG^{-1}_{{\bf k}}(0^+) \Phi_{{\bf k}}
&= \half \sum_{{\bf k}} \Psi^\dagger_{\bf k} \begin{pmatrix}
    \lamb^{-1}_{\bf k} & 0 \\ 0 & \eta\lamb^{-1}_{\bf k}
\end{pmatrix} \Psi_{{\bf k}} \nonumber \\
&= \sum_{\bf k} \psi^*_{\bf k} \lamb^{-1}_{\bf k} \psi_{\bf k} ,
\end{align}
where $\Psi^\dagger_{{\bf k}}=(\psi_{\bf k}^*,\psi_{-\bf k})$,
$\lambda_{\bf k}\geq 0$ and the sum runs over all momenta ${\bf k}$ of the
first Brillouin zone.

We now first consider the case where $\psi_{\bf k}$ is a bosonic
field. We note that in this case the correlations of $\psi_{\bf k}$ may
be thought as those of a free charged scalar fields whose
Hamiltonian is given by
\begin{eqnarray}
\hat H_{\rm scalar} &=& \sum_{\bf k} \left(\hat \Pi_{\bf k}^{\dagger}
\hat\Pi_{\bf k} + \frac{1}{4\lambda_{\bf k}^2}  \hat\psi^{\dagger}_{\bf k} \hat\psi_{\bf k} \right) , \label{harmonicham}
\end{eqnarray}
where $[\hat\psi_{\bf k}, \hat\Pi_{{\bf k}'}]=[\hat\psi^\dagger_{\bf
k}, \hat\Pi^\dagger_{{\bf k}'}]=i \delta_{{\bf k},-{\bf k}'}$ (and
all other commutators vanishing). From~(\ref{harmonicham}) one
easily obtains $\mean{\hat\psi_{\bf k}\hat\psi^\dagger_{{\bf
k}'}}=\delta_{{\bf k},{\bf k}'}\lambda_{\bf k}$ and
$\mean{\hat\Pi_{\bf k}\hat\Pi^\dagger_{{\bf k}'}}=\delta_{{\bf
k},{\bf k}'}/4\lambda_{\bf k}$. We note that since the action of the
system is Gaussian, the correlations are completely specified by the
$\lambda_{\bf k}$'s; thus all the quantities including
entanglement entropy $S$ corresponding to such an action are
identical to that obtained from $\hat H_{\rm scalar}$.

It is convenient to rewrite $\hat\psi^{(\dagger)}$ and
$\hat\Pi^{(\dagger)}$ in terms of four real operators $\hat\psi_i$,
$\hat\Pi_i$ ($i=1,2$) defined by \beq
\begin{split}
\hat\psi_{\bf k} = \frac{1}{\sqrt{2}} ( \hat\psi_{1{\bf k}} + i \hat\psi_{2{\bf k}} ) , \\
\hat\Pi_{\bf k} = \frac{1}{\sqrt{2}} ( \hat\Pi_{1{\bf k}} - i \hat\Pi_{2{\bf k}} ) ,
\end{split}
\eeq and $\hat\psi_{i{\bf k}}^\dagger=\hat\psi_{i-{\bf k}}$,
$\hat\Pi_{i{\bf k}}^\dagger=\hat\Pi_{i-{\bf k}}$, $[\hat\psi_{i{\bf
k}},\hat\Pi_{j{\bf k}'}]=i\delta_{ij}\delta_{{\bf k},-{\bf k}'}$.
The Hamiltonian can then be written as the sum of two uncoupled
harmonic oscillators, \beq \hat H_{\rm scalar} = \half \sum_{i=1}^2
\sum_{\bf k}  \left(\hat \Pi_{i-{\bf k}} \hat\Pi_{i\bf k} +
\frac{1}{4\lambda_{\bf k}^2}  \hat\psi_{i-\bf k} \hat\psi_{i\bf k}
\right) . \label{harmosc} \eeq The computation of the von Neumann
entropy $S$ for each oscillator is straightforward.\cite{huertorev}
One chooses a subsystem $A$ which is a part of the full system and
construct the covariance matrix $\Lambda_{jj'}$ with $j,j' \in A$
given by \beq
\begin{split}
\Lambda_{jj'} &= \left(\begin{array} {cc}  \langle \psi_j \psi_{j'}
\rangle & 0 \\ 0 & \langle \Pi_{j} \Pi_{j'} \rangle
\end{array} \right) = \left(\begin{array} {cc} A_{jj'}  & 0 \\ 0 & B_{j j'}
\end{array} \right) \label{corrmat} \\
A_{jj'} &=  \int_{\bf k} \lambda_{\bf k} \, e^{i {\bf k}\cdot ({\bf
r}_j-{\bf r}_{j'})}, \\
B_{jj'} &= \int_{\bf k}
\frac{1}{4\lambda_{\bf k}} \, e^{i {\bf k}\cdot ({\bf r}_j-{\bf
r}_{j'})} ,
\end{split}
\eeq
where the integral $\int_{\bf k}=\int\frac{d^dk}{(2\pi)^d}$ is over the first Brillouin zone. The symplectic
eigenvalues $\nu_{\ell}$ of the covariance matrix $\Lambda$ then
yields the boson entanglement entropy \cite{pleniorev,huertorev}
\begin{align}
S_b &= -2 \sum_\ell [ n_\ell \ln n_\ell -(n_\ell+1) \ln(n_\ell +1) ]
\nonumber \\
&=2 \sum_{\ell} \sum_{s=\pm 1} s \left( \nu_\ell + \frac{s}{2} \right) \ln \left( \nu_\ell +
\frac{s}{2} \right) , \label{bosonvn}
\end{align}
where $n_\ell=1/(e^{\epsilon_\ell}-1)$ and $\epsilon_\ell=2\, {\rm
arcoth}(2\nu_\ell)$ denotes the eigenvalues of the entanglement
Hamiltonian. A factor of 2 has been introduced in~(\ref{bosonvn}) to
take into account both oscillators of the
Hamiltonian~(\ref{harmosc}).

For fermionic fields, the procedure is simpler. We first note for
such systems $\hat\Pi=\hat\psi^\dagger$ and hence $B$ is identical
to $A$. Thus the only non-trivial correlation function is
$\mean{\hat\psi_{\bf k}\hat\psi^\dagger_{\bf k}}$ and there is no
need to consider the harmonic oscillator~(\ref{harmonicham}). The
entanglement Hamiltonian is determined by requiring that it
reproduces the correlation matrix $A_{jj'}$ of the subsystem, which
gives
\begin{eqnarray}
S_f &=& -\sum_{\ell} [\nu_{\ell} \ln \nu_{\ell} +(1-\nu_{\ell})
\ln(1-\nu_{\ell}) ] , \label{fermionvn}
\end{eqnarray}
where $\nu_\ell=1-n_\ell$ denotes the eigenvalues of $A$ and
$n_\ell=1/(e^{\epsilon_\ell}+1)$, {\it i.e.}, $\epsilon_\ell= 2\,
{\rm arctanh}(2\nu_\ell-1)$.

We note that this procedure also allows us to compute the R\'enyi
entropies for Gaussian systems. In the basis where the entanglement
Hamiltonian $\hat H_e=\sum_\ell \epsilon_\ell \hat a^\dagger_\ell
\hat a_\ell$ is diagonal, the reduced density matrix reads
\begin{eqnarray}
\hat\rho_{\rm red}= \frac{e^{-\hat H_e}}{Z} =
\prod_\ell (1 -\eta e^{-\epsilon_{\ell}})^{\eta} e^{-\epsilon_{\ell} \hat a^\dagger_\ell \hat a_\ell} . \label{eham}
\end{eqnarray}
The occupation number operator $\hat a^\dagger_\ell \hat a_\ell$ has
all positive integer eigenvalues for bosons and $0,1$ for fermions.
The $n^{\rm th}$ R\'enyi entropy~(\ref{sendef}) is given
by\cite{pleniorev, huertorev}
\begin{equation}
S_n = - \frac{\eta}{1-n}
\sum_{\ell} \ln \frac{1-\eta e^{-n \epsilon_{\ell}}}{(1-\eta
e^{-\epsilon_{\ell}})^n} . \label{renyiexp}
\end{equation}

One can also compute the entanglement negativity $N_b$ for bosonic systems
from the covariance matrix constructed in Eq.\ \eqref{corrmat}. To
this end, we note that for computing
$N_b$ for any bosonic many-body systems one needs to
identity two subsystems $L_1$ and $L_2$. One then creates a partial
transposed density matrix, $\rho^T$, which is defined as
\cite{eisertref1}
\begin{eqnarray}
\rho^T_{fi} &=& \langle \phi_f^{L_1}, \phi_{i}^{L_2 \ast}|\hat \rho|
\phi_{i}^{L_1} \phi_{f}^{L_2 \ast} \rangle  .\label{ptdm1}
\end{eqnarray}
Such a partial transposition has been studied for a harmonic chain and
it is known that it amounts to the transformation of the canonical
momenta in subsystem $L_2$: $\Pi^{L_2} \to -\Pi^{L_2}$. Thus one
needs to change the sign of the momentum correlators between the two
subsystems. The covariance matrix can now be written as
\begin{eqnarray}
\Lambda^T &=& \left( \begin{array} {cccc} 0 & 0 & A_{jj'}^{L_1 L_1}&
A_{jj'}^{L_1 L_2} \\ 0 & 0 & A_{jj'}^{L_2 L_1} & A_{jj'}^{L_2 L_2}
\\ B_{jj'}^{L_1 L_1} &- B_{jj'}^{L_1 L_2} & 0 & 0 \\ -B_{jj'}^{L_2
L_1} & B_{jj'}^{L_2 L_2} & 0 & 0 \end{array} \right) , \label{ptdm2}
\end{eqnarray}
where $A_{jj'}^{L_1,L_2} [B_{j j'}^{L_1,L_2}]$ involve field
(canonical momentum) correlators defined in Eq.~(\ref{corrmat}) with
$j \in L_1$ and $j'\in L_2$. The eigenvalues of this covariance
matrix are denoted by $\nu_{\ell}^T$. From Eq.\ \eqref{bosonvn}, we
find $\nu_{\ell}^T <1$ occurs if $\rho^T$ has negative eigenvalues.
Thus, in terms of these eigenvalues, one can define the entanglement
negativity for bosons as \cite{eisertref1}
\begin{equation}
N_b= - \sum_{\ell} \ln ({\rm Min}[1,\nu_{\ell}^T]) .
\label{pten1}
\end{equation}

For fermions, partial transposition is more complicated than for bosons.
Indeed it has been shown in Ref.\ \onlinecite{eisertref2} that a
transposition of fermionic fields with a generic Gaussian action does
not keep the action Gaussian. Moreover such a transformation,
performed twice, does not lead to the starting Gaussian Hamiltonian or
action. To remedy this, a different transformation which amounts to
replacement of the partial transformation by partial time reversal
operation has been defined in Ref.\ \onlinecite{ryu1}. However, to
the best of our knowledge, it is not fully understood if this new
measure accurately reflects the presence of entanglement for generic
fermionic systems. We shall not address this issue further in this
work.

\subsection{Non-equilibrium dynamics}
\label{noneq}

In this subsection, we shall outline the application of our approach
to quantum systems taken out of equilibrium {\it via} a drive with a
given protocol. To this end, we shall compute $U_{fi}(t,0)= \langle
\phi_f | \hat U(t,0)|\phi_i\rangle$ where $\hat U$ is given by Eq.\
\eqref{unievol1}. To compute $U_{fi}$ using the formalism developed in
Sec.\ \ref{genform1}, we analytically continue to imaginary time
using $t' \to -i\tau$. Denoting $t=-i \beta \hbar$, we find
\begin{eqnarray}
U_{fi}(\beta,0) &=& \langle \phi_f |{\mathcal T}_{\tau}
e^{-\int_0^{\beta \hbar} d \tau \bar  H(\tau)} |\phi_i\rangle
\label{unievol3}
\end{eqnarray}
where $\bar H(\tau) = \hat H(t'=-i\tau)$ is the analytically
continued Hamiltonian. We now retrace the steps outlined in Sec.\
\ref{genform1}. The difference that arises in such a procedure is
that the Hamiltonian can in principle be $\tau$ dependent. However,
this issue does not lead to any major complication. One finds that
the evolution operator can be written as
\begin{multline}
U_{fi}(\beta,0) = e^{\eta\phi^*_f\phi_i} \int \calD[\lamb^*,\lamb]
\int \calD[\phi^*,\phi]   \exp \biggl\{ -S[\phi^*,\phi] \\ +
\sum_\alpha \Bigl[ \eta \lamb^*_\alpha (\phi_\alpha(0) -
\phi_{i\alpha}) - (\phi^*_\alpha(\beta)
- \phi^*_{f\alpha})
\lamb_\alpha ) \Bigr] \biggr\} \label{dmexp4}
\end{multline}
with $\phi^{(*)}(\beta)=\eta \phi^{(*)}(0)$ and where the Euclidean action $S[\phi^*,\phi]$ can now be explicitly time dependent.

The time-evolution operator $U_{fi}(\beta,0)$ can be explicitly computed when the dynamics of the system is governed by a Gaussian action,
\begin{align}
S[\phi^*,\phi] ={}& \frac{1}{2} \int dt dt' \sum_{\alpha,\alpha'} \bigl(\phi^*_{\alpha}(t),\phi_{\alpha}(t)\bigr) \non \\ & \times \calG^{-1}_{\alpha\alpha'}(t,t')
\begin{pmatrix}  \phi_{\alpha'}(t') \\ \phi^*_{\alpha'}(t') \end{pmatrix} .
\end{align}
Denoting by ${\cal G}(\tau,\tau')$ the analytically-continued Green function, one then finds
\begin{equation}
\begin{split}
U_{fi}(\beta,0) &= {\cal N} e^{\frac{\eta}{2} \sum_{\alpha,\alpha'} \Phi^{\dagger}_{\alpha}
\bar L_{\alpha\alpha'}(\beta) \Phi_{\alpha'}} , \\
\bar L(\beta) &= I_0 \delta_{\alpha\alpha'} - I_0 [{\cal
G}_{\alpha\alpha'}(\tau=0^+,\tau'=0)]^{-1} I_0 ,
\end{split}
\label{unievol4}
\end{equation}
where ${\cal N}=(\det {\cal G}^{-1})^{-\eta/2} (\det {\cal
G}(0^+,0))^{-\eta/2}$. The two component field $\Phi_\alpha$ and the
matrix $I_0$ are defined in Eq.~(\ref{quadres2a}). The temperature
dependence of $\bar L(\beta)$ comes {\it via} the
dependence of $\cal G$ on $\beta$.  The analytic continuation to
real time finally gives
\begin{eqnarray}
U_{fi}(t,0) &=& {\cal N} e^{\frac{\eta}{2} \sum_{\alpha\alpha'} \Phi^{\dagger}_{\alpha}
L_{\alpha\alpha'}(t) \Phi_{\alpha'} }
\label{unievol5}
\end{eqnarray}
where $L(t) = \bar L(\beta=i t)$. We note that
the field $\Phi_\alpha$ mixes $\phi_i$ and $\phi_f$. In
Sec.~\ref{rpcs} we shall see how one can express the evolution
operator in a more natural way in terms of
$\Phi_{i\alpha}=(\phi_{i\alpha},\phi^*_{i\alpha})^T$ and
$\Phi_{f\alpha}^\dagger=(\phi_{f\alpha}^*,\phi_{f\alpha})$.

Next, we note a couple of features of our computation. First, we
point out that the computation of $U_{fi}$ does not require the use
of a Keldysh contour even though we are addressing the dynamics of a
non-equilibrium system with time dependent Hamiltonian. This can be
understood as follows. A computation of $U_{fi}$ involves matrix
element of $\hat U$ between two {\it different} coherent states. It
does not require these two states to be identical; consequently, we
do not need to construct a contour for implementing this
restriction. Second, our method requires an analytic continuation to
real time at the end of the calculation. This can be trivially done
if ${\cal G}(0^+,0)$ is analytically known; however, it is a
significant challenge to carry this out numerically. This
constitutes one of the main difficulties concerning the application
of our formalism to non-equilibrium systems.

To demonstrate the application of the above-mentioned steps, in
Appendix~\ref{app1b} we consider a single degree of freedom with a
time-dependent Hamiltonian. This example demonstrates, albeit for a
very simple model Hamiltonian, that the analytic continuation
between real and imaginary times along with the formalism developed
in Sec.\ \ref{genform1} allows us to compute matrix elements of the
evolution operator between any two arbitrary coherent states in a
driven system without using Keldysh formalism. Indeed for time
evolution following sudden quenches an analogous formulation for 1D
quantum systems has been presented in Ref.~\onlinecite{cardy2}. Our
formulation generalizes the work of Ref.~\onlinecite{cardy2} to
arbitrary quench protocols and for quantum systems in arbitrary
dimensions. However, we note that the formalism is only effective
when the time-dependent Green function of the system is analytically
known. Thus it can be most easily used in practice for piecewise
constant drive protocol such as periodic square pulses or kicks,
where the time ordering involved in the definition of $\hat U$ does
not preclude analytical treatment. We shall use this formalism to
compute experimentally relevant quantities of periodically driven
integrable many-body systems in Sec.\ \ref{rpcs}.

\section{Bose-Hubbard Model}
\label{bhmodel}

In this section, we shall apply the results obtained in Sec.\
\ref{genform2} to compute several entanglement measures of the
two-dimensional Bose-Hubbard model which can be described by a
quadratic action within a strong coupling mean-field theory
\cite{dupuis1}. In what follows, we shall obtain entanglement
measures for this model {\color {blue} on a bipartite 2D square
lattice} both in its pristine form and in the presence of artificial
gauge fields \cite{kush1,sinha1}. The former topic shall be
discussed in Sec.\ \ref{bhm1} while the latter shall be addressed in
Sec.\ \ref{bhm2}.

\subsection{Entanglement for the pristine Bose-Hubbard model}
\label{bhm1}

The Hamiltonian of the Bose-Hubbard model is given by $\hat H=\hat
H_0 +\hat H_1$,\cite{bhrefs}
\begin{equation}
\begin{split}
\hat H_1 &= - t \sum_{\langle {\bf r} {\bf r}'\rangle} \left( \hat
b_{\bf
r}^{\dagger} \hat b_{{\bf r}'} + {\rm h.c.} \right) , \\
\hat H_0 &= \sum_{{\bf r}} \left[- \mu \hat b^\dagger_{{\bf r}}
\hat b_{{\bf r}} + \frac{U}{2}\hat b^\dagger_{{\bf r}} \hat b_{{\bf
r}}  (\hat b^\dagger_{{\bf r}} \hat b_{{\bf r}}-1)
\right] ,
\end{split}
\label{bham}
\end{equation}
where $\hat b_{{\bf r}}$ denotes the boson anihiliation operator at
site ${\bf r}$, $t$ is the nearest-neighbor hopping amplitude for
the bosons, $U$ is their on-site interaction potential, and $\mu$ is
the chemical potential. For $t/U \ll 1$, the ground state of the
model is a Mott insulating state of bosons. At $t=t_c$, the bosons
undergo a superfluid-insulator transition and the ground state for
$t>t_c$ is a correlated superfluid. The precise value of $t_c/U$ is
well-known from quantum Monte-Carlo (QMC) studies \cite{qmcref1}; it
turns out that $t_c/U \ll 1$ {\color {blue} for $d>1$}. Thus the
study of the Mott and the superfluid phases near the transition
point requires addressing the properties of the model in the
strongly correlated regime where $U \gg t$.

An analytic treatment of the Bose-Hubbard model usually involves
standard mean-field theory where the kinetic energy term is treated
within mean-field approximation \cite{mftref}, projection operator
approach \cite{projref}, and slave boson technique \cite{sbosonref}.
All of these methods use the local nature of the interaction and
involves treatment of the kinetic energy term using different
approximations. However, none of them allows for a direct access to
the momentum-space properties near the superfluid-insulator critical
point, either in the Mott or the superfluid phases. Such information
is particularly relevant for computation of the
momentum-distribution function which can be directly measured
experimentally \cite{bhmexp1}. In contrast, the strong coupling
expansion technique, which was developed in Ref.\
\onlinecite{dupuis1} and applied to Bose-Hubbard model in the
presence of artificial gauge fields in
Refs.~\onlinecite{kush1,sinha1}, provides a direct access to the
momentum-space Green function in the strong coupling regime. The
momentum distribution function computed using this technique
provides a near-exact match with experimental measurements carried
out in the superfluid and Mott phases near the transition
\cite{bhmexp1}. In what follows, we shall use this technique to
obtain the action of the bosons and use it to obtain several
entanglement measures both in the Mott and the SF phases near the
critical point.

It was shown in Ref.\ \onlinecite{dupuis1} that the effective action
for the Bose-Hubbard model can be written at zero temperature, after
a series of Hubbard Stratonovitch transformations, as $S= S_0 +S_1$,
where
\begin{eqnarray}
S_0 &=& \frac{1}{\beta} \sum_{{\bf k},\omega_n}
\phi^{\ast}(\omega_n, {\bf k}) \left[- G_0^{-1}(\omega_n) +
\epsilon_{{\bf k}} \right] \phi(
\omega_n, {\bf k}) , \nonumber\\
S_1 &=& \frac{g_0}{2} \int_0^\beta d \tau \int d^d r  |\phi({\bf
r},\tau)|^4 , \label{bhac1}
\end{eqnarray}
where $\epsilon_{{\bf k}}=-2t(\cos k_x+\cos k_y)$ is the boson
kinetic energy, $g_0$ is the coefficient of the quartic interaction term,
and $G_0$ is the local single-particle zero-temperature
Green function in the Mott limit ($t=0$) given by
\begin{eqnarray}
G_0(\omega_n) &=& \frac{-n_0}{i\omega_n +E_{h}(n_0)} +
\frac{n_0+1}{i\omega_n -E_p(n_0)} .
\label{blocgreen}
\end{eqnarray}
Here $n_0 \equiv n_0(\mu/U)$ is the ground state boson occupation
number in the Mott limit and $E_p(n_0)= -\mu + U n_0$ and $E_h= \mu
-U(n_0-1)$ denotes the energy cost of adding (removing) a boson to
(from) the ground state at $t=0$. In what follows, to comply with
the notations of Ref.~\onlinecite{dupuis1}, we define the Green
function as $G_0(\omega_n)=-\mean{\psi(\omega_n)\psi^*(\omega_n)}$
for the local propagator. Following Ref.~\onlinecite{dupuis1} we
approximate $S_1$ within a mean-field theory as
\begin{equation}
S_1^{\rm MF} =  \frac{g_0\Delta_0^2}{2} \int_0^\beta d \tau \int d^d r
[ \phi({\bf r},\tau)^2 + \phi^*({\bf r},\tau)^{2} + 4 |\phi({\bf
r},\tau)|^2 ]
 \label{mfteq1}
\end{equation}
where $\Delta_0^2=\mean{\phi({\bf r},\tau)}^2 = [G_0^{-1}(0) +z
t]/g_0$ in the superfluid phase and vanishes in the Mott phase, and
$z=2d=4$ is the coordination number of the square lattice. We note that the
mean-field theory used here differs from its standard weak-coupling
counterpart since $S_0$ captures the effect of strong correlation
through $G_0$. The strong coupling mean-field theory therefore
allows one to describe both the superfluid and the Mott phases on
the same footing by using a Gaussian action.

In the Mott phase, $S_1^{\rm MF}=0$ and the action of the bosons is
given by $S_0$ [Eq.~(\ref{bhac1})]. The Green function of the bosons
can be read off from~(\ref{bhac1}) and is given by
\begin{eqnarray}
&& G(\omega_n,{\bf k}) = \frac{z_{{\bf k}}}{i \omega_n - E_{{\bf
k}}^+} + \frac{1-z_{\bf
k}}{i \omega_n - E_{{\bf k}}^-} , \nonumber \\
E_{{\bf k}}^{\pm} &=&  -\mu_0 + \frac{1}{2} \left(\epsilon_{{\bf k}}
\pm \sqrt{\epsilon_{{\bf k}}^2+ U^2 + 4 U \epsilon_{{\bf k}}
(n_0+1/2)}\right)  ,\nonumber\\
z_{{\bf k}} &=& [E_{{\bf k}}^+ + \mu_0 +U(n_0+1/2)]/(E_{{\bf k}}^+ -
E_{{\bf k}}^-)  , \label{mgreen}
\end{eqnarray}
where $\mu_0= \mu-U(n_0-1/2)$. A straightforward calculation thus
yields $G(\tau=0^+,{\bf k}) =z_{{\bf k}}$. Using the results of
Sec.~\ref{genform1} we therefore find
\begin{eqnarray}
\rho_{fi}^{\rm BH}  &=&  K e^{\sum_{{\bf k}} \phi_{f {\bf k}}^{\ast}
(1-z_{{\bf k}}^{-1}) \phi_{i {\bf k}}} , \label{bhmatel1}
\end{eqnarray}
where $K=[\det\,G(0^+)]^{-1}=\prod_{\bf k}z_{\bf k}^{-1}$, and
\begin{eqnarray}
\langle \phi_{{\bf k}} \phi_{{\bf k}}^{\ast} \rangle &=& z_{{\bf k}}
= \lambda_{{\bf k}}^{\rm Mott} . \label{corrmot}
\end{eqnarray}
From Eq.~(\ref{corrmat}), one can now construct the covariance
matrix $\Lambda_{jj'}$. A numerical diagonalization of this matrix
yields the eigenvalues $\nu_{\ell}^{\rm Mott}$. The von Neumann
entropy $S_b$ and the $n^{\rm th}$ R\'enyi entropy can then be
computed using Eqs.~(\ref{bosonvn}) and (\ref{renyiexp})
respectively. One can also compute the entanglement negativity by
numerically diagonalizing the partial transposed density
matrix~(\ref{ptdm2}) and then using~(\ref{pten1}). We emphasize that
all the measures of entanglement, within the strong-coupling
mean-field theory, can be directly linked to $z_{{\bf k}}$ and thus
in turn to the Matsubara Green functions of the bosons.
\begin{figure}
\vspace{0.1 in} \centering {\ing[width=0.8 \linewidth]{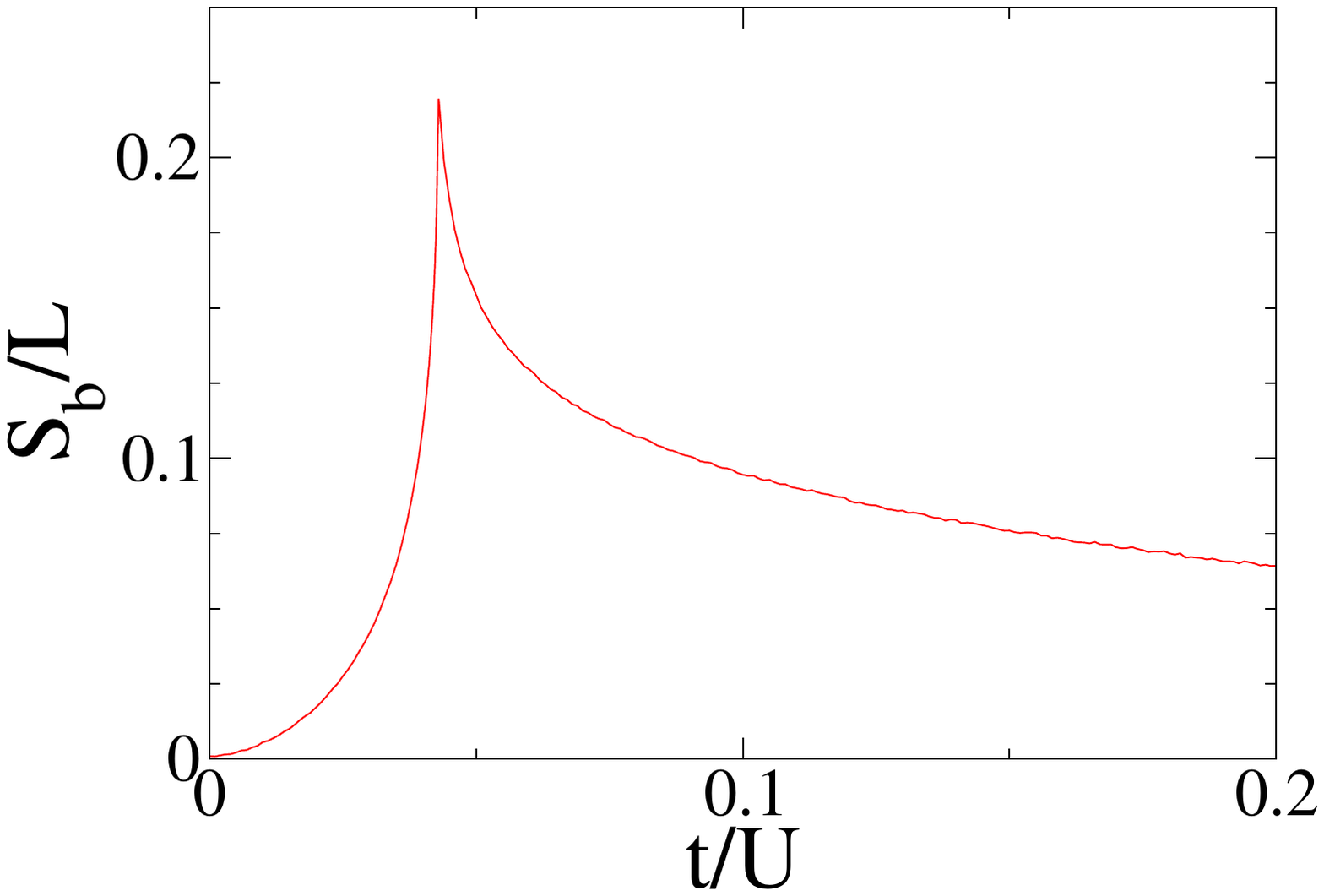}}
\caption{Plot of $S_b$ as a function of $t/U$ for the Bose-Hubbard
model. We have considered a cylindrical subsystem with $L \equiv
L_x/a=30$ and $\mu/U = 2t/U+(n_0-1/2)$ is chosen so that we follow a
line of constant boson density as we vary $t/U$ and access the
superfluid phase at the tip of the Mott lobe. See text for details.}
\label{fig1a}
\end{figure}
\begin{figure}
\centering {\ing[width=0.49 \linewidth]{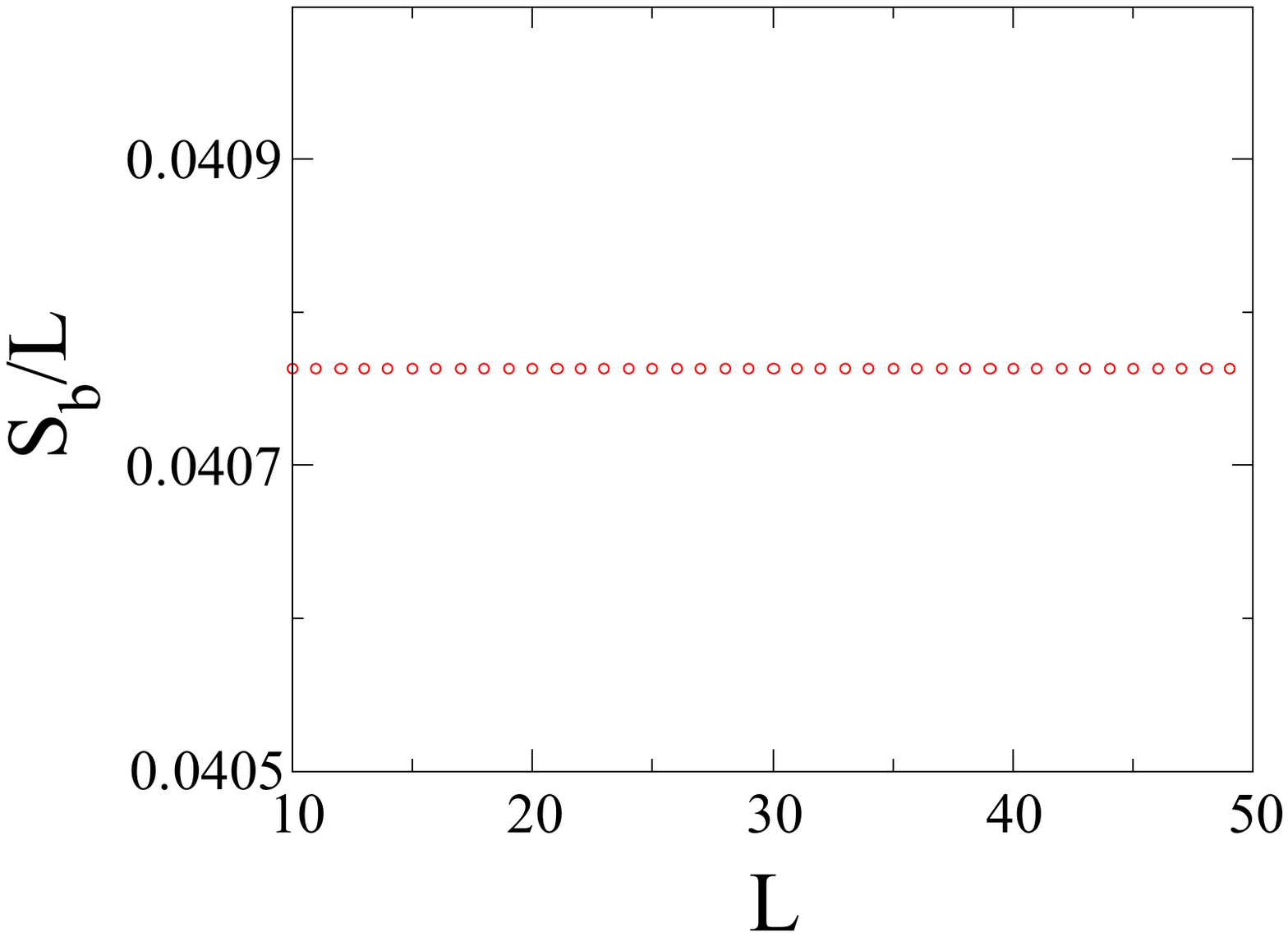}}
\centering{\ing[width=0.49 \linewidth]{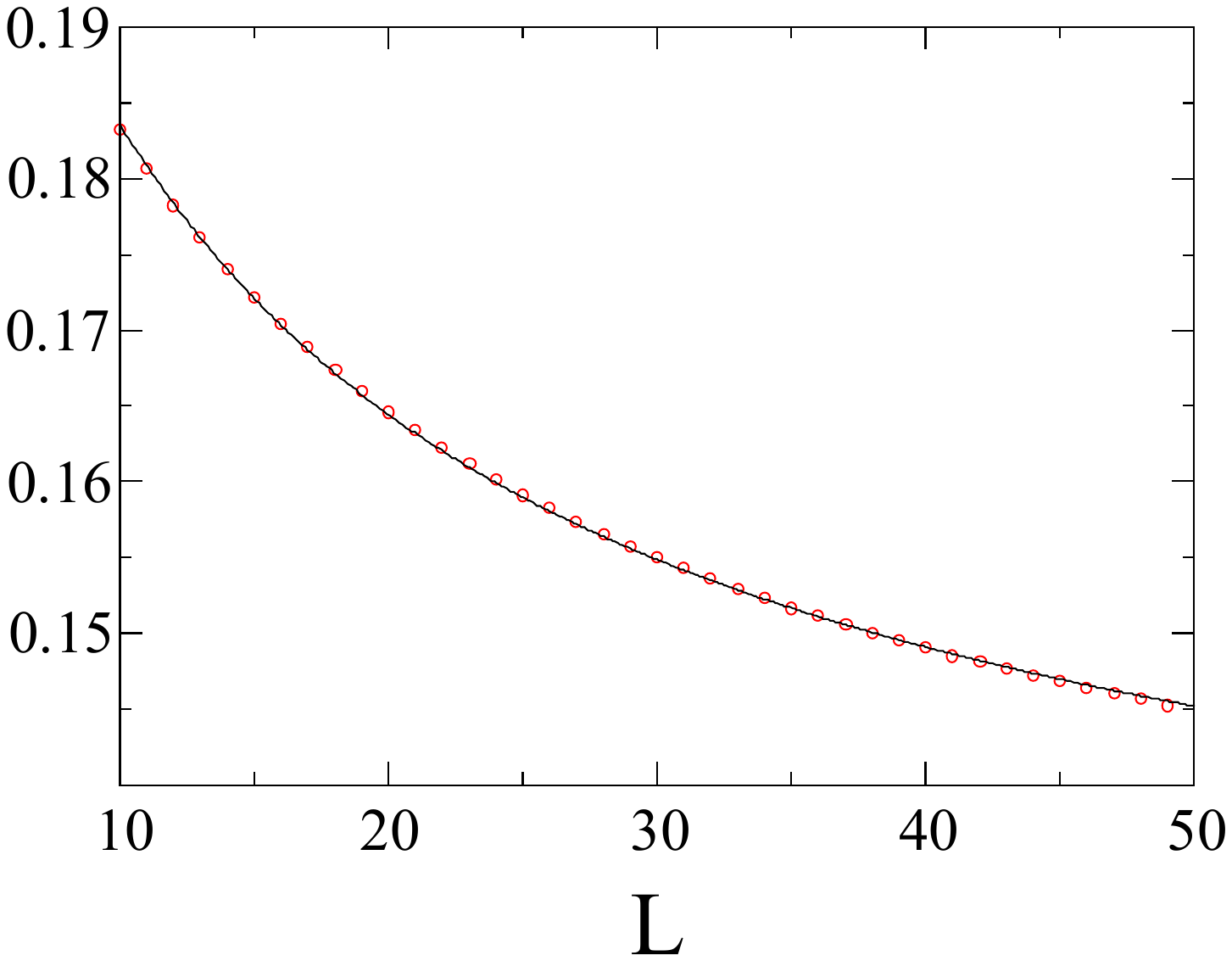}} \caption{Plot
of $S_b$ as a function of the subsystem size $L_x/a \equiv L$ in the
Mott (left panel) and superfluid (right panel) phases. The left
panel correspond to $t/U=0.03$ while the right panel to $t/U=0.05$.
All other parameters are same as in Fig.\ \ref{fig1a}. The red dots
correspond to numerical data points and the black line is generated
from a fitting function $S_b=0.122 L_x/a + 0.335 \ln(L_x/a)$ for the
right panel. See text for details.} \label{fig1b}
\end{figure}
In the superfluid phase, the boson action is approximated by the
quadratic action $S=S_0+S^{\rm MF}_1$. The Green function
reads\cite{dupuis1}
\begin{widetext}
\begin{eqnarray}
G^s(\omega_n, {\bf k}) &=& \frac{[i \omega_n +\mu_0
+U(n_0+1/2)](i\omega_n -z_{{\bf k}}^+)(i \omega_n -z_{\bf
k}^-)}{(\omega_n^2+E_{{\bf k}}^{+ 2})(\omega_n^2+E_{{\bf k}}^{-
2})}, \quad  F^s(\omega_n, {\bf k}) =  g_0 \Delta_0^2 \frac{
\omega^2_n + [\mu_0 + U(n_0+1/2)]^2}
{(\omega_n^2+E_{{\bf k}}^{+ 2})(\omega_n^2+E_{{\bf k}}^{- 2})}\nonumber\\
{E_{{\bf k}}^{\pm}}^2 &=& -\frac{B_{{\bf k}}}{2} \pm \frac{1}{2}\sqrt{B_{{\bf
k}}^2- 4 C_{{\bf k}}} , \quad
z_{{\bf k}}^{\pm} = \frac{\tilde
A_{\bf k}}{2} \pm \frac{1}{2}\sqrt{\tilde
A_{{\bf k}}^2-4 \tilde B_{{\bf k}}} , \quad \tilde A_{{\bf k}} = 2 \mu_0 -\epsilon_{{\bf k}} - 2g_0 \Delta_0^2  \nonumber\\
\tilde B_{{\bf k}} &=& \mu_0^2 -U^2/4 -[\mu_0+
U(n_0+1/2)](\epsilon_{ {\bf k}}+ 2g_0 \Delta_0^2 ), \quad B_{{\bf
k}} = 2\tilde B_{{\bf k}}
- \tilde A^2_{{\bf k}} + g_0^2 \Delta_0^4, \nonumber\\
C_{{\bf k}} &=&  \tilde B_{{\bf k}}^2 - g_0^2 \Delta_0^4[\mu_0 +
U(n_0+1/2)]^2, \quad g_0 \Delta_0^2 =\frac{\mu_0^2
-U^2/4}{\mu_0+U(n_0+1/2)}+zt ,
\label{sfgreen}
\end{eqnarray}
\end{widetext}
where $G^s$ and $F^s$ denotes the diagonal and off-diagonal parts of$C_{jj'}=\langle \hat
c_j^{\dagger} \hat c_{j'} \rangle = [G(0^-)]_{jj'}$.
the boson Green function ${\cal G}^{\rm SF}$ in the superfluid phase
and $E_{{\bf k}}^{\pm}$ is chosen to be a positive quantity for all
${\bf k}$. The Green function ${\cal G}^{\rm SF}$, computed within
the strong coupling mean-field analysis, has four poles at $\pm
E_{{\bf k}}^{\pm}$ corresponding to two different types of mode: a
gapless sound mode $\omega=\pm E^-_{\bf k}$ with a linear dispersion
at small ${\bf k}$ and a gapped Higgs mode $\omega=\pm E^+_{\bf
k}$.\cite{dupuis1} We note that the latter is absent in a simple
Bogoliubov theory in the weak coupling regime. The corresponding
boson correlation functions in momentum space are given by
\begin{eqnarray}
C_{{\bf k}}^{\rm SF} &=& \langle \psi_{{\bf k}} \psi_{\bf
k}^{\ast}\rangle = 1 - q_{{\bf k}}^+ - q_{{\bf k}}^-  , \nonumber\\
F_{{\bf k}}^{\rm SF} &=& \langle \psi_{{\bf k}} \psi_{-\bf k}\rangle
= -g_0 \Delta_0^2 \left( \frac{1-p_{{\bf k}}}{2 E_{{\bf k}}^-}
+ \frac{p_{{\bf k}}}{2E_{{\bf k}}^+} \right) ,  \nonumber\\
q_{{\bf k}}^{\sigma} &=& \frac{[E_{{\bf k}}^{\sigma}-\mu_0 -
U(n_0+1/2)](E_{{\bf k}}^{\sigma}+z_{{\bf k}}^+)(E_{{\bf
k}}^{\sigma}+ z_{{\bf k}}^-)}{2E_{{\bf k}}^{\sigma }(E_{{\bf
k}}^{\sigma  2}-E_{\bf k}^{\bar \sigma
2})} , \nonumber\\
p_{{\bf k}} &=& \frac{E_{{\bf k}}^{+ 2}-[\mu_0 +
U(n_0+1/2)]^2}{E_{{\bf k}}^{+ 2}-E_{{\bf k}}^{- 2}} ,
\label{corrsfeq}
\end{eqnarray}
where $\sigma=\pm$. From these equations, following the method of
Sec.~\ref{genform2}, one easily finds
\begin{equation}
\lambda_{{\bf k}}^{\rm SF} = \sqrt{{C_{{\bf k}}^{\rm SF}}^2 -
{F_{{\bf k}}^{\rm SF}}^2} . \label{eigensf1}
\end{equation}
In the absence of superfluidity, $\Delta_0=0$ and $F_{\bf k}=0$, we
recover the expression obtained in the Mott phase: $\lambda_{\bf
k}=C_{\bf k}=\mean{\phi_{\bf k}\phi^*_{\bf k}}$. We are now in a
position to construct the covariance matrix~(\ref{corrmat}), find
its eigenvalues $\nu^{\rm SF}_\ell$ and deduce the VonNeuman and
R\'enyi entropies as well as the entanglement negativity in the SF
phase (see Sec.~\ref{genform2}).

\begin{figure}
\centering {\ing[width=\linewidth]{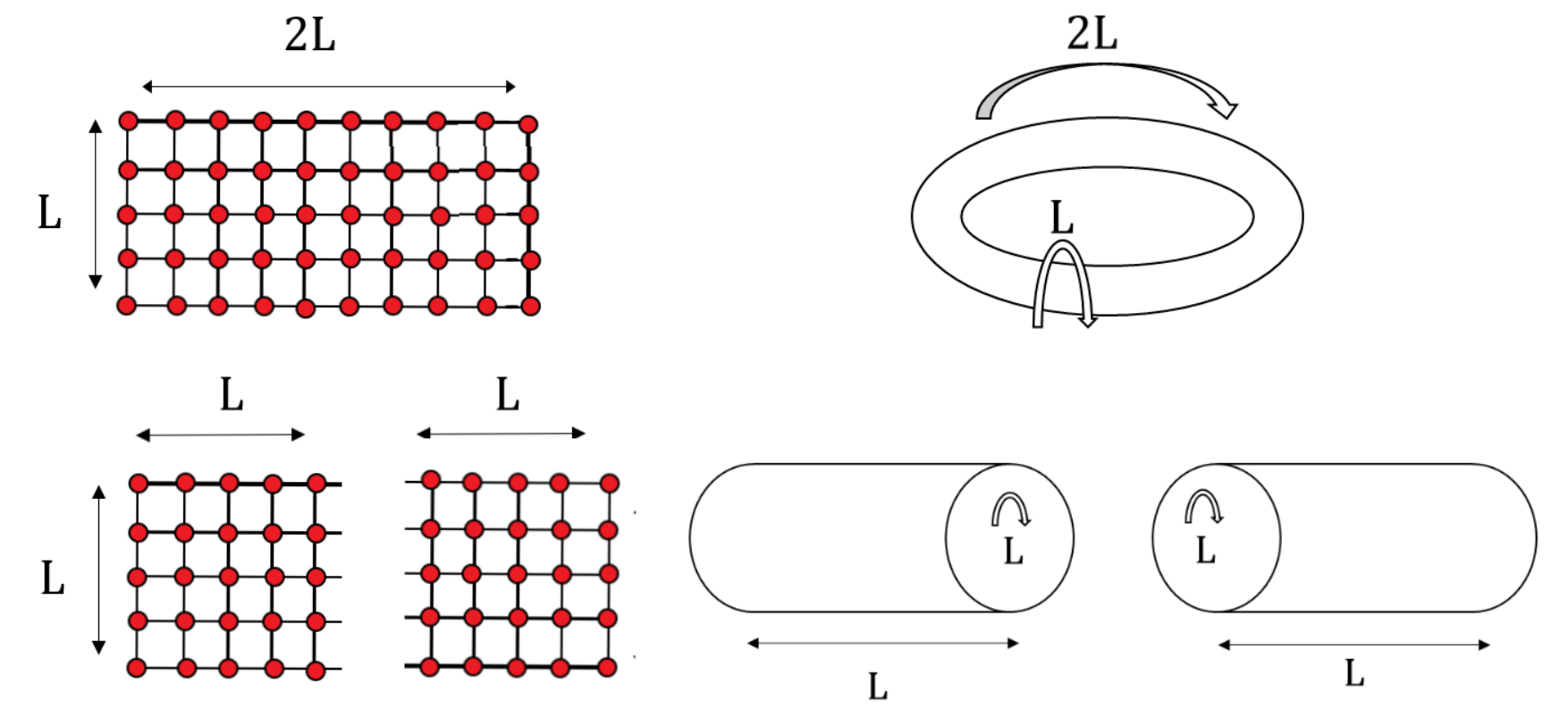}} \caption{Geometry of
the Bose-Hubbard model. The left shows the square lattice
configuration of the full system (above) and the subsystems (below).
The same with periodic boundary condition used in this work is shown
in the right side for the full system (above) and the subsystem
(below). In this work, we have chosen $R=L=30a$, where $a$ denotes
the lattice spacing and $R$ is the circumference of the cylinder.
See text for details.} \label{figzero}
\end{figure}

A plot of the VonNeuman entropy of the pristine bosons, $S_b$,
obtained from the method detailed above is shown in Figs.\
\ref{fig1a} and \ref{fig1b}. As shown in Fig.\ \ref{figzero}, we
have chosen a subsystem in the form of a cylindrical region with
circumference $R=L_y/a=30$; this ensures one has periodic boundary
condition along $y$. The length of the cylinder $L_x = L a$ is
chosen to be $L=30$ ($a$ is the lattice spacing), where the system
size corresponds to $L_x=2 L_y=60 a$. The chemical potential $\mu/U$
for all the plots is chosen so that we follow a line of constant
boson density as we change $t/U$ \cite{dupuis1}. We find that our
results reproduce the expected behavior of $S_{b}$ as a function of
$t/U$. $S_b$ peaks at the critical point and obeys an area law, \beq
\begin{split}
S_b = \left\{
\begin{array}{ll}
    A  \left(\frac{L_x}{a}\right) & \mbox{(Mott phase),} \\
    A' \left(\frac{L_x}{a}\right) + B' \ln \left(\frac{L_x}{a}\right) & \mbox{(superfluid phase),}
\end{array}
\right.
\end{split}
\label{Sb}
\eeq
in accordance with standard expectation.

\begin{figure}
    \centering {\ing[width=0.8\linewidth]{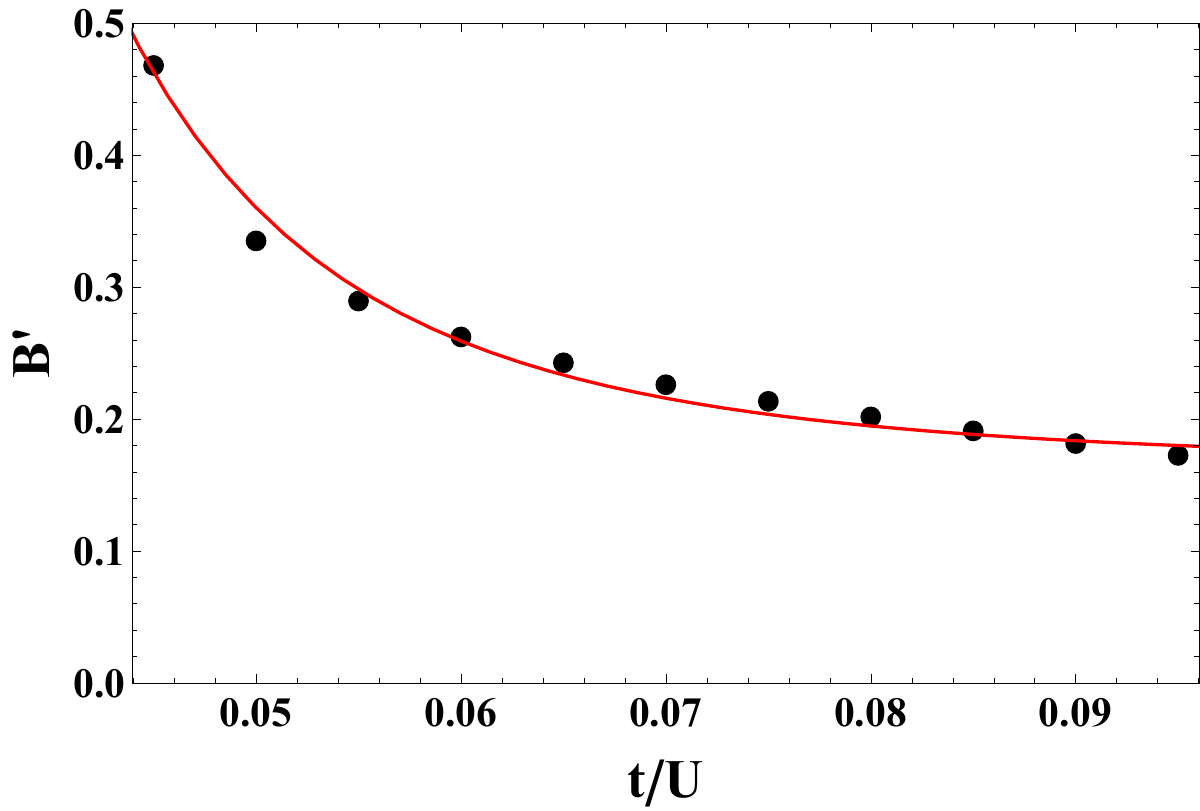}} \caption{Plot of $B'$
        as a function of $t/U$ in the superfluid phase. All other parameters are same in Fig.\ \ref{fig1a}. The
        black dots indicate numerical data points while the red line denotes
        the fit $B'=c_1 + c_2 (U/t)^4$ with $c_1\simeq 0.165$ and $c_2 \simeq
        1.23 \times 10^{-6}$. We find that $\chi^2=0.99$ for the fit. See text for
        details.} \label{fig1c}
\end{figure}

\begin{figure}
    \centering {\ing[width=0.8\linewidth]{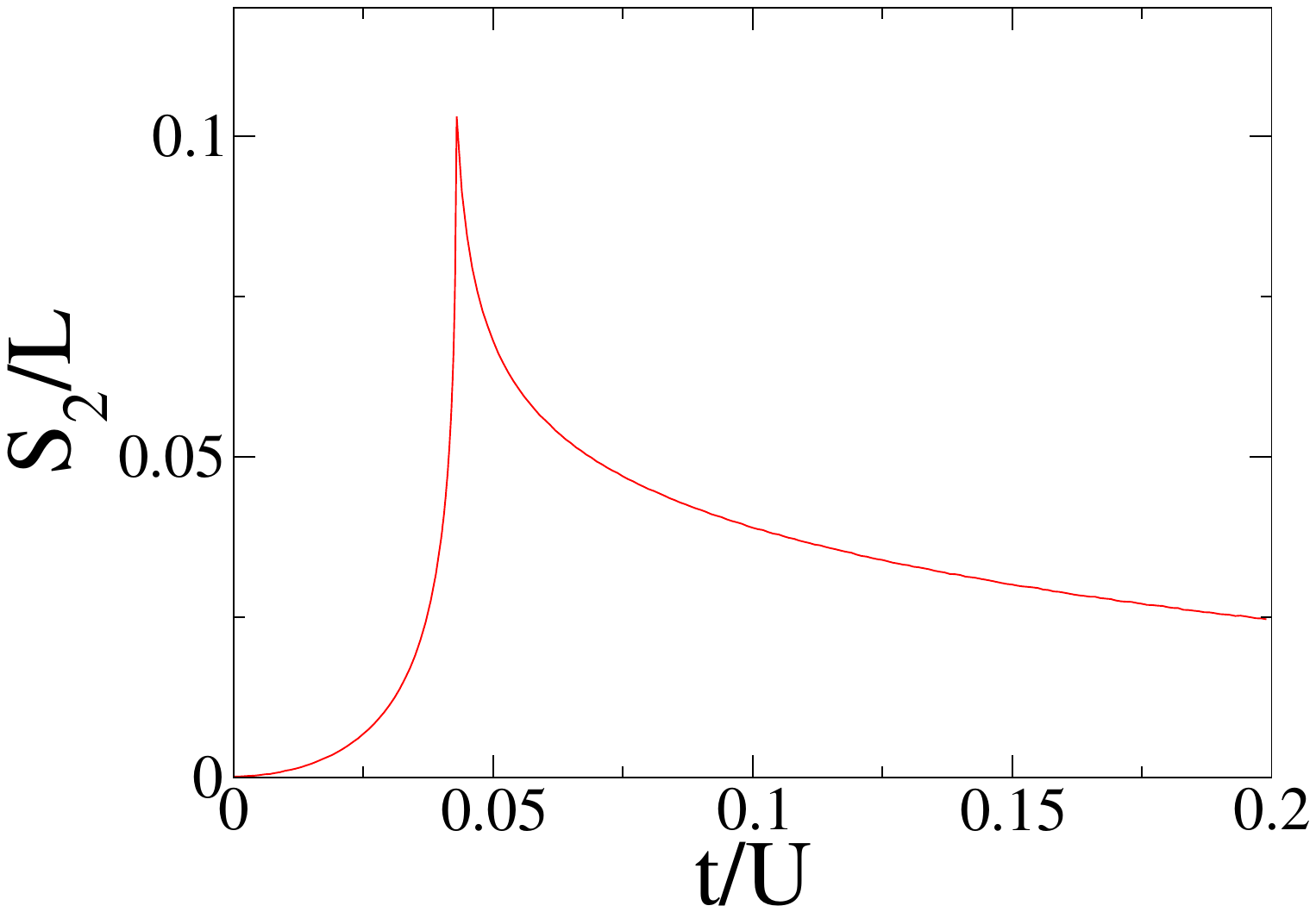}} \caption{Plot of the
        second R\'enyi entropy as a function of $t/U$ for $L \equiv L_x/a=30$.
        We find that the behavior of $S_2$ is qualitatively similar to
        $S_b$. All other parameters are same as in Fig.\ \ref{fig1a}. See
        text for details.} \label{fig1d}
\end{figure}

Our work here qualitatively agrees with that of Refs.\
\onlinecite{bentang1,entro1} with one important difference. This
difference can be understood from Fig.\ \ref{fig1c} where the
subleading part of $S_b$ in the superfluid phase is plotted as
function of $t/U$. It is well-known that in a superfluid whose
excitations are given by Goldstone modes, $S_b = A' (L_x/a) + B' \ln
(L_x/a)+ ...$ where the ellipsis denotes $O(1/L)$ terms as shown in
the right panel of Fig.\ \ref{fig1b}. Moreover, within the gapless
Bogoliubov approximation, $b'$ is expected to be a universal
constant independent of $t/U$. However, in our case, for correlated
superfluid near the critical point which has both gapped and gapless
modes, $B'$ turns out to be a monotonic function of $t/U$; its
dependence on $t/U$ is shown in Fig.\ \ref{fig1c}. We note that the
deviation of $B'$ from its universal value {\color {blue} (at the
critical point where $B'\simeq 1/2$ in the absence of any gapped
mode)} is a signature of the presence of gapped modes in the system
which arise due to strong correlations.

\begin{figure}
\centering {\ing[width=0.51 \linewidth]{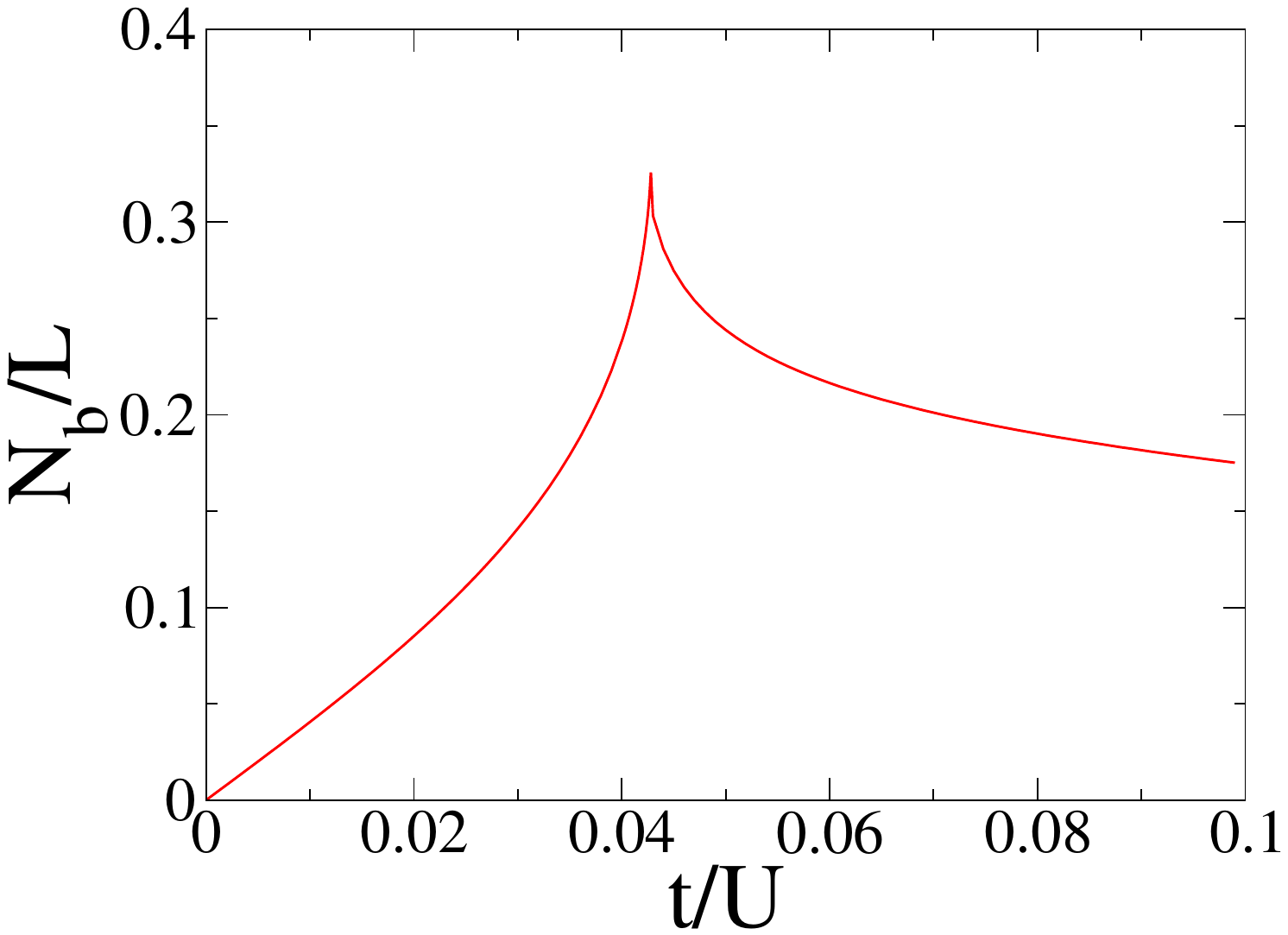}} \centering
{\ing[width=0.47 \linewidth]{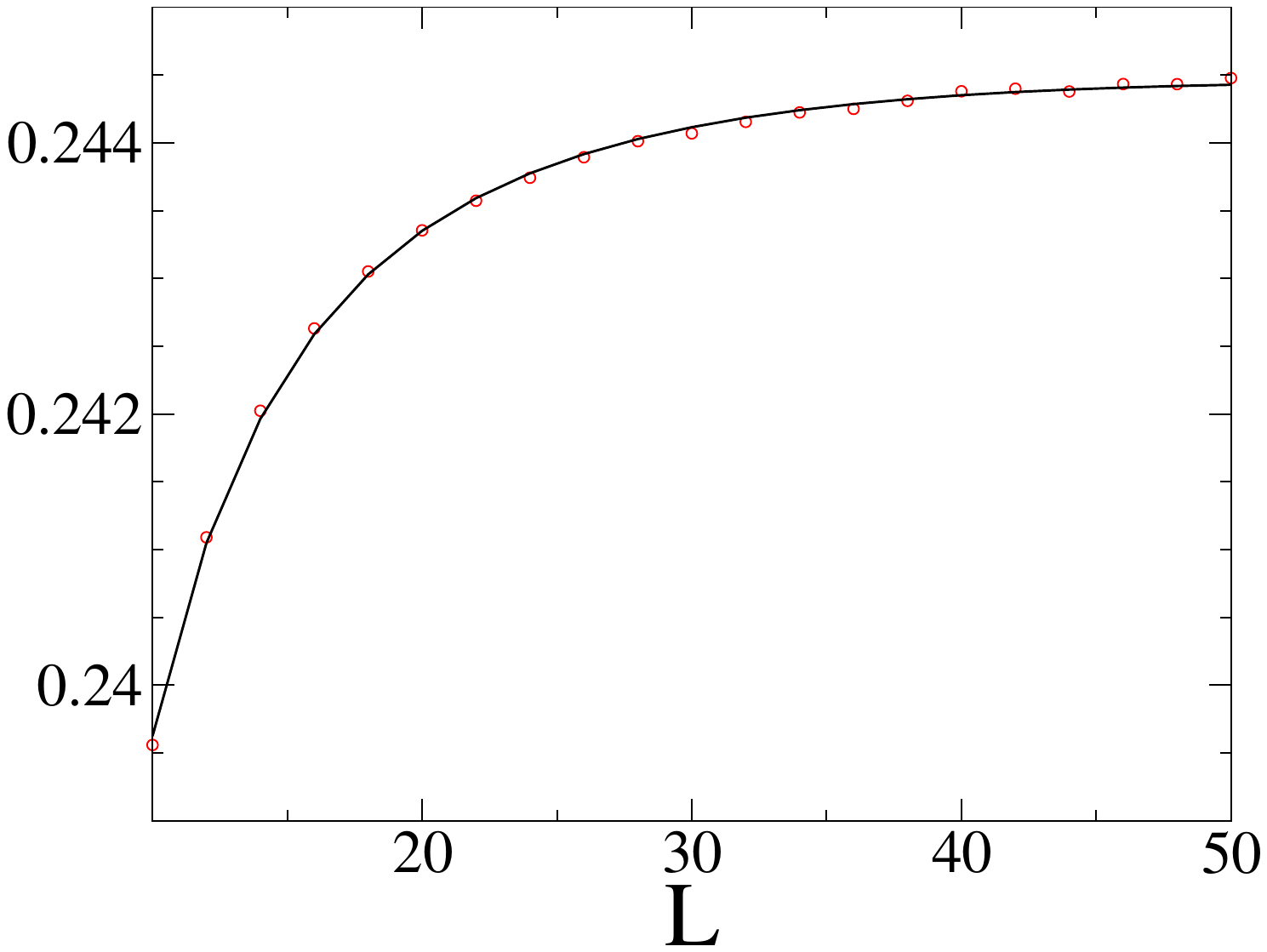}} \caption{Left Panel: Plot
of the entanglement negativity $N_b$ as a function of $t/U$ for $L
\equiv L_x/a=30$. Right Panel: The functional dependence of $N_b$ as
a function of $L_x$ is found to be much weaker and opposite to that
of $S_b$ as shown for $t/U=0.05$. The red dots correspond to
numerical data points while the black line is the generated from a
fitting function $N_b= 0.244 L_x/a + 0.05 \ln(L_x/a)$. All other
parameters are same as in Fig.\ \ref{fig1a}. See text for details.}
\label{fig1e}
\end{figure}

Next, we plot the second R\'enyi entropy $S_2$ as a function of
$t/U$ in Fig.\ \ref{fig1d}. We find that $S_2$ has qualitatively
similar feature as the von Neumann entropy and peaks at the
transition. This behavior can in principle be verified in
experiments since $S_2$ has been experimentally measured for bosonic
systems \cite{rislam1}. Finally, in Fig.\ \ref{fig1e}, we plot the
entanglement negativity $N_b$ of the Bose-Hubbard model as a
function of $t/U$ both in the SF and MI phases. We find, from the
left panel of Fig.\ \ref{fig1e}, that $N_b$ also exhibits a peak at
the critical point and displays the expected decay to zero as we
approach the Mott limit $t/U\to 0$. Its behavior as a function of
$t/U$ is found to be similar to that of $S_b$ and $S_2$. However,
the subsystem size ($L_x$) dependence of $N_b$ is found to be
opposite to and much weaker than that of $S_b$ as can be seen by
comparing the right panels of Figs.\ \ref{fig1b} and \ref{fig1e}.

\subsection{Abelian gauge fields}
\label{bhm2}

In this section, we shall discuss several entanglement measures for
the Mott phases of strongly correlated bosonic systems in the
presence of artificial Abelian gauge fields. These fields are
typically generated using additional Raman lasers coupled to the
bosons \cite{gaugeref} and can be either Abelian or non-Abelian. In
this subsection, we  shall discuss the Bose-Hubbard Hamiltonian for
the bosons in a 2D optical lattice the presence of an Abelian gauge
field; the case of non-Abelian gauge fields will be discussed in
Sec.\ \ref{bhm2na}. It is well-known that such a Hamiltonian is
given by \cite{gaugeref,sinha1}
\begin{eqnarray}
\hat H &=& \hat H_0 - \sum_{\langle {\bf r} {\bf r}'\rangle} \left(
{\hat b}_{\bf r}^{\dagger} {\hat b}_{{\bf r}'} t e^{i e^{\ast}
\int_{{\bf r}}^{{\bf r}'} d{\vec l} \cdot {\vec A} /\hbar c} + {\rm
h.c.} \right) . \label{bhamab}
\end{eqnarray}
Here $\vec A_{{\bf r}}= B_0(0,x_i) $ (where ${\bf r}_i= (x_i,y_i)$)
is the synthetic vector potential and $e^{\ast} B_0$ is the
effective magnetic field whose strength can be tuned by varying the
intensity of the Raman lasers \cite{gaugeref}. In what follows, we
shall treat the cases for which the effective magnetic field
corresponds to flux of $p \Phi_0/q$ through the lattice:
$e^{\ast}B^{\ast} a^2/(h c) = 2 \pi p/q$, where $\Phi_0=hc/e^{\ast}$
is the flux quanta and $c$ denotes speed of light.
We shall also choose $p/q$ to be a rational fraction. We note that the
creation of synthetic gauge fields amounts to
creation of the vector potential $\vec A_{{\bf r}}$ using light-atom
coupling; this is in contrast to standard electromagnetic fields
where the fields are the physical entities and choosing vector
potentials to describe them involves freedom of gauge choice.

\begin{figure}
    \centering {\ing[width=0.8\linewidth]{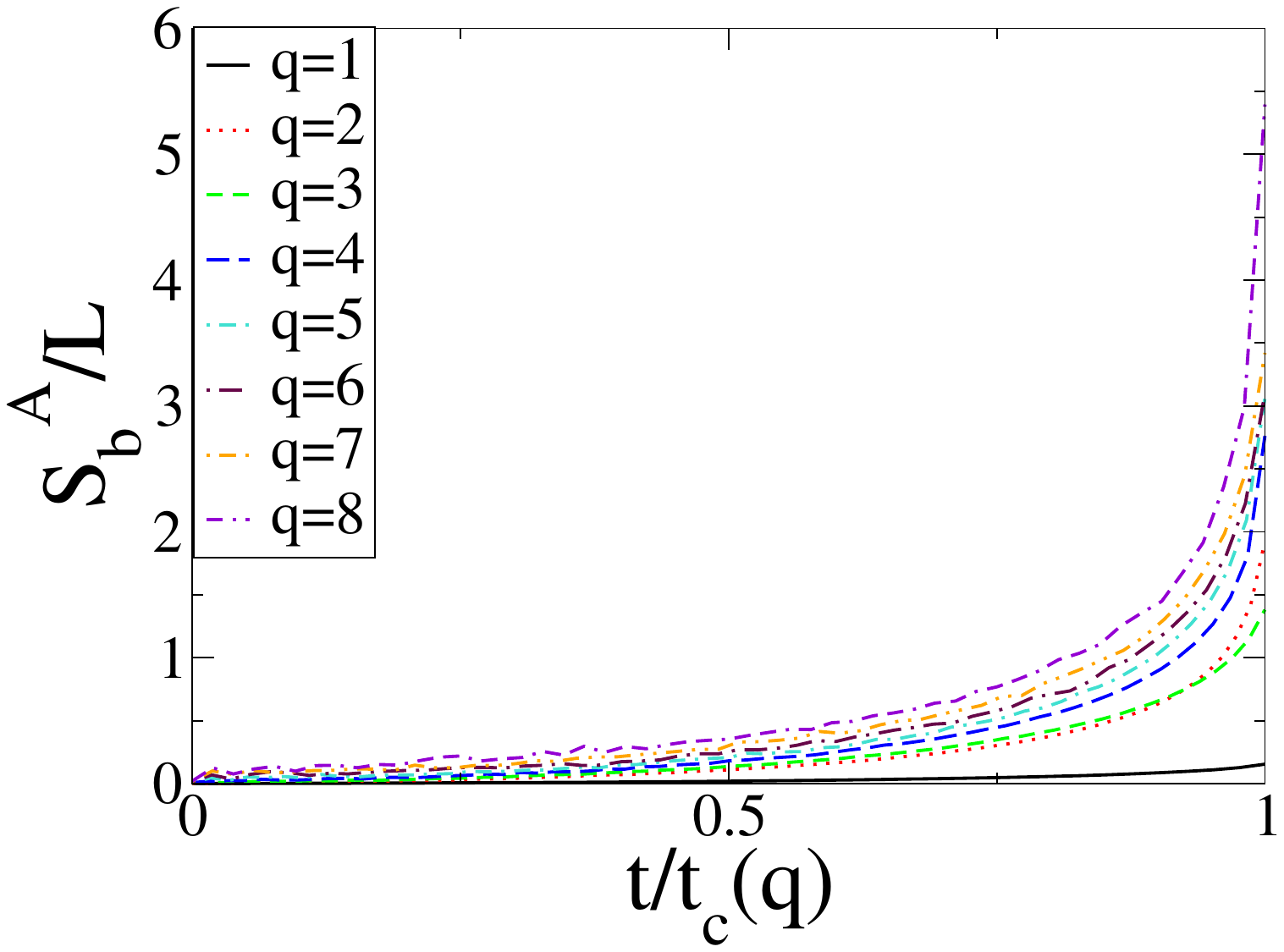}}\\ \centering
    {\ing[width=0.8 \linewidth]{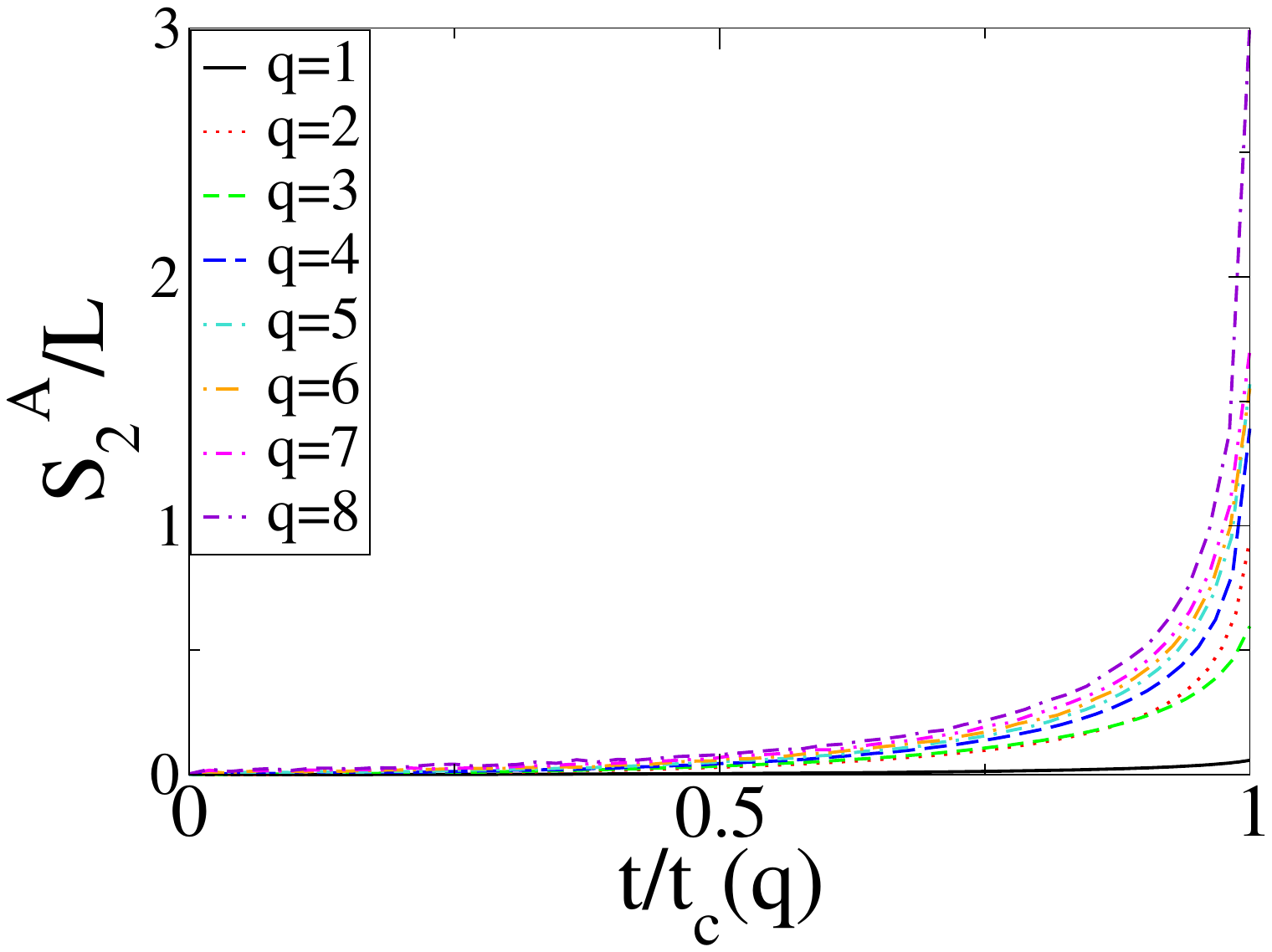}} \caption{
        Plot of the VonNeuman entropy $S_b^A$ (top panel) and the second
        R\'enyi entropy $S_2^A$ (bottom panel) in the presence of an
        Abelian gauge field as a function of $t/t_c(q)$ for
        $L \equiv L_x/a=30$. The case $q=1$ corresponds to zero effective magnetic
        field and reproduces the corresponding results for pristine bosons.
        All other parameters are chosen as in Fig.\ \ref{fig1}. See text for
        details.}
    \label{fig2a}
\end{figure}

The strong coupling expansion developed in Ref.\
\onlinecite{dupuis1} can be used to obtain the boson Green function
in their Mott phase in the presence of such Abelian gauge fields. As
shown in Ref.\ \onlinecite{sinha1}, the action of such a system can
be written in terms of a $q$ component bosonic field $\Psi(\omega_n,
{\bf k}) = (\psi_0(\omega_n,{\bf k}), \psi_1(\omega_n,{\bf k}) ...
\psi_{q-1}(\omega_n, {\bf k}))^T \equiv \psi(k)$ where
$\psi_{\alpha}(\omega_n,{\bf k})= \psi(\omega_n,{\bf k} + 2 \pi
\alpha/q \hat x)$. In the Mott phase, this action is given by
\begin{equation}
S_0 = \frac{1}{\beta} \sum_{\omega_n} \int \frac{d^2k}{(2\pi)^2}
\Psi^{\dagger}(k) \left[ - G_0^{-1}(\omega_n) I + J_q({\bf k}) \right] \Psi(k) .
\label{abelac}
\end{equation}
In this notation ${\bf k}= (k_x, k_y)$ lies within the magnetic
Brillouin zone defined by $-\pi/a \le k_y \le \pi/a$ and $-\pi/(qa) \le k_x \le
\pi/(q a)$,  $J_q({\bf k})$ is a $q \times q$ tridiagonal matrix
whose upper off-diagonal (diagonal) components are $-t e^{-i k_y a}
(-t \cos[k_x a + 2 \pi \alpha/q] )$ for $\alpha=0,1.. q-1$, and $I$
is the $q\times q$ unit matrix. The diagonalization of $J_q({\bf
k})$ leads to $q$ bands with energy dispersion
$\epsilon_{\alpha}^q({\bf k})$. For example, in the simplest case
where $q=2$ (corresponding to half flux quanta through each lattice
plaquette) there are two bands with energy dispersion
$\epsilon_{0(1)}^{(2)}( {\bf k})= +(-) t \sqrt{\cos^2 k_x + \cos
k_y^2}$. These bands lead to $q$ energy minima within the magnetic
Brillouin zone of the bosons. The Green function of the bosons in
the Mott phase can be written as $G(\omega_n, {\bf k}) = [G_0^{-1}
(\omega_n) I - J_q({\bf k})]^{-1}$. Noting that $G_0^{-1}$ is
independent of ${\bf k}$, one can write
\begin{eqnarray}
G(\omega_n, {\bf k}) &=& U_q({\bf k})^{\dagger} G_D(\omega_n,{\bf
k})
U_q({\bf k}) , \label{abelgreen} \\
G_D(\omega_n, {\bf k}) &=& \sum_{\alpha=0}^{q-1} \left(
\frac{1-z_{\alpha}^q({\bf k})}{i \omega_n -E_{\alpha}^{q-}({\bf k})}
+ \frac{z_{\alpha}^q({\bf k})}{i \omega_n -E_{\alpha}^{q+}({\bf k})}
\right) , \nonumber
\end{eqnarray}
where $U_k(q)$ is the matrix which diagonalizes $J_q({\bf k})$ and
$z_{\alpha}^q({\bf k})$ and $E_{\alpha}^{q \pm}({\bf k})$ are
obtained from Eq.\ \eqref{mgreen} by substituting $\epsilon_{{\bf k}}
\to \epsilon_{\alpha}^q({\bf k})$. Using the fact that $U_q({\bf
k})$ is independent of $\omega_n$, one finds
\begin{eqnarray}
G(\tau=0^+, {\bf k}) = U_q({\bf k})^{\dagger} Z_q({\bf k})  U_q({\bf
k}) , \label{abelg0plus}
\end{eqnarray}
where $Z_q$ is a $q \times q$ dimensional diagonal matrix whose
diagonal elements are given by $z_{\alpha}^q({\bf k})$. Using Eqs.\
\eqref{abelg0plus} and \eqref{quadres1}, we find
\begin{eqnarray}
\rho_{fi}^{(1)} &=& e^{\sum_{{\bf k}} \sum_{\alpha=0,1..q-1}
\psi^{'\ast}_{\alpha}( {\bf k}) [1- 1/z_{\alpha}^q({\bf k})]
\psi'_{\alpha}({\bf k})} , \label{dabeldenmat}
\end{eqnarray}
where $\Psi'({\bf k}) = U_q({\bf k}) \Psi({\bf k})$ and the sum over
${\bf k}$ is restricted to the magnetic Brillouin zone.
Following earlier analysis of pristine bosons in Sec.\ \ref{bhm1},
we find that for bosons in the presence of the Abelian gauge fields,
one has
\begin{eqnarray}
\langle \psi'_{\alpha}({\bf k}) \psi^{' \ast}_{\alpha}({\bf k})
\rangle &=& z_{\alpha}^q ({\bf k}) = \lambda_{\alpha}^{q \rm
Mott}({\bf k})  .\label{corrabmot}
\end{eqnarray}
The covariance matrix can then be formed by using Eq.\ \eqref{corrmat}
and all measures of entanglement may be computed from its
eigenvalues using Eqs.\ \eqref{bosonvn}, \eqref{eham}, and \eqref{pten1}.

\begin{figure}
\centering {\ing[width=0.8\linewidth]{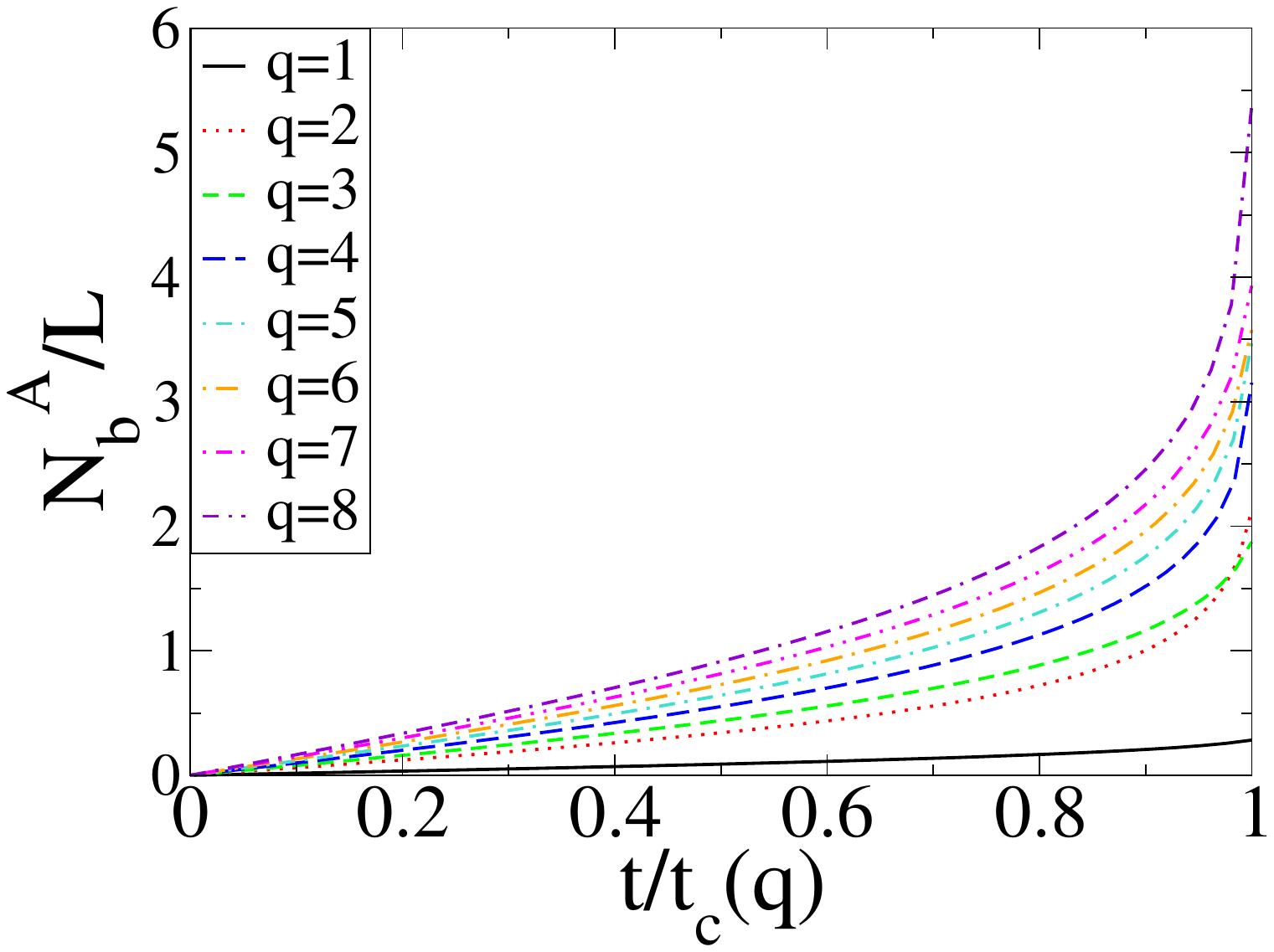}} \caption{
Plot of the entanglement negativity $N_b^A$ as a function of
$t/t_c(q)$. All other parameters are same as in Fig.\ \ref{fig2a}.
See text for details.} \label{fig2b}
\end{figure}

The plot of the von Neumann and second R\'enyi entropies in the
presence of Abelian gauge field, $S_b^{A}$ and $S_2^{A}$, is shown
in Fig.\ \ref{fig2a} as a function of $t/U$ for several values of
$q$ and $L_x/a=30$. We find that these entropies show qualitatively
similar behavior as a function of $t/U$; in particular they display
peaks around the critical point. The entanglement negativity, shown
in Fig.\ \ref{fig2b}, exhibits a similar behavior. It is to be
noted, however, that $N_b^A$ shows a slight stronger dependence on
$t/U$ in the Mott phase. In all of the plots, we have chosen $\mu$
to correspond to the tip of the Mott lobe for all $q$.

The increase of all measures of entanglement with increasing $q$ for
a given $t/U$ in the Mott phase can be qualitatively understood as
follows. The presence of the magnetic field with a flux $ 2\pi/q$
per plaquette leads to $q$ equal-amplitude peaks of the momentum
distribution $n_{{\bf k}}$ at ${\bf k}= {\bf Q}_n= (0, 2 \pi n/q)$
with $n=0,1,..q-1$. This indicates that the real space correlation
of bosons $n_{{\bf r}}= \langle \hat b_{{\bf r}}^{\dagger} \hat
b_{{\bf r}}\rangle$, for any given $t \ne 0$, receives its main
contribution from $q$ wave-vectors (${\bf Q}_n$). Thus with
increasing $q$, $n_{{\bf r}}$ receives contribution from smaller
non-zero momenta. This leads to longer-ranged boson correlations in
real space with increasing $q$. Therefore one expects to have a strongly
entangled boson system with increasing $q$ leading to larger value
of $S_b$, $S_2$ or $N_b$ for larger $q$. This expectation is
corroborated by the results shown in Figs.\ \ref{fig2a} and
\ref{fig2b}.

\subsection{Non-Abelian gauge fields}
\label{bhm2na}

In this section, we consider the Mott phase of bosons in the
presence of a non-Abelian gauge field. There are concrete proposals
of realization for $^{87}{\rm Rb}$ atomic gases by
inducing an effective spin-orbit coupling between two hyperfine
states of the atoms \cite{gaugeref,nonabref1}. The effective
Hamiltonian in the presence of such Raman lasers can
be written as
\begin{eqnarray}
\hat H'_0 &=& \sum_{{\bf r} a} [- \mu {\hat n}_{ {\bf r} a} + U
{\hat n}_{ {\bf r} a}( {\hat n}_{ {\bf r} a}-1)/2] +\lambda U
\sum_{{\bf r}}  {\hat n}_{ {\bf r} 1}  {\hat n}_{{\bf r} 2} \nonumber\\
&& -  \sum_{\langle {\bf r} {\bf r}' \rangle a} t_a \hat b_{ {\bf r}
a}^{\dagger} \hat b_{ {\bf r}' a} , \label{ham0}
\end{eqnarray}
where $\hat b_{ {\bf r} a}$ denotes the bosons annihilation operator
on the site with coordinates ${\bf r}=(x,y)$ on a 2D square lattice,
$a=1,2$ correspond to the index of the hyperfine states, $ {\hat
n}_{ {\bf r} a} = \hat b_{ {\bf r} a}^{\dagger} \hat b_{{\bf r} a}$
is the boson number operator, $U(\lambda U)$ is the interaction
strength between the bosons in same (different) hyperfine states,
and $t_a$ (with $t_1=t$ and $t_2= \eta t$) denotes the nearest
neighbor hopping amplitudes. In the presence of the Raman lasers
inducing a Rashba spin-orbit coupling, the additional terms in the
boson Hamiltonian are given, in terms of a two component boson field
$\hat {\Psi}_{{\bf r}}= (\hat b_{{\bf r} 1},\hat b_{{\bf r}
2})^{T}$, by
\begin{equation}
\hat H'_{1} = i \gamma \sum_{\langle {\bf r} {\bf r}'\rangle}
{\hat \Psi}_{{\bf r}}^{\dagger}\, {\hat z} \cdot \left( {\vec
\sigma} \times {\bf d}_{{\bf r} {\bf r}'} \right) {\hat \Psi}_{{\bf
r}'}
 - \sum_{{\bf r}} \Omega {\hat \Psi}_{{\bf r}}^{\dagger} \sigma_z
{\hat \Psi}_{{\bf r}}.  \label{ham1}
\end{equation}
The first term represents the lattice analogue of the Rashba
spin-orbit coupling generated by the Raman lasers \cite{nonabref1},
${\bf d}_{{\bf r} {\bf r}'}$ is a unit vector along the $x-y$ plane
between the neighboring sites ${\bf r}$ and ${\bf r}'$, $\Omega$ is
the hyperfine state dependent shift in the chemical potential of the
bosons. Here we have neglected additional on-site terms $\sim
\sigma_y$ arising due to boson-laser interaction; these terms can
always be made small by adjusting the detuning of the lasers.

Following exactly similar analysis as that for the pristine bosons,
it is possible to develop a strong-coupling expansion for the bosons
in the presence of non-Abelian gauge fields. Such an analysis was
carried out in Ref.\ \onlinecite{kush1}. The on-site single-particle
Green function, computed from the on-site terms in $H_0$
for $n_0=1$, is given by\cite{kush1}
\begin{eqnarray}
G_{1} (\omega_n) &=& \frac{-1}{i\omega_n +E_1} +  \frac{2}{i\omega_n
+E_1
-U}, \nonumber\\
G_{2}(\omega_n) &=& \frac{1}{i\omega_n - E_2}, \nonumber\\
E_1 &=& \mu +\Omega, \quad E_2 = \mu - \Omega - \lambda U.
\label{gfmott0}
\end{eqnarray}
Using this one can write the action of the bosons in the Mott phase
as\cite{kush1}
\begin{align}
S_{\rm Mott}^{NA} ={}& - \sum_{a,b=1,2} \frac{1}{\beta}
\sum_{\omega_n} \int \frac{d^2k}{(2\pi)^2} \Phi^{\dagger} (\omega_n,
{\bf k}) \nonumber\\
& \times G^{-1}(\omega_n,{\bf k}) \Phi(\omega_n, {\bf k}) ,
\label{gfmott1} \\
G_{\rm NA}^{-1}(\omega_n,{\bf k}) ={}& \left( \begin{array}{cc}
-G_1^{-1}(\omega_n)
+\epsilon_{{\bf k}} & \Delta_{{\bf k}}\\
\Delta_{{\bf k}} & -G_2^{-1}(\omega_n) +\epsilon_{{\bf k}}
\end{array} \right) , \nonumber
\end{align}
where $\epsilon_{{\bf k}}= -2t[\cos(k_x)+\cos(k_y)]$ and
$\Delta_{{\bf k}}=  2\gamma [i \sin(k_x) + \sin(k_y)]$,
$\Phi=(\phi_{\uparrow},\phi_{\downarrow})^T$ is the two component
boson field, and we have set the lattice spacing to unity. The
corresponding boson Green function can be written as
\begin{widetext}
\begin{equation}
G_{\rm NA}(\omega_n, {\bf k}) = \frac{i \omega_n +U+
E_1}{\prod_{i=0..2} (i\omega_n- \Lambda_{i {\bf k}})}\left(
\begin{array}{cc} -G_2^{-1}(\omega_n)
+\epsilon_{{\bf k}} & \Delta_{{\bf k}}\\
\Delta_{{\bf k}} & -G_1^{-1}(\omega_n) +\epsilon_{{\bf k}}
\end{array} \right) ,
\label{gfmott2}
\end{equation}
\end{widetext}
where $\Lambda_i$, for $i=0,1,2$, are solutions of the cubic equation
$(G_1^{-1}-\epsilon_{{\bf k}})(G_2^{-1}-\epsilon_{\bf
k})-|\Delta_{{\bf k}}|^2=0$. We have checked numerically that in the
Mott phase, this equation has three real roots out of which two are
positive. In the Mott limit, these two positive roots correspond to
energies $U_0-E_1$ and $E_2$ as can be read off from Eq.\
\eqref{gfmott0}. Denoting these positive roots by $\lambda_1$ and
$\lambda_2$, one obtains, after performing a Matsubara sum over
$\omega_n$,
\begin{widetext}
\begin{eqnarray}
G_{\rm NA}(0^+, {\bf k}) &=& \sum_{j=1,2}  \frac{\Lambda_{j {\bf k}}
+U+ E_1}{(\Lambda_{j {\bf k}}- \Lambda_{j+1 {\bf k}})(\Lambda_{j \bf
k}-\Lambda_{j+2, {\bf k}})}\left(
\begin{array}{cc} -G_2^{-1}(\Lambda_{j,{\bf k}})
+\epsilon_{{\bf k}} & \Delta_{{\bf k}}\\
\Delta_{{\bf k}} & -G_1^{-1}(\Lambda_{j {\bf k}})
+\epsilon_{{\bf k}}
\end{array} \right) = \left( \begin{array}{cc} S^{\uparrow \uparrow}_{{\bf k}} & S^{\uparrow \downarrow}_{{\bf k}}\\
S^{\downarrow \uparrow}_{{\bf k}}& S^{\downarrow \downarrow}_{{\bf
k}}
\end{array} \right) ,
\label{gfmott3}
\end{eqnarray}
\end{widetext}
where $j$ is a cyclic variable with $j \in Z  \, {\rm Mod }\,3$ and
$S^{\uparrow \downarrow}_{{\bf k}}=(S^{\downarrow \uparrow}_{\bf
k})^{\ast}$. Using this, and following the method outlined in Sec.\
\ref{genform1}, we find that for any two arbitrary boson coherent
states $|\Phi_f\rangle$ and $|\Phi_i\rangle$
\begin{eqnarray}
\rho_{fi}^{\rm NA} &=& e^{ \sum_{{\bf k}} \Phi_{f}^{\dagger}({\bf
k}) \left(I- [G(0^+,{\bf k})]^{-1} \right) \Phi_{i}({\bf k})} ,
\end{eqnarray}
where $\phi_{f}^{\dagger}({\bf k}) = [\phi_{f \uparrow}^{\ast} ({\bf
k}), \phi_{f \downarrow}^{\ast}({\bf k})]$ and $\Phi_i({\bf k})$ are
two component Bose fields. Using this density matrix one obtains the
Boson correlation matrix in the presence of a non-Abelian gauge
field as
\begin{equation}
\langle \phi_{\sigma}^{\ast}({\bf k}) \phi_{\sigma'}({\bf k})
\rangle = S^{\sigma \sigma'}_{{\bf k}} . \label{corrmatna}
\end{equation}
Using this correlation matrix, one can now use the procedure of
Sec.\ \ref{genform2} to obtain the covariance matrices of the bosons
and hence different measures of entanglement entropies. The
eigenvalues of the correlation matrix $G_{\rm NA}(0^+, {\bf k})$, as
discussed in Sec.\ \ref{genform2}, yield the frequency
$\lambda_{{\bf k}}$ of the effective oscillator Hamiltonian. A
straightforward but cumbersome calculation shows that
\begin{equation}
\lambda_{{\bf k}}^{(1)} = -1+ \sum_{\sigma} S^{\sigma \sigma}_{\bf
k}, \quad \lambda_{{\bf k}}^{(2)}=1 . \label{eugenvalna}
\end{equation}
Note that only one of these eigenvalues depends on system parameters
such as $t/U$, $\gamma/U$. This property can be easily verified in
the Mott limit and holds for all values of $t/U$ and $\gamma/U$ in
the Mott phase. Using these eigenvalues, one can construct the
covariance matrix using Eq.~\eqref{corrmat} and numerically compute
the different entanglement measures using Eqs.\ \eqref{bosonvn},
\eqref{renyiexp}, and \eqref{pten1}.

\begin{figure}
\centering {\ing[width=0.48 \linewidth]{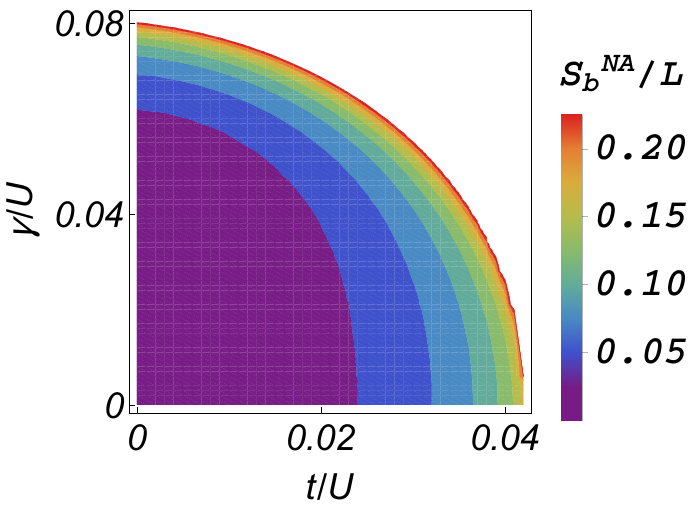}}
\centering{\ing[width=0.48 \linewidth]{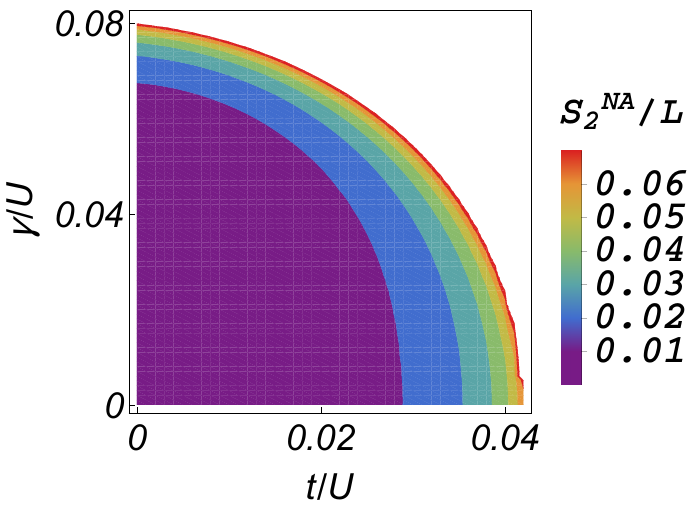}} \caption{Plot
of the von Neumann entropy $S_b^{\rm NA}$ (left panel) and the
second R\'enyi entropy $S_2^{\rm NA}$ (right panel) in the presence
of a non-Abelian gauge field as a function of $t/U$ and $\gamma/U$
for $L \equiv L_x/a=30$, $ \Omega=0.01$ and $\lambda=0.04$ in the
Mott phase. All other parameters are same as in Fig.\ \ref{fig2a}.
See text for details.} \label{fig3a}
\end{figure}

\begin{figure}
\centering {\ing[width=0.7\linewidth]{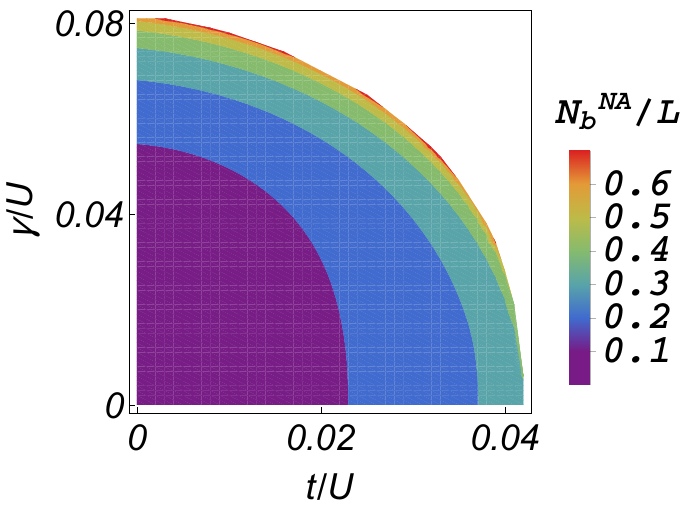}} \caption{
Plot of the entanglement negativity $N_b^{\rm NA}$ as a function of
$t/U$ and $\gamma/U$. All other parameters are same as in Fig.\
\ref{fig3a}. See text for details.} \label{fig3b}
\end{figure}

The result of such computation is shown in Figs.\ \ref{fig3a} and
\ref{fig3b}. In Fig.\ \ref{fig3a}, we plot the von Neumann
($S_b^{\rm NA}$) and the R\'enyi ($S_2^{\rm NA}$) entropies as a
function of both $\gamma/U$ and $t/U$. We find that the entanglement
peaks as one approaches the critical point either by increasing $t$
or $\gamma$. A similar feature is seen for entanglement negativity
$N_b^{\rm NA}$. We note that our computation indicates that for
correlated boson systems the entanglement negativity has a
qualitatively similar characteristic to R\'enyi and von Neumann
entropies.

\section{Non-equilibrium dynamics for integrable models}
\label{rpcs}

In this section, we demonstrate the application of our method to a
class of integrable fermionic quantum models driven out of
equilibrium by a periodic square pulse array. In Sec.\ \ref{rpcsa},
we define the model and obtain its Floquet Hamiltonian under
periodic drive. This is followed by the computation of the return
probability in Sec.\ \ref{rpcsb} and counting statistics of number
operator in Sec.\ \ref{rpcsc}.

\subsection{Floquet Hamiltonian}
\label{rpcsa}

The Hamiltonian of free fermion models obeying Dirac-like equation
in $d-$dimensions can be written as
\begin{eqnarray}
\hat H &=& \sum_{{\bf k}} \hat \psi^{\dagger}_{{\bf k}} \hat H_{{\bf
k}} \hat \psi_{\bf k},
\nonumber\\
\hat H_{{\bf k}} &=& (g(t)-b_{{\bf k}}) \sigma_3 + \Delta_{{\bf k}}
\sigma_1 \label{hamint1}
\end{eqnarray}
where the sum over ${\bf k}$ extends over half the Brillouin zone
(defined, e.g., by $k_x\geq 0$). $\vec \sigma =(\sigma_1,\sigma_2,
\sigma_3)$ denotes Pauli-matrices in particle-hole space, and
$g(t)$, $b_{{\bf k}}$ and $\Delta_{{\bf k}}$ are parameters of the
Hamiltonian whose precise forms depend on the system that $\hat H$
represents. We note that Eq.~\eqref{hamint1} provides a
representation of the low-energy theory for Dirac quasiparticles in
graphene and on surface of topological insulators; moreover, it gives
an exact fermionic representation of spin models such as
Ising model in $d=1$ and Kitaev model in $d=2$. The two-component
field $\hat \psi_{\bf k}$ is either of the form $\hat \psi_{{\bf
k}}= (\hat c_{{\bf k}}, \hat c_{\bf -k})^T$ (for the case of
graphene or TI quasiparticles and 2D Kitaev model) or of the form
$\hat \psi_{{\bf k}}= (\hat c_{{\bf k}}, \hat c^\dagger_{\bf -k})^T$
for superconductors and Ising model in a transverse field. In the
second case, we perform a particle-hole transformation $\hat c_{-\bf
k}\leftrightarrow \hat c^\dagger_{-\bf k}$ thus allowing us to
return to the first case. In what follows, we shall study the
properties of this model under a periodic drive protocol,
\begin{equation}
g(t) = \left\{
\begin{array}{ccc} g_1 & \mbox{if} & 0\leq  t\leq T/2 , \\
g_2 & \mbox{if} & T/2\leq  t\leq T ,
\end{array}
\right.
\label{prot1}
\end{equation}
where $T= 2\pi/\omega_D$ is the drive period and $\omega_D$ is the
drive frequency. Our aim is to show that our method enables one to
semi-analytically compute counting statistics and return probability
of such a system in terms of Floquet eigenvalues and eigenvectors.

We begin by computing the Floquet Hamiltonian of the model by using
the formalism developed in Sec.\ \ref{noneq}. To this end, we first
note that the evolution operator of the system for the protocol
given by Eq.\ \eqref{prot1}, at $t=T$, can be written as
\begin{eqnarray}
\hat U(T,0) &=& \hat U(T,T/2) \hat U(T/2,0)  \nonumber\\
&=& e^{-i \hat H[g_2] T/(2\hbar)} e^{-i \hat H[g_1]T/(2\hbar)} ,
\label{evolint1}
\end{eqnarray}
where $H[g_{1(2)}]$ is given by Eq.\ \eqref{hamint1} with $g(t)=g_{1
(2)}$. To compute the Floquet Hamiltonian we first obtain the matrix
elements of $\hat U(T/2,0)$ and $\hat U(T,T/2)$ between two coherent
states $|\phi_1\rangle$ and $|\phi_2\rangle$. To this end, we note
that after a Wick rotation $T/2 \to -i \beta \hbar$, we find, using
Eq.\ \eqref{unievol4},
\begin{eqnarray}
U_{12}(\beta,0) &=& \langle \phi_1|e^{- \hat H[g_1] \beta }
|\phi_2\rangle  \label{ueq1} \\
&=& e^{-\sum_{{\bf k}}  \Phi_{1 {\bf k}}^{\dagger} (\sigma_3 -
\sigma_3 [G({\bf k},0^+)]^{-1} \sigma_3) \Phi_{2 {\bf k}} }\nonumber
\end{eqnarray}
where the sum over ${\bf k}$ extends over half the Brillouin zone
and $\Phi_{i{\bf k}} = (\phi_{i{\bf k}}, \phi_{i-{\bf k}})^T$
($i=1,2$) is a two-component Grassmann field.

To obtain this matrix element, we must compute the Green function
$G(0^+)$ corresponding to $\hat H$. To this end, we first note that
$\hat H_{{\bf k}}[g_1]$ can be brought into a diagonal form via the
transformation $\hat H_{{\bf k}}^D = \Lambda^{\dagger}_{{\bf k}}
\hat H_{{\bf k}}[g_1] \Lambda_{{\bf k}} = E_{{\bf k}}(g_1) \tau_3$
where
\begin{equation}
\begin{split}
\Lambda_{{\bf k}} &= \left(\begin{array}{cc} u_{{\bf k}} & -v_{{\bf k}} \\
v_{{\bf k}} & u_{{\bf k}} \end{array} \right), \\
E_{\bf k}(g_1) &=\sqrt{(g_1-b_{{\bf k}})^2+ \Delta_{{\bf k}}^2} , \\
u_{{\bf k}} [v_{{\bf k}}] &=  \frac{1}{{\sqrt 2}} \sqrt{1+[-]
\frac{g_1-b_{{\bf k}}}{E_{{\bf k}}}} .
\end{split}
\end{equation}
In the diagonal basis, it is easy to see that the Green function
$G_D ({\bf k}; i\omega_n) = -(i \omega_n - \tau_3 E_{{\bf k}})^{-1}$
which leads to
\begin{eqnarray}
G_D({\bf k}; \tau=0^+) &=& \frac{1}{\beta} \sum_{i\omega_n} G_D({\bf
k}
; i \omega_n) e^{-i \omega_n 0^+} \nonumber\\
&=& (I+ \exp[-\sigma_3 \beta E_{{\bf k}}(g_1)]^{-1}.
\label{gzeroplus1}
\end{eqnarray}
Rotating back to the original basis, we find
\begin{widetext}
\begin{eqnarray}
&& [G({\bf k}; \tau=0^+)]^{-1} = \Lambda [G_D({\bf k};
\tau=0^+)]^{-1} \Lambda^{\dagger} = \left( \begin{array}{cc} 1+
u_{{\bf k}}^2 e^{-\beta E_{{\bf k}}(g_1)} + v_{{\bf k}}^2 e^{\beta
E_{\bf
k}(g_1)} & 2u_{{\bf k}} v_{{\bf k}} \sinh(\beta E_{{\bf k}}(g_1)) \\
2u_{{\bf k}}  v_{{\bf k}} \sinh(\beta E_{{\bf k}}(g_1)) & 1+ v_{\bf
k}^2 e^{-\beta E_{{\bf k}}(g_1)} + u_{{\bf k}}^2 e^{\beta E_{\bf
k}(g_1)}
\end{array} \right) .
\label{greenexp1} \nonumber\\
\end{eqnarray}
\end{widetext}
Substituting Eq.\ \eqref{greenexp1} in Eq.\ \eqref{ueq1} and rotating
back to real time we find
\begin{widetext}
\begin{eqnarray}
U_{12}(T/2,0) &=& e^{-\sum_{{\bf k}} \Phi_{1 {\bf k}}^{\dagger}
{\mathcal L}_{\bf k}(g_1,T) \Phi_{2 {\bf k}} },\quad  {\mathcal
L}_{\bf k}(g_1,T) = \left(
\begin{array} {cc} -\cos \varphi^{(1)}_{{\bf k}} + i n_{1{\bf k}}^{(1)}
\sin \varphi^{(1)}_{{\bf k}} & i n_{3 {\bf k}}^{(1)} \sin
\varphi^{(1)}_{{\bf k}} \\ i n_{3 {\bf k}}^{(1)} \sin
\varphi^{(1)}_{{\bf k}}& -\cos \varphi^{(1)}_{{\bf k}} - i n_{1\bf
k}^{(1)} \sin \varphi^{(1)}_{{\bf k}} \end{array} \right) ,
\label{uexp1}
\end{eqnarray}
\end{widetext}
where $\varphi_{{\bf k}}^{(1)} = E_{{\bf k}}(g_1) T/(2 \hbar)$ and $
\vec n_{ {\bf k}}^{(1)} = (n_{1 {\bf k}}^{(1)}, 0,n_{3 {\bf
k}}^{(1)}) = ( (g_1-b_{{\bf k}})/E_{{\bf k}}(g_1),0, \Delta_{{\bf
k}}/E_{\bf k}(g_1))$.

Similarly one can carry out the evaluation of $U_{12}(T,T/2)$. It is
easy to see that the result is given by Eq.\ \eqref{uexp1} with $g_1
\to g_2$. One then obtains
\begin{eqnarray}
U_{fi}(T,0) &=& \int D\Phi^{\dagger} D \Phi  \, e^{- \sum_{{\bf k}}
|\Phi_{{\bf  k}}|^2 } \nonumber\\
&& \times \langle \Phi_f|\hat U(T,T/2)|\Phi\rangle \langle
\Phi|\hat U(T/2,0) | \Phi_i \rangle \nonumber\\
&=& e^{ \sum_{{\bf k}} \Phi^{\dagger}_{f {\bf k}} {\mathcal
L}_{\bf k}(g_2,T) {\mathcal L}_{\bf k}(g_1,T) \Phi_{i {\bf k}}} ,
\end{eqnarray}
 where $|\Phi_{{\bf k}}|^2 = \Phi_{\bf k}^{\dagger} \Phi_{\bf k}$,
we have performed the Gaussian integral over $\Phi$ and
$\Phi^{\dagger}$ and used Eq.\ \eqref{uexp1} to arrive at the
last line. The product matrix $ {\mathcal L}_{\bf k}(g_2,T)
 {\mathcal L}_{\bf k} (g_1,T)$ provides an analytic expression
for the evolution operator and plays a central role in our analysis.
It is given by
\beq {\mathcal L}_{\bf k}(g_2,T){\mathcal  L}_{\bf k}(g_1,T) =
\left( \begin{array}{cc}
    \alpha_{{\bf k}} & \beta_{{\bf k}}
    \\ -\beta_{{\bf k}}^{\ast} & \alpha_{{\bf k}}^{\ast} \end{array}
\right),
\label{L2L1}
\eeq
where
\begin{align}
\beta_{{\bf k}} ={}& -i [ n_{3 {\bf k}}^{(1)} \sin
\varphi_{{\bf k}}^{(1)} (\cos\varphi_{{\bf k}}^{(2)} - i n_{1 \bf
k}^{(2)} \sin \varphi_{{\bf k}}^{(2)}) \nonumber \\ & + n_{3 {\bf k}}^{(2)} \sin
\varphi_{{\bf k}}^{(2)} (\cos\varphi_{{\bf k}}^{(1)} + i n_{1 \bf
k}^{(1)} \sin
\varphi_{{\bf k}}^{(1)})] , \nonumber \\
\alpha_{{\bf k}} ={}& \cos\varphi^{(1)}_{{\bf k}} \cos
\varphi^{(2)}_{{\bf k}} - \vec n^{(1)}_{{\bf k}} \cdot \vec n_{{\bf
k}}^{(2)} \sin \varphi^{(1)}_{{\bf k}} \sin \varphi^{(2)}_{{\bf k}}
\nonumber \\ &  - i (n_{1 {\bf k}}^{(1)} \sin \varphi^{(1)}_{{\bf
k}} \cos \varphi^{(2)}_{{\bf k}} + n_{1 \bf k}^{(2)} \cos
\varphi^{(1)}_{{\bf k}} \sin \varphi^{(2)}_{{\bf k}}) .
\label{mat12}
\end{align}
Its eigenvalue and eigenvectors can be easily found to be
\begin{eqnarray}
\nu_{{\bf k}}^{a} &=&  e^{i a \epsilon_{{\bf k}}^F T} ,
\nonumber\\
\frac{u_{{\bf k}}^{a}}{v_{{\bf k}}^{a}} &=&  i\frac{ \beta_{{\bf
k}}}{{\rm Im}(\alpha_{\bf k})-a\sin[\epsilon^F_{{\bf k}} T]},
\label{evec1}
\end{eqnarray}
where $a=\pm$ and the Floquet eigenvalues $\epsilon_{{\bf k}}^F$ are
given by
\begin{eqnarray}
\epsilon^F_{{\bf k}} T &=& \arccos[\cos \varphi^{(1)}_{\bf k}
\cos\varphi^{(2)}_{\bf k} \nonumber\\
&& -\vec n_{ {\bf k}}^{(1)} \cdot \vec n_{ {\bf k}}^{(2)} \sin
\varphi^{(1)}_{\bf k} \sin \varphi^{(2)}_{\bf k}] .
\label{flen1}
\end{eqnarray}
These results agree with those derived in Ref.\ \onlinecite{asen1}
from standard Hamiltonian methods. We shall use these results in
Secs.\ \ref{rpcsb} and \ref{rpcsc} to compute the return probability
and counting statistics of the number operator.

\subsection{Return Probability}
\label{rpcsb}

In this section, we consider the computation of the return
probability using our method. This quantity, for a quench protocol,
is identical to the Lochsmidt echo ${\mathcal G}(t)= \langle \hat
U(t,0) \rangle$ computed for transverse field Ising model in Ref.\
\onlinecite{silva1} and is related to the work statistics of the
driven quantum system. Here, we consider the fermion system
introduced in the previous section and whose dynamics is governed by
the Hamiltonian~\eqref{hamint1}. We assume that the initial state is
$\ket{n}\equiv\ket{\{n_{\bf k},n_{-{\bf k}}\}}$ where $n_{\bf k}$
and $n_{-{\bf k}}$ are the occupation numbers of the single-particle
states with momentum ${\bf k}$ and $-{\bf k}$. Note that in the case
of superconductor, where the two-component field $\hat\psi_{\bf
k}=(\hat c_{\bf k},\hat c^\dagger_{-\bf k})^T$ becomes
$\hat\psi_{\bf k}=(\hat c_{\bf k},\hat c_{-\bf k})^T$ after the
particle-hole transformation $\hat c_{-\bf k}\leftrightarrow\hat
c^\dagger_{-\bf k}$, the occupation number $n_{-\bf k}$ actually
refers to the number of holes: $n_{-\bf k}=0$ (1) if the electron
state with momentum $-\bf k$ is occupied (empty).

The probability amplitude that the system returns to its initial state after a time $t$ is given by
\begin{align}
{\mathcal P}_n(t) ={}& \langle n | \hat U(t,0)|n \rangle \nonumber\\
={}& \int D\Phi_{1}^{\dagger} D \Phi_{1}  D\Phi_{2}^{\dagger} D
\Phi_{2} \braket{n}{\Phi_1} \, e^{-\sum_{\bf k} (|\Phi_{1\bf k}|^2 + |\Phi_{1\bf k}|^2) } \nonumber \\ & \times
\bra{\Phi_1} \hat U(t,0) \ket{\Phi_2} \braket{\Phi_2}{n} .
\end{align}
Using\cite{not1}
\beq
\begin{split}
\langle \phi_1|\hat U(t,0)|\phi_2\rangle &=
e^{ -\sum_{{\bf k}} \Phi_{1 {\bf k}}^{\dagger}
{\mathcal M}_{{\bf k}} (t)\Phi_{2 {\bf k}} }  , \\
\braket{n}{\Phi} &= \prod_{\bf k}\phi_{\bf k}^{n_{\bf k}} \phi_{-\bf k}^{n_{-\bf k}} ,
\end{split}
\eeq
we obtain
\begin{multline}
{\mathcal P}_n(t) = \int D\Phi_{1}^{\dagger} D \Phi_{1}  D\Phi_{2}^{\dagger} D\Phi_{2}
\prod_{{\bf k}_1,{\bf k}_2}
{\Phi^1_{1{\bf k_1}}}^{n_{{\bf k}_1}^1}
{\Phi^2_{{1\bf k}_1}}^{n_{{\bf k}_1}^2} \\  \times
{\Phi^{2*}_{2{{\bf k}_2}}}^{n_{{\bf k}_2}^2}
{\Phi^{1*}_{2{\bf k}_2}}^{n_{{\bf k}_2}^1}
e^{-\sum_{\bf k} ( |\Phi_{1\bf k}|^2 + |\Phi_{2\bf k}|^2 + \Phi^\dagger_{1\bf k} {\cal M}_{\bf k}(t) \Phi_{2\bf k} )} ,
\label{rp1}
\end{multline}
where we use the notation $\Phi^1_{\bf k}=\phi_{\bf k}$, $\Phi^2_{\bf k}=\phi_{-\bf k}$, $n_{\bf k}^1=n_{\bf k}$ and $n_{\bf k}^2=n_{-\bf k}$.
The functional integral in~(\ref{rp1}) is Gaussian and can be performed using Wick's theorem. Let us write the ``action'' as
\beq
\sum_{\bf k} ( \Phi^{1*}_{1\bf k} , \Phi^{2*}_{1\bf k} , \Phi^{1*}_{2\bf k} ,  \Phi^{2*}_{2\bf k} )
 \calD_{\bf k}^{-1}(t) \begin{pmatrix} \Phi^{1}_{1\bf k} \\ \Phi^{2}_{1\bf k} \\ \Phi^{1}_{2\bf k} \\  \Phi^{2}_{2\bf k}
 \end{pmatrix} ,
 \label{action}
\eeq
where the matrix
\beq
 \calD_{\bf k}^{-1}(t) = \begin{pmatrix}
    1 & 0 & {\cal M}_{{\bf k},11}(t) & {\cal M}_{{\bf k},12}(t) \\
    0 & 1 & {\cal M}_{{\bf k},21}(t) & {\cal M}_{{\bf k},22}(t) \\
    0 & 0 & 1 & 0 \\
    0 & 0 & 0 & 1
 \end{pmatrix}
 \eeq
satisfies $\det\,\calD_{\bf k}^{-1}(t)=1$ and $\calD_{\bf k}(t)$ is
simply deduced from $\calD_{\bf k}^{-1}(t)$ by changing ${\cal
M}_{{\bf k},ij}(t)$ into $-{\cal M}_{{\bf k},ij}(t)$. We then obtain
\begin{align}
{\mathcal P}_n(t)
={}& \prod_{\bf k} \mean{ {\Phi^1_{1\bf k}}^{n_{\bf k}^1} {\Phi^2_{1\bf k}}^{n_{\bf k}^2} {\Phi^{2*}_{2\bf k}}^{n_{\bf k}^2} {\Phi^{1*}_{2\bf k}}^{n_{\bf k}^1} } \nonumber \\
={}& \prod_{\bf k} \bigl[ (-{\cal M}_{{\bf k},11}(t))^{n^1_{\bf k}} (-{\cal M}_{{\bf k},22}(t))^{n^2_{\bf k}} \nonumber \\ &
- \delta_{n^1_{\bf k},1}\delta_{n^2_{\bf k},1} {\cal M}_{{\bf k},12}(t) {\cal M}_{{\bf k},21}(t) \bigr] ,
\end{align}
where $\mean{\cdots}$ denotes an average with the action~(\ref{action}).

Thus the return probability after a period $T$ of the drive
protocol~(\ref{prot1}) is determined by the matrix ${\mathcal
M}_{{\bf k}}(T) = -{\mathcal L}_{\bf k}(g_2,T){\mathcal L}_{\bf k}(g_1,T)$ defined
by~(\ref{L2L1}). For the state defined by $n_{\bf k}=a$ and $n_{-\bf
k}=b$ ($a,b=0,1$), we find that the probability ${\cal P}_{ab}(T)$
takes the simple expression
\beq
\begin{split}
    {\cal P}_{00}(T) &= 1, \quad {\cal P}_{11}(T) = \prod_{\bf k} \det\,{\cal M}_{\bf k}(T) = 1 , \\
    {\cal P}_{10}(T) &= \prod_{\bf k} \alpha_{\bf k}, \quad {\cal P}_{01}(T) = \prod_{\bf k} \alpha^*_{\bf k} .
\end{split}
\eeq We note that the states $|\{0_{\bf k}, 0_{-\bf{k}}\}\rangle$
and $|\{1_{\bf k}, 1_{-\bf {k}}\}\rangle$, which correspond to the
empty and maximum density states, respectively, do not evolve in time. In
the case of a superconductor, the state $|\{0_{\bf k},
0_{-\bf{k}}\}\rangle$ has all electron states with momentum $\bf k$
empty while all states with momentum $-\bf k$ are occupied (since
$0_{-\bf k}$ denotes the number of holes with momentum $-\bf k$).
Being the state with minimal total momentum, and the total momentum
being conserved, this state does not evolve in time (a similar
analysis applies to the state of maximum total momentum, $|\{1_{\bf
k},1_{-\bf{k}}\}\rangle$).

Thus we find that the return probability of a fermionic system with
Gaussian action driven by an arbitrary protocol can be obtained from
its thermal Green function $G(0^+)$ via its analytic continuation
to real time. The probability amplitude depends on $\alpha_{\bf k}$
and $\alpha_{\bf k}^*$; thus our analysis ties the return probability to the matrix elements
of the unitary evolution operator.

\begin{figure}
\centering {\ing[width=0.85\linewidth]{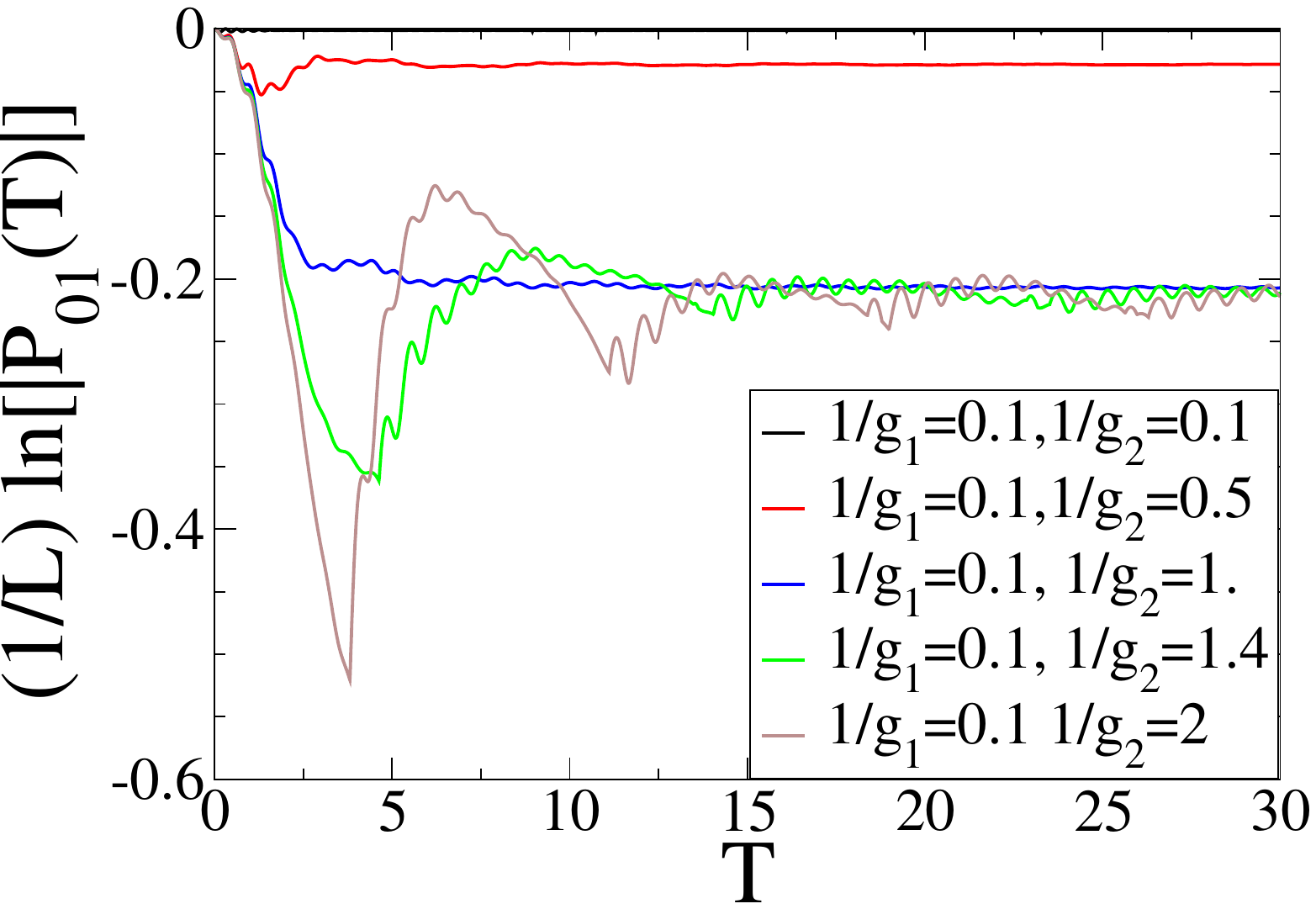}} \caption{Plot of $\ln[
|{\mathcal P}_{01}(T)|]/L$ for the $d=1$ transverse field Ising model as
a function of the drive period $T$ for several transverse fields
$g_1$ and $g_2$. The plot clearly shows the difference in behavior
at large $T$ between protocols which do and do not cross the
critical point. See text for details.} \label{fig1}
\end{figure}

In Fig.\ \ref{fig1} we show \beq \ln |{\cal P}_{01}(T)| = L^d \int
\frac{d^d k}{(2\pi)^d}  \ln |\alpha_{\bf k}| \eeq for the specific
case of the one-dimensional transverse field Ising model for which
$b_{k}=\cos k$ and $\Delta_{k}=\sin k$. We assume the starting state
to correspond to $g=g_1=10$ which is the ground state deep inside
the ferromagnetic phase. The dynamics of
the model is studied using the square pulse protocol defined in Eq.\
\eqref{prot1} with several $g_2$. We find that the return
probability computed using this protocol falls into two distinct
categories. For protocol where the dynamics never crosses the
critical point ($g_2 \ge g_c=1$), the return probability has a
higher value for slow enough protocols. This is shown in Fig.\
\ref{fig1} where $\ln [|{\mathcal P}_{01}(T)|]/L$ is plotted as a
function of the drive time $T= 2 \pi /\omega_D$. In contrast, the
return probability reaches a much lower value for dynamics which
cross the critical point ($g_2 \le g_c=1$).  Moreover for $g_2 <
g_c$, the return probability shows a stronger non-monotonic behavior
as a function of $T$.

These features can be understood by noting that for both near-adiabatic
and sudden protocols, the system is expected to remain
close to its original ground state when it does not pass through a
critical point. Thus the return probability should be close to unity
for such drives. However, for finite $\omega_D$,
where the drive frequency matches the energy gap at some $k$, one
expects significant excitation production leading to weak
non-monotonicity of $\ln |{\mathcal P}_{01}|$.

In contrast, for drives which take the system through the critical
point, there is no adiabaticity and $|{\mathcal P}_{01}(T \gg 1)|
\le |{\mathcal P}_{01}(T \ll 1)|$. However, even in this case,
excitation production is maximal at intermediate frequency, where
one can produce excitations at maximal number of $k$ modes. Thus the
return probability shows a dip close to $\omega_D \simeq 1$; the
precise position of this dip depends on $g_2$.

\begin{figure}
\centering {\ing[width=0.85\linewidth]{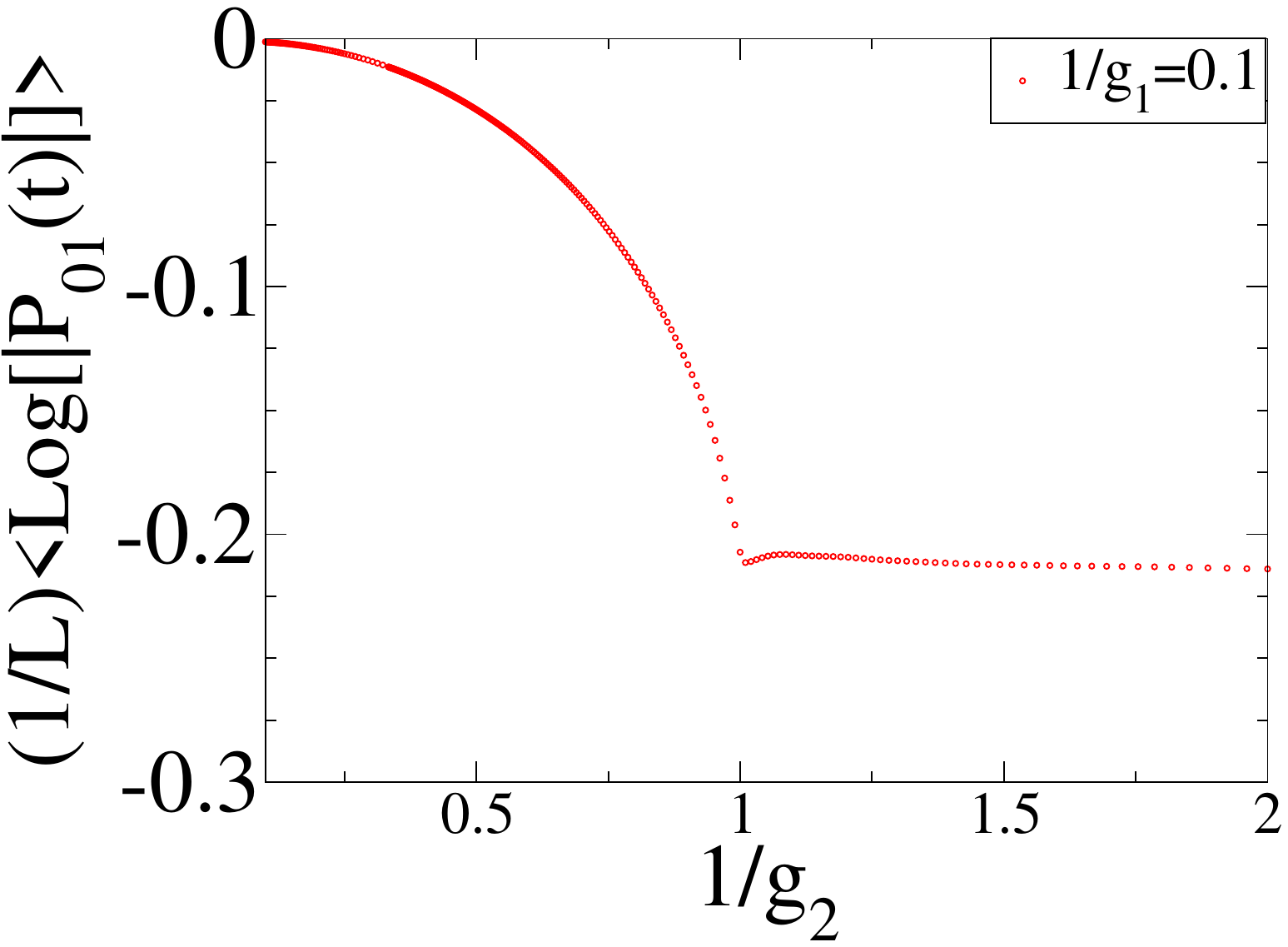}} \caption{Plot of
$\langle \ln[|{\mathcal P}_{01}(T)|]\rangle/L$ as a function of
$1/g_2$ for a fixed $g_1=10$ indicating the presence of critical
point at $g_2=1$. See text for details. \label{fig2}}
\end{figure}

We note that our results indicate that the position of a critical
point in a quantum system can be inferred from its return
probability as a function of the drive amplitude for slow enough
drive frequencies. To elucidate this feature, we consider
\begin{eqnarray}
\langle \ln[ |{\mathcal P}_{01}(T)|]\rangle &=& \frac{1}{T_f-T_i}
\int_{T_i}^{T_f} dT \ln [|{\mathcal P_{01}(T)}|] , \label{avprob}
\end{eqnarray}
where $T_i$ and $T_f$ are the lowest and highest time periods
between which the return probability is summed over. For our plots
$T_f \simeq 200$ and $T_i=20$ so that $\langle \ln[ |{\mathcal
P}_{01}(T)|]\rangle$ reflects the drive time averaged value of the
return probability for large $T$. A plot of $\langle \ln[ |{\mathcal
P}_{01}(T)|]\rangle/L$ as a function of $1/g_2$, shown in Fig.\
\ref{fig2}, clearly indicates the difference between drives with
$g_2 >1$ and $g_2<1$. We note that for large $g_2>1$, the system is
gapped; thus the large $T$ limit of the return probability remains
constant as $g_2$ is increased. The gapped nature of the system
ensures that all excitations are produced near the critical point at
large $T$. Therefore increasing $g_2$ does not change the value of
$\langle \ln[ |{\mathcal P}_{01}(T)|]\rangle/L$.

\subsection{Probability distribution of operators: Counting Statistics}
\label{rpcsc}

In this section, we compute the probability distribution of a
quadratic operator $\hat O$ for the fermionic model considered in
the preceding sections. The probability distribution is given by
$P(O,t) = {\rm Tr}[\hat \rho(t) \delta(O- {\hat O})]$, where $\hat
\rho(t)$ denotes the density matrix of the system. The corresponding
characteristic function is defined by\cite{cref1,cref2,cref3}
\begin{equation}
\begin{split}
P(O,t) &= \int_{-\infty}^{\infty} \frac{d f}{2 \pi} e^{- i
f  O} C(f,t) , \\
C(f,t) &= \int_{-\infty}^{\infty} dO \, e^{ifO} P(O,t) .
\end{split}
\label{POdef}
\end{equation}
For a system in a pure state,
$\hat\rho(t)=\ket{\Psi(t)}\bra{\Psi(t)}$, the characteristic
function can be written as
\begin{align}
C(f,t) &=  \bra{\Psi(t)} e^{if\hat O} \ket{\Psi(t)} \nonumber \\
&= \bra{\Psi(0)} \hat U^\dagger(t,0) e^{if\hat O} \hat U(t,0) \ket{\Psi(0)} ,
\label{cexp1}
\end{align}
where $\hat U(t,0)$ is the time evolution operator and
$\ket{\Psi(0)}$ the initial state at time $t=0$.

Let us first compute the characteristic function associated with the
density operator $\hat n_{{\bf k}} = \hat c_{{\bf k}}^{\dagger} \hat
c_{{\bf k}} - \hat c_{-{\bf k}}^{\dagger} \hat c_{-{\bf k}}+1$
(which in the case of a superconductor corresponds to the electron
density $\hat c_{{\bf k}}^{\dagger} \hat c_{{\bf k}} + \hat c_{-{\bf
k}}^{\dagger} \hat c_{-{\bf k}}$ before the particle-hole
transformation). The matrix element of the operator ${\hat O}_{\pm
{\bf k}}= \pm \hat c_{\pm {\bf k}}^{\dagger} \hat c_{\pm {\bf k}}$
(with $\bf k$ restricted to half the Brillouin zone, e.g. $k_x\geq
0$) is given by
\begin{eqnarray}
\langle \Phi_1|e^{i f \hat O_{{\bf k}}}|\Phi_2 \rangle &=&
e^{ \sum_{{\bf k}'\neq {\bf k}} \Phi^\dagger_{1{\bf k}'} \Phi_{2{\bf k}'} +  \Phi_{1 {\bf k}}^{\dagger} O_{{\bf k}} \Phi_{2 {\bf k}}} , \nonumber \\
\langle \Phi_1|e^{i f \hat O_{- {\bf k}}}|\Phi_2 \rangle &=&
e^{ \sum_{{\bf k}'\neq {\bf k}} \Phi^\dagger_{1{\bf k}'} \Phi_{2{\bf k}'} + \Phi_{1 {\bf k}}^{\dagger} O_{- {\bf k}} \Phi_{2 {\bf k}}} ,
\label{matel2}
\end{eqnarray}
where  \beq O_{{\bf k}} = \begin{pmatrix}  e^{if} & 0 \\ 0 & 1
\end{pmatrix} , \quad O_{-{\bf k}} = \begin{pmatrix} 1 & 0 \\ 0 &
e^{-if} \end{pmatrix} \eeq are $2 \times 2$ matrices. This implies
\begin{equation}
\begin{split}
\langle \Phi_1|e^{i f \hat n_{{\bf k}}}|\Phi_2 \rangle &=
e^{ if + \sum_{{\bf k}'\neq {\bf k}} \Phi^\dagger_{1{\bf k}'} \Phi_{2{\bf k}'} + \Phi_{1 \bf
k}^{\dagger} O_{\hat n_{\bf k}} \Phi_{2 {\bf k}}} , \\
O_{\hat n _{\bf k}} &= \left( \begin{array}{cc} e^{i f} & 0 . \\
0 & e^{-i f} \end{array} \right) .
\end{split}
\label{denel1}
\end{equation}

The characteristic function~(\ref{cexp1}) for the operator $\hat O=\hat n _{{\bf k}}$ can be written as
\begin{align}
C_{\hat  n_{{\bf k}}}(f,t) ={}& \int \prod_{\ell=1}^4 D\Phi^\dagger_\ell D\Phi_\ell \, e^{-\sum_{\ell,{\bf k}'} |\Phi_{\ell {\bf k}'}|^2 }
\braket{\Psi(0)}{\Phi_1} \nonumber \\ & \times \bra{\Phi_1} \hat U^\dagger(t,0) \ket{\Phi_2} \bra{\Phi_2} e^{if\hat n_{{\bf k}}} \ket{\Phi_3}  \nonumber \\ & \times \bra{\Phi_3} \hat U(t,0) \ket{\Phi_4} \braket{\Phi_4}{\Psi(0)} .
\end{align}
Assuming that the system is initially in the number state $\ket{n}\equiv \ket{\{n_{\bf k},n_{-\bf k}\}}$, we find
\begin{multline}
C_{\hat n_{{\bf k}}}(f,t) = \int \prod_{\ell=1}^4 D\Phi^\dagger_\ell D\Phi_\ell  \,
\prod_{{\bf k}_1,{\bf k}_2} {\Phi_{1{\bf k}_1}^1}^{n^1_{{\bf k}_1}} {\Phi_{1{\bf k}_1}^2}^{n^2_{{\bf k}_1}}
{\Phi_{4{\bf k}_2}^{2*}}^{n^2_{{\bf k}_2}} \\ \times {\Phi_{4{\bf k}_2}^{1*}}^{n^1_{{\bf k}_2}}
e^{-\sum_{{\bf k}'} [ \sum_\ell |\Phi_{\ell {\bf k}'}|^2 + \Phi^\dagger_{1{\bf k}'} {\cal M}^\dagger_{{\bf k}'}(t) \Phi_{2{\bf k}'} ]}
\\ \times
e^{-\sum_{{\bf k}'} [\Phi^\dagger_{3{\bf k}'} {\cal M}_{{\bf k}'}(t) \Phi_{4{\bf k}'} - \Phi^\dagger_{2{\bf k}'} (1+\delta_{{\bf k}',{\bf k}} O_{\hat n_{{\bf k}}} ) \Phi_{3{\bf k}'} ] } ,
\end{multline}
where the matrix $O_{\hat n_{{\bf k}}}$ is defined by~(\ref{denel1}) and we use the same notations as in Sec.~\ref{rpcsb}. Integrating out $\Phi_2$ and $\Phi_3$, we obtain
\begin{multline}
C_{\hat n_{{\bf k}}}(f,t) = \int D\Phi^\dagger_1 D\Phi_1 D\Phi^\dagger_4 D\Phi_4 \,
\prod_{{\bf k}_1,{\bf k}_2} {\Phi_{1{\bf k}_1}^1}^{n^1_{{\bf k}_1}} {\Phi_{1{\bf k}_1}^2}^{n^2_{{\bf k}_1}}
 \\ \times {\Phi_{4{\bf k}_2}^{2*}}^{n^2_{{\bf k}_2}} {\Phi_{4{\bf k}_2}^{1*}}^{n^1_{{\bf k}_2}}
e^{-\sum_{{\bf k}'} [ |\Phi_{1{\bf k}'}|^2 + |\Phi_{4{\bf k}'}|^2 ] } \\ \times
e^{ - \sum_{{\bf k}'} \Phi^\dagger_{1{\bf k}'} ( 1+ \delta_{{\bf k}',{\bf k}} {\cal M}^\dagger_{{\bf k}}(t) O_{\hat n_{\bf k}} {\cal M}_{{\bf k}}(t)  ) \Phi_{4{\bf k}'} } .
\label{Cn}
\end{multline}
Thus the probability distribution after a period $T$ of the drive
protocol~(\ref{prot1}) is determined by the matrix ${\mathcal
M}_{{\bf k}}(T) = -{\mathcal L}(g_2,T){\mathcal L}(g_1,T)$ defined
by~(\ref{L2L1}). The functional integral in~(\ref{Cn}) is Gaussian
and can be easily performed (see Sec.~\ref{rpcsb}). One finds \beq
\begin{split}
    C^{00}_{\hat n_{\bf k}}(f,T) &= C^{11}_{\hat n_{\bf k}}(f,T) = e^{if} , \\
    C^{10}_{\hat n_{\bf k}}(f,T) &= |\alpha_{\bf k}^2| e^{2if} + |\beta_{\bf k}^2| , \\
    C^{01}_{\hat n_{\bf k}}(f,T) &= |\alpha_{\bf k}^2| + |\beta_{\bf k}^2| e^{2if} ,
\end{split}
\label{C1} \eeq where $C^{00}$ denotes the characteristic function
for the initial state $\ket{\Psi(0)}=\ket{\{0_{\bf k},0_{-\bf
k}\}}$, etc. From~(\ref{POdef}) we finally obtain
\beq
\begin{split}
P^{00}_{\hat n_{\bf k}}(n_{\bf k},T) &= P^{11}_{\hat n_{\bf k}}(n_{\bf k},T) = \delta(n_{\bf k}-1) , \\
P^{10}_{\hat n_{\bf k}}(n_{\bf k},T) &= |\alpha_{\bf k}^2| \delta(n_{\bf k}-2) + |\beta_{\bf k}^2|  \delta(n_{\bf k}) , \\
P^{01}_{\hat n_{\bf k}}(n_{\bf k},T) &= |\alpha_{\bf k}^2| \delta(n_{\bf k}) + |\beta_{\bf k}^2|  \delta(n_{\bf k}-2) .
\end{split}
\label{P1}
\eeq
The trivial expression of $P^{00}_{\hat n_{\bf
k}}(n_{\bf k},T)$ and $P^{11}_{\hat n_{\bf k}}(n_{\bf k},T)$ comes
from the fact that the states $|\{0_{\bf k},0_{-\bf{k}}\}\rangle$
and $|\{1_{\bf k}, 1_{-\bf {k}}\}\rangle$ have no dynamics (see
Sec.~\ref{rpcsb}).  By contrast the states  $|\{1_{\bf
k},0_{-\bf{k}}\}\rangle$ and $|\{0_{\bf k}, 1_{-\bf {k}}\}\rangle$
have  nontrivial dynamics since $\ket{1_{\bf k},0_{-\bf k}}=\hat
c^\dagger_{\bf k}|\{0_{{\bf k}'},0_{-{\bf k}'}\}\rangle$ and $\ket{0_{{\bf
k}},1_{-\bf k}}=\hat c^\dagger_{-\bf k}|\{0_{{\bf
k}'},0_{-{\bf{k}'}}\}\rangle$ are not eigenstates of the Hamiltonian and
mix in the time evolution. Noting that these two states are
eigenstates of $\hat n_{\bf k}$ with eigenvalues 2 and 0,
respectively, the expression of $P^{10}_{\hat n_{\bf k}}(n_{\bf
k},T)$ and $P^{01}_{\hat n_{\bf k}}(n_{\bf k},T)$  can then directly
be deduced from the return probability computed in Sec.~\ref{rpcsb}
(e.g., $|\alpha_{\bf k}^2|$ is the probability that the system
initially prepared in the state $|1_{\bf k}, 0_{-\bf {k}}\rangle$
returns to the same state after time $T$).

In the more general case where the initial state is defined by
\beq
\ket{\Psi(0)} = \prod_{\bf k} ( a_{\bf k} \hat c^\dagger_{\bf k} + b_{\bf k} \hat c^\dagger_{-\bf k} ) \ket{\{0_{{\bf k}'},0_{-{\bf k}'}\}}
\label{Psiab}
\eeq
(with $|a_{\bf k}^2|  + |b_{\bf k}^2|=1$), the characteristic function is given by
\beq
C_{\hat n_{\bf k}}(f,T) = |a_{\bf k}\beta^*_{\bf k} - b_{\bf k}\alpha^*_{\bf k}|^2
+ e^{2if} |a_{\bf k}\beta^*_{\bf k} + b_{\bf k}\alpha^*_{\bf k}|^2 .
\label{C1a}
\eeq

A similar analysis can be done for the order parameter $\hat
\Delta_{{\bf k}} = \hat c^{\dagger}_{{\bf k}} \hat c_{-{\bf k}}$
(which in the case of a superconductor corresponds to $\hat
\Delta_{{\bf k}} = \hat c^{\dagger}_{{\bf k}}\hat c^{\dagger}_{-{\bf
k}}$ before the particle-hole transformation). It is convenient to
consider the real and imaginary parts of $\hat\Delta_{\bf k}$, \beq
\begin{split}
\hat\Delta'_{\bf k} &= \frac{1}{2} (\hat \Delta_{\bf k} + \hat \Delta^\dagger_{\bf k} ) , \\
\hat\Delta''_{\bf k} &= \frac{1}{2i} (\hat \Delta_{\bf k} - \hat \Delta^\dagger_{\bf k} ) ,
\end{split}
\eeq
which have real expectation values. For the computation of the characteristic functions, the relevant matrices are
\beq
O_{\hat\Delta'_{\bf k}} = \begin{pmatrix} 1 & if/2 \\ if/2 & 1 \end{pmatrix} , \quad
O_{\hat\Delta''_{\bf k}} = \begin{pmatrix} 1 & f/2 \\ -f/2 & 1 \end{pmatrix} .
\eeq
Assuming that the system is initially in the state~(\ref{Psiab}), one finds
\beq
\begin{split}
    C_{\hat \Delta'_{\bf k}}(f,T) ={}& 1 + if \Re[ (|b_{\bf k}^2|  - |a_{\bf k}^2|) \alpha_{\bf k}\beta_{\bf k} \\ & +  a^*_{\bf k} b_{\bf k} (\alpha^*_{\bf k}{}^2 - \beta_{\bf k}^2 ) ] , \\
    C_{\hat \Delta''_{\bf k}}(f,T) ={}& 1 + if \Im[ (|a_{\bf k}^2|  - |b_{\bf k}^2|) \alpha_{\bf k}\beta_{\bf k} \\ & +  a^*_{\bf k} b_{\bf k} (\alpha^*_{\bf k}{}^2 + \beta_{\bf k}^2 ) ] ,
\end{split}
\label{C2}
\eeq
and in turn
\beq
\begin{split}
P_{\hat\Delta'_{\bf k}}(\Delta'_{\bf k},T) ={}& \delta(\Delta'_{{\bf k}}) - \delta'(\Delta'_{{\bf k}})  \Re[ (|b_{\bf k}^2|  - |a_{\bf k}^2|) \alpha_{\bf k}\beta_{\bf k} \\ & +  a^*_{\bf k} b_{\bf k} (\alpha^*_{\bf k}{}^2 - \beta_{\bf k}^2 ) ] , \\
P_{\hat\Delta''_{\bf k}}(\Delta''_{\bf k},T) ={}& \delta(\Delta''_{{\bf k}}) - \delta'(\Delta''_{{\bf k}}) \Im[ (|a_{\bf k}^2|  - |b_{\bf k}^2|) \alpha_{\bf k}\beta_{\bf k} \\ & +  a^*_{\bf k} b_{\bf k} (\alpha^*_{\bf k}{}^2 + \beta_{\bf k}^2 ) ] .
\end{split}
\label{P2}
\eeq
The expectation value of the order parameter at time $T$ is therefore given by
\begin{align}
\bra{\Psi(T)} \hat\Delta'_{\bf k} \ket{\Psi(T)} ={}& \int d\Delta'_{\bf k} \, P_{\hat\Delta'_{\bf k}}(\Delta'_{\bf k},T) \Delta'_{\bf k} \nonumber \\
={}&  \Re[ (|b_{\bf k}^2|  - |a_{\bf k}^2|) \alpha_{\bf k}\beta_{\bf k} \nonumber \\ & +  a^*_{\bf k} b_{\bf k} (\alpha^*_{\bf k}{}^2 - \beta_{\bf k}^2 ) ]
\end{align}
and
\begin{align}
\bra{\Psi(T)} \hat\Delta''_{\bf k} \ket{\Psi(T)} ={}& \int d\Delta''_{\bf k} \, P_{\hat\Delta''_{\bf k}}(\Delta''_{\bf k},T) \Delta''_{\bf k} \nonumber \\
={}&  \Im[ (|a_{\bf k}^2|  - |b_{\bf k}^2|) \alpha_{\bf k}\beta_{\bf k}  \nonumber\\ & +  a^*_{\bf k} b_{\bf k} (\alpha^*_{\bf k}{}^2 + \beta_{\bf k}^2 ) ] ,
\end{align}
or, equivalently,
\beq
\bra{\Psi(T)} \hat\Delta_{\bf k} \ket{\Psi(T)} =
(|b_{\bf k}^2|  - |a_{\bf k}^2|) \alpha^*_{\bf k}\beta^*_{\bf k}  +  a^*_{\bf k} b_{\bf k} \alpha^*_{\bf k}{}^2
- a_{\bf k} b^*_{\bf k} \beta^*_{\bf k}{}^2  .
\eeq

Equations~(\ref{C1},\ref{P1},\ref{C1a}) and (\ref{C2},\ref{P2}) represent the
main result of this section. They yield the characteristic functions
and probability distributions of the quadratic operators $\hat
n_{\bf k}$ and $\hat\Delta_{\bf k}$ in terms of the elements of the
Floquet Hamiltonian of the system for any arbitrary state
$|\psi(0)\rangle$ that can be constructed out of superposition of
$|1_{{\bf k}}, 0_{-{\bf k}} \rangle$ and $|0_{{\bf k}},1_{-{\bf
k}}\rangle$.

\begin{figure}
\centering {\ing[width=\linewidth]{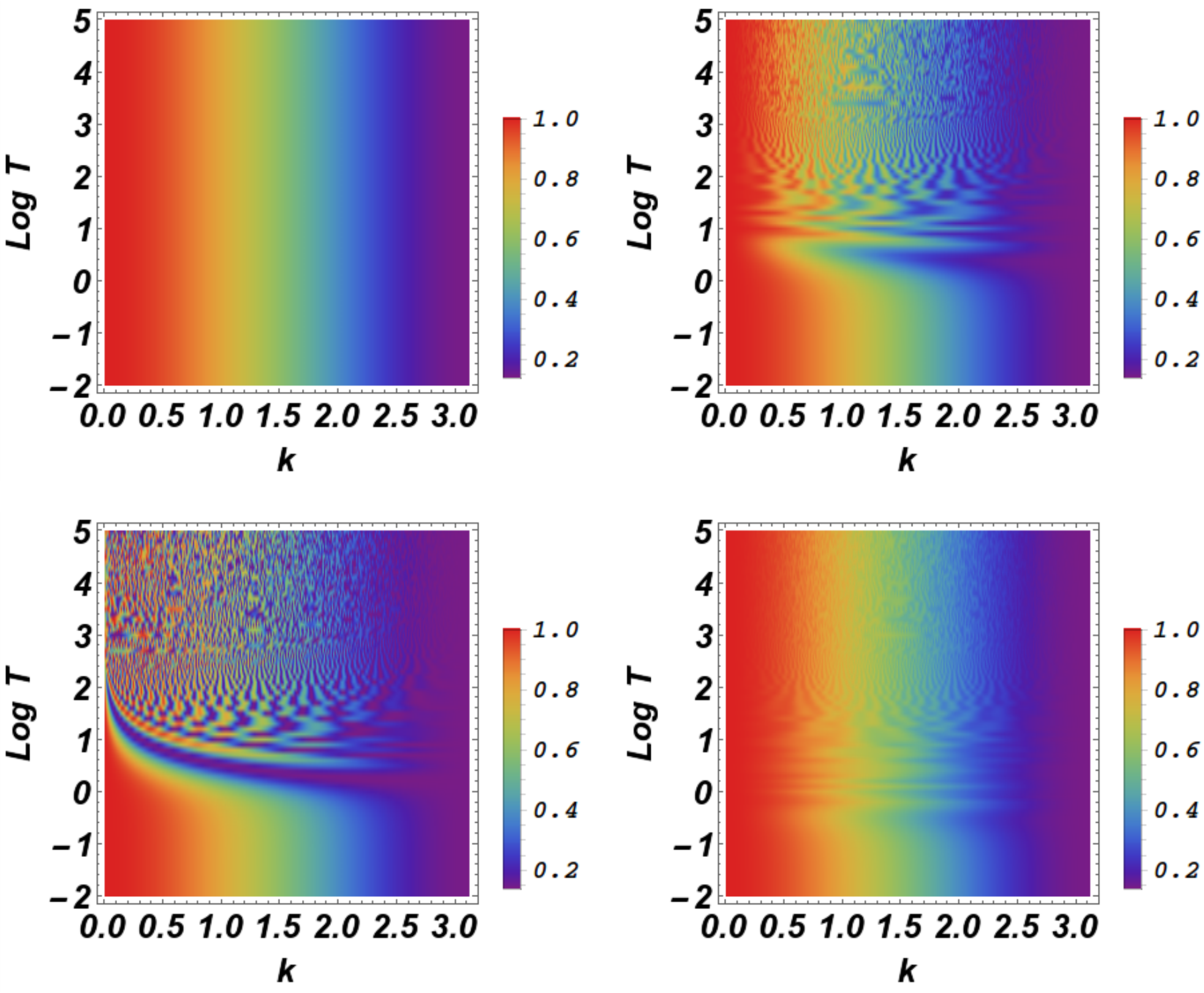}} \caption{ $C_{\hat
n_k}(f=i,T)$ for the one-dimensional Ising model as a function of
the drive period $T=2\pi/\omega_D$ (in units of $\hbar/J$) and
momentum $k$ (in units of $a^{-1}$). Each of the plots corresponds
to $g_1=0.01$ and the initial state is taken to be the ground state of $H(g_1)$.
The top left (right) panel corresponds to $g_2=0.01 (0.4)$ and the
bottom left(right) panel to $g_2=0.99 (10)$.} \label{fig5a}
\end{figure}

\begin{figure}
    \centering {\ing[width=\linewidth]{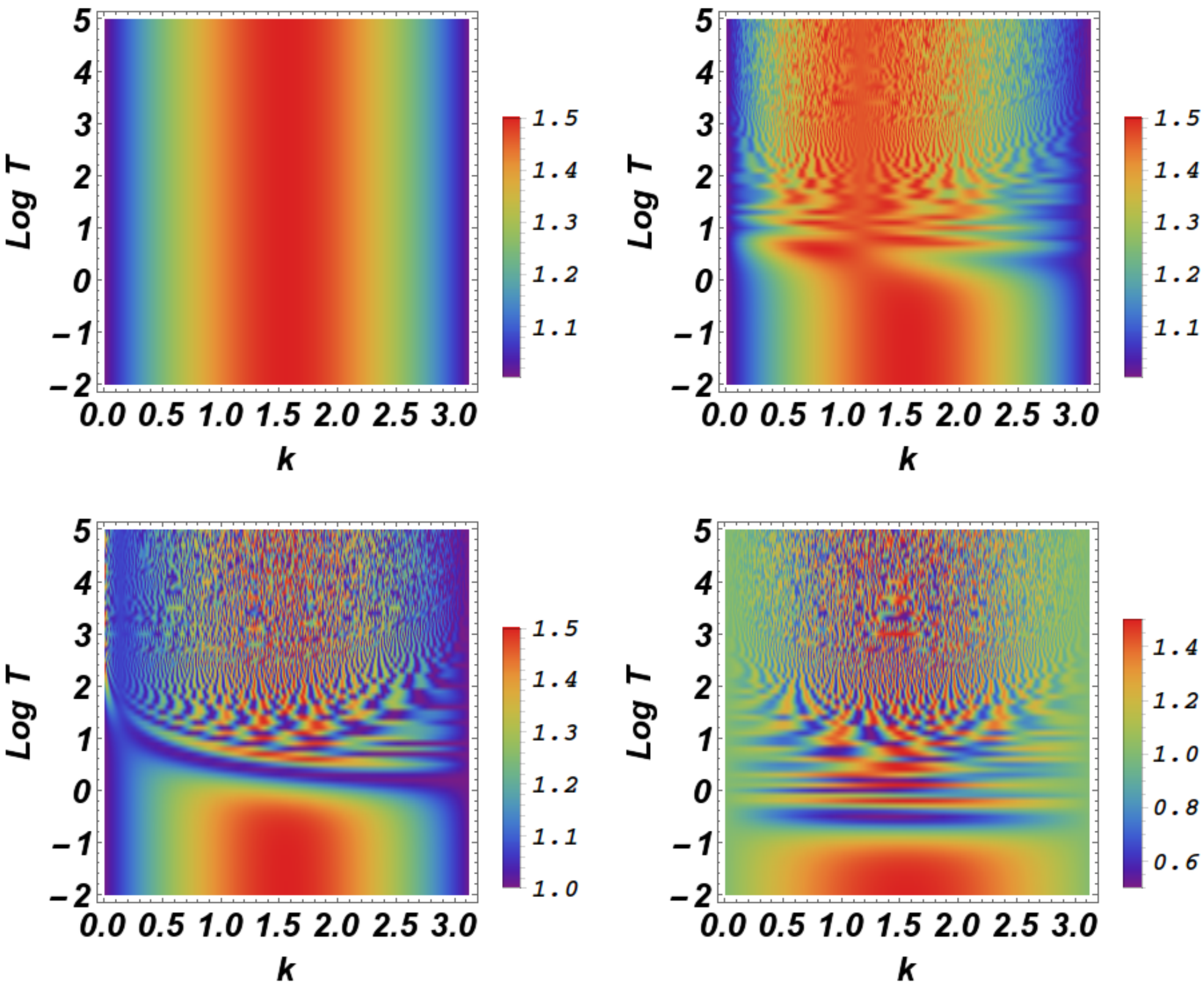}} \caption{$C_{\hat\Delta^{\prime}_k}(f=i,T)$ for the one-dimensional Ising
        model as a function of the drive period $T=2\pi/\omega_D$ (in units
        of $\hbar/J$) and momentum $k$ (in units of $a^{-1}$). All other
        parameters are same as in Fig.\ \ref{fig5a}. } \label{fig5b}
\end{figure}

Finally we show that the characteristic functions exhibit a clear
signature of the dynamics of the system. In Figs.~\ref{fig5a} and
\ref{fig5b} we plot $C_{\hat n_k}(f=i,T)$
and $C_{\hat\Delta'_k}(f=i,T)$ for the one-dimensional transverse field Ising model
as a function of $k$ (measured in units of the lattice spacing $a$)
and the drive period $T=2 \pi/\omega_D$ (in units of $\hbar /J$).
The square pulse
protocol has $g_1=0.01$  while $g_2$ is varied between 0.01 and 10 as shown in Figs.\
\ref{fig5a} and \ref{fig5b}.
The initial state is assumed to be given by~(\ref{Psiab}) with
\beq
\begin{split}
a_k &= \frac{E_k(g_1)-g_1+\cos k}{{\cal N}E_k(g_1)} , \\
b_k &= \frac{\sin k}{{\cal N}E_k(g_1)}
\end{split}
\eeq ($\cal N$ is a normalization constant), which corresponds to
the ground state deep in the paramagnetic phase. For small $T$, i.e.
for very fast drive ($\hbar \omega_D \gg J$), the system has no time
to react and the characteristic functions are similar to their $T=0$
values. Similarly
if $g_2 \simeq g_1$, the system evolves according to an almost time
independent Hamiltonian ($H[g_1]$) for the entire drive cycle and
thus remains in its starting ground state. By contrast, when these
conditions ($\hbar \omega_D \gg J$ and $g_2\simeq g_1$) are not
fulfilled, the characteristic functions show an oscillatory,
irregular, behavior as a function of $k$. In the case where $\hbar \omega_D \ll J$,
this is consistent with the fact that lowering the drive
frequency makes the elements of the Floquet Hamiltonian $\alpha_{k}$
and $\beta_{k}$ a much more convoluted function of $k$; this is a
well-known signature of the increasing complexity of the Floquet
Hamiltonian with decreasing drive frequency \cite{asen1}.

These irregular oscillations are however not present in $C_{\hat
n_k}$ when $\hbar \omega_D \ll J$ and $g_2\gg 1$. This can be
understood as follows. For $g_2 \gg g_c$ (ferromagnetic phase;
$g_c=1$ being the critical point) and $g_1 \ll g_c$ (paramagnetic
phase), both cycles of the square pulses correspond to a gapped
Hamiltonian with a large gap. As a result, the system does not
absorb significant energy during the drive. Therefore, the
drive-induced excitations of the system are mostly fluctuations of
the phase of $\langle \hat \Delta_k \rangle$. Thus $\langle \hat n_k
\rangle$, which does not depend on this phase, does not show
oscillations at low frequencies. In contrast, the order parameter
receives contribution from these phase fluctuations, which explains
the irregular oscillations observed in the bottom right plot of
Fig.~\ref{fig5b}.

\section{Discussion}
\label{diss}

We have presented a formalism for computing the matrix
element of the density matrix of a many-body quantum system between
two arbitrary coherent states. This formalism shows that it is
possible, at least in principle, to express such matrix elements in terms of
the Matsubara correlation functions of the system. In practice,
such computation can be straightforwardly carried out for systems
with a Gaussian action. This does not necessarily require a
quadratic Hamiltonian; for example, the Bose-Hubbard model treated
within strong coupling mean-field theory can be represented by a
Gaussian action but not a quadratic Hamiltonian. One of our main results is
that for systems with Gaussian actions these matrix
elements are completely determined by the Matsubara Green function
$G(0^+)$.

For systems with Gaussian actions, we are thus able to express
several measures of entanglement entropy in terms of their Matsubara
Green functions. Our results in this respect reproduce the ones
derived earlier for systems with Gaussian Hamiltonians
\cite{peschelref, henleyref}. We stress however that our method
allows us to compute these measures for systems with Gaussian action
which may not have a quadratic Hamiltonian. As an example, we have
considered strongly interacting bosons in the framework of the
Bose-Hubbard model, both in its pristine form and in the presence of
Abelian and non-Abelian gauge fields. Our results not only
demonstrate the usefulness of our methods but also provide, to the
best of our knowledge, the first calculation of entanglement entropy
of bosons in their Mott phase in the presence of synthetic gauge
fields. We note that the second R\'enyi entropy of pristine bosons
has recently been measured \cite{rislam1}; we therefore expect our
computations to be useful for similar experiments carried out on
bosons in the strong coupling regime in the presence of synthetic
gauge fields.

Our analysis also allows us to compute the matrix elements of the evolution
operator $\hat U(t,0)$ for a driven Gaussian system using analytic
continuation $t \to -i \beta \hbar$. In systems where one can
compute the Green function $G(0^+)$ at any finite temperature $T_0$
analytically so that the Wick rotation can be carried out in a
straightforward manner, we are therefore able to compute the dynamical
correlation functions without resorting to Keldysh formalism. In
contrast, if the Green function of the system is only known
numerically, it might be difficult to carry out the Wick rotation
and this is one of the drawbacks of this approach. To illustrate our approach we
have computed the return probability and the
counting statistics of density and order parameter operators for
Dirac fermions subjected to a periodic drive with square pulse
protocol. Our analysis allows us to obtain analytic expressions
in terms of elements of the Floquet Hamiltonian. We
reproduce earlier results as special
cases and also find that these measures, in the presence
of a suitable drive protocol, can point out the location of the
critical point.

The analysis carried out in this work is expected to find
applications in several other theoretical models. Examples include
quantum rotor models treated with the large-$N$ approximation. We
note that while there is some progress in understanding the dynamics
of these models in the paramagnetic and ferromagnetic phases
\cite{dyn1,dyn2}, the behavior of the driven system near its
critical point is still not resolved. Second, the R\'enyi entropy
near the critical point has not been understood analytically.
Moreover, much of the formalism developed here for Dirac fermions
may be applied to a class of models used to describe spin-liquids
within RVB mean-field theories;\cite{vish1} in particular we expect
our method to yield analytical results concerning the dynamics.
Third, we hope to address the out-of equilibrium dynamics and
entanglement measures in Luttinger liquids whose action can be
written in a quadratic form in terms of bosonic degrees of freedom
\cite{kop1}.

{\color {blue} The extension of our analysis to open quantum systems
which generates noise and dissipation may be another subject of
future work. It is easy to see that our approach could be easily
extended to integrable fermionic/bosonic systems coupled linearly to
quadratic fermionic/bosonic baths. However for general case where
integrating out the bath leads to non-Gaussian terms in the
effective action, we expect things to be more complicated. Such a
situation warrants a more detailed and separate study.}

In conclusion, we have developed a formalism which allows one to
express the matrix elements of the density matrix of a many-body
quantum system in terms of its Matsubara correlation functions. This
allows us to compute several entanglement measures and address
out-of-equilibrium dynamics of several fermionic and bosonic systems
with Gaussian actions. In particular, we have used our formalism to
compute entanglement measures in strongly interacting bosons described
by the Bose-Hubbard model in the presence of synthetic gauge fields
within a strong-coupling mean-field approach. We have also obtained
semi-analytic expressions for return probability and counting
statistics of Gaussian operators for Dirac fermions subjected to a
periodic drive in terms of elements of the Floquet Hamiltonian. We
note that our analysis may be useful for entanglement related
experiments on bosons and may also be used for analysis of
non-equilibrium dynamics and entanglement entropy of several other
systems as discussed above. 

\begin{acknowledgments}
The work of A.S. is partly supported through the Max Planck Partner
Group program between the Indian Association for the Cultivation of
Science, Kolkata and the Max Planck Institute for the Physics of
Complex Systems, Dresden. R.G. acknowledges CSIR SPM fellowship for
support and N.D. thanks I. Fr\'erot for correspondence.
\end{acknowledgments}

\appendix

\section{Density matrix for a single degree of freedom}

\subsection{Equilibrium case}
\label{app1a}

We consider a single, bosonic or fermionic, degree of freedom
described by the Hamiltonian $\hat H=\w \hat c^\dagger \hat c$ with
$\hat c\hat c^\dagger-\eta \hat c^\dagger \hat c=1$. Using $|\phi\rangle=e^{\eta\phi
\hat c^\dagger}|{\rm vac}\rangle$, where $|{\rm vac}\rangle$ is the
vacuum state ($\hat c|{\rm vac}\rangle=0$), a straightforward calculation
gives
\begin{equation}
\rho_{fi} =\frac{1}{Z} \bra{\phi_f} e^{-\beta\hat  H} \ket{\phi_i} =
\frac{1}{Z} \exp\left\{ e^{-\beta \w}\phi^*_f \phi_i \right\}
\label{app1}
\end{equation}
with the partition function $Z=(1-\eta e^{-\beta\w})^{-\eta}$. Let
us now reproduce this result from the path integral formalism
discussed in Sec.~\ref{genform1}. For the problem at hand, the
discrete-time path integral~(\ref{dmexp2a})  reads
\begin{align}
U_{fi} =&{} \frac{1}{Z} e^{\eta\phi^*_f\phi_i}   \int
\prod_{k=1}^{N} d\phi_k^* d\phi_k \int d\lambda^* d\lambda \non \\ &
\times e^{ - \sum_{k,k'=1}^{N} \phi_{k}^* S_{kk'} \phi_{k'} -
(\phi^*_{N}-\phi^*_{f}) \lambda + \lambda^* (\phi_{N}-\eta \phi_{i})
} ,
\end{align}
where
\begin{equation}
S = \left(
\begin{matrix} 1 &   0    & \cdots & 0    & -\eta a \\
-a &   1    & 0      &              & 0        \\
0 &  -a    & \ddots      & \ddots &        \vdots   \\
\vdots & \ddots & \ddots &    \ddots & 0    \\
0 & \cdots   & 0 & -a     & 1
\end{matrix}
\right)
\end{equation}
is a $N\times N$ matrix and $a=1-(\beta/N)\w$. Integrating out
$\phi_k$ and $\phi^*_k$, we find
\begin{align}
U_{fi} &= (\det S)^{-\eta}\frac{e^{\eta\phi^*_f\phi_i}}{Z}  \int
d\lambda^* d\lambda
\, e^{-\lamb^* S^{-1}_{NN} \lamb + \phi^*_{f} \lambda - \eta \lambda^* \phi_{i} } \non \\
&= \frac{1}{Z} (\det S)^{-\eta} (S^{-1}_{NN})^{-\eta} \, e^{
\eta\phi^*_f\phi_i [ 1- (S^{-1}_{NN})^{-1} ]} .
\end{align}
Using $\det S=1-\eta a^N$ and
\begin{equation}
S^{-1} = \frac{1}{1-\eta a^N} \left(
\begin{matrix}
1      & \eta a^{N-1}& \eta a^{N-2} & \cdots        &  \eta a   \\
a      &   1          & \ddots &           & \vdots \\
a^2    &   \ddots          &  \ddots            & \ddots & \vdots   \\
\vdots &        & \ddots & 1                      & \eta a^{N-1} \\
a^{N-1} & \cdots          & \cdots        & a             & 1
\end{matrix}
\right) ,
\end{equation}
we finally obtain
\begin{align}
U_{fi} = \frac{1}{Z} \, e^{ a^N \phi^*_f\phi_i } ,
\end{align}
since $\lim_{N\to\infty}a^N=e^{-\beta\w}$,
in agreement with~(\ref{app1}). The present derivation makes it
clear that the path integral involves the Matsubara Green function.
In the continuum-time limit $N\to\infty$, $S^{-1}_{kk'}$ becomes the
imaginary-time propagator $G(\tau,\tau')\equiv G(\tau-\tau')$. We
also see that $U_{fi}$ depends on the equal-time correlation
function $S^{-1}_{NN}$, which should be interpreted as
$G(0^+)$.\cite{norref1} The preceding path-integration calculation
is exact since it is based on the discrete-time
expression~(\ref{dmexp2a}). The same result can however be obtained
from the continuous-time expression given by~(\ref{dmexp3}).

\subsection{Out-of-equilibrium case}
\label{app1b}

We now consider the case where the energy $\omega(t)$ is
time-dependent: $\hat H(t)=\omega(t)\hat c^\dagger \hat c$. The evolution operator
satisfies $\hat H(t)\hat  U(t,0) = i\partial_t \hat U(t,0)$ and is given by
\beq
\hat U(t,0) =e^{-f(t)\hat c^\dagger \hat c} , \quad f(t) = i \int_0^t dt'
\omega(t') . \eeq A straightforward computation then gives \beq
U_{fi}(t,0) = \langle \phi_f| \hat U(t,0) |\phi_i\rangle = \exp\{
e^{-f(t)} \phi^*_f \phi_i \} .
\label{eqapp2}
\eeq

In the path-integral formalism, the dynamics of the system is
governed by the real-time action \beq S = \int dt \, \phi^*(t)
[i\partial_t - \omega(t)] \phi(t) . \eeq The imaginary-time Green
function satisfies \beq [\partial_\tau + \bar\omega(\tau)]
G(\tau,\tau') = \delta(\tau-\tau') , \eeq the solution of which is
\beq G(\tau,\tau') = e^{-\int_{\tau'}^\tau d\tau''
\bar\omega(\tau'')} [ \theta(\tau-\tau') (1+\eta n) +
\theta(-\tau+\tau') \eta n ] , \eeq where $\bar\omega(\tau)$ is the
analytic continuation of $\omega(t)$ to imaginary time $\tau=it$.
The value of $n$ is determined from the boundary conditions
$G(\tau,\beta)=\eta G(\tau,0)$ and $G(\beta,\tau')=\eta G(0,\tau')$
(which follow from $\phi^{(*)}(\beta)=\eta \phi^{(*)}(0)$ in the
Euclidean path integral), i.e. \beq n = \frac{1}{e^{\bar
f(\beta)}-\eta} , \quad \bar f(\beta) = \int_0^\beta d\tau \,
\bar\omega(\tau) . \eeq Equation~(\ref{unievol4}) then gives \beq
U_{fi}(\beta,0) = {\cal N} \exp\{ e^{-\bar f(\beta) \phi^*_f \phi_i
} \} \eeq and therefore Eq.~(\ref{eqapp2}) after analytic continuation
to real time. We do not compute the normalization factor $\cal N$
which does play an essential role in the dynamics of the system.

\section{Relation to previous results}
\label{app2}

In this appendix, we show that our formulation reproduces
well-known earlier results on
entanglement measures of many-body systems with a Gaussian
Hamiltonian.\cite{peschelref,henleyref,cardyref1,callan}

First, we show that Eqs.\ \eqref{quadres1} and \eqref{quadres2} lead
to the well-known expressions of the $n^{\rm th}$ R\'enyi entropy
$S_n$.\cite{huertorev} For this we use the replica field technique
introduced in Ref.\ \onlinecite{cardyref1}. We start with our
expression of the density matrix for a $U(1)$ unbroken symmetry
Hamiltonian,
\begin{align}
\rho_A(\phi_f, \phi_i)&= \langle \phi_f|\hat\rho|\phi_i
\rangle = \frac{1}{Z} e^{\eta \sum_{jj'} \phi_j^{f
\ast}[I-G^{-1}(0^+)]_{jj'}\phi_{j'}^i}
\nonumber\\
&= \frac{1}{Z} e^{\sum_{jj'} \phi_j^{f \ast} \left(C[1+\eta
C]^{-1}\right)_{jj'} \phi_{j'}^i} ,
\end{align}
where $\eta=1[-1]$ for bosons[fermions] and
$C_{jj'}=\eta(G_{jj'}(0^+)-\delta_{jj'})$.

To compute $S_n$ we now introduce $n$ replica fields $\phi_1
...\phi_n$. Using these one can write
\begin{widetext}
\begin{align}
S_n &= \frac{1}{1-n} \ln \Tr(\hat\rho^n)= \frac{1}{1-n} \ln
\left[\frac{1}{Z^n}\int d\phi_1^* d\phi_1 \hdots d\phi_n^* d\phi_n
\langle {\eta \phi_1}|\hat\rho|{\phi_2}\rangle \hdots\langle
{\phi_n}|\hat\rho|{\phi_1}\rangle e^{-\sum_{i=1}^n |\phi_i|^2} \right]
\nonumber\\
&= \frac{1}{1-n} \ln \left\{\frac{1}{Z^n}\int d\phi_1^* d\phi_1\hdots
d\phi_n^* d\phi_n \, e^{ -\Phi^\dagger
\mathcal{M} \Phi } \right\} ,
\label{master1}
\end{align}
\end{widetext}
where $\Phi^\dagger=(\phi_1^*,\cdots,\phi^*_n)$ and the
matrix $\mathcal{M}$ (in block matrix form) is given by
\begin{align}
\mathcal{M}_{ij} ={}& I \delta_{ij}+\eta \frac{C}{I+\eta C}
\delta_{i,1}\delta_{j,2}+\frac{C}{I+\eta C} \nonumber\\
& \times \delta_{i,
j-1}(1-\delta_{i,1}\delta_{j,2})+\frac{C}{I+\eta C}\delta_{n, i}
\delta_{j,1} .
\end{align}
Also it can be seen that $Z={\rm det}(1+\eta C)^{\eta}$. So Eq.\
\eqref{master1} can finally be written as
\begin{equation}
S_n = \frac{1}{1-n} \ln \left\{  \frac{{\rm det}[I-\eta(C/(I+\eta
C))^n]^{-\eta}}{{\rm det}(1+\eta C)^{n\eta}} \right\} .
\end{equation}
This expression can be simplified to yield
\begin{equation}
S_n=\frac{1}{1-n} {\rm Tr}\ln[(I+\eta C)^n-\eta C^n] ,
\label{nren}
\end{equation}
which is the R\'enyi entropy expression obtained in Ref.\
\onlinecite{huertorev}. The results for the $U(1)$ symmetry broken
case may also be obtained by carrying out exactly the same analysis
after switching to a basis which diagonalizes $G$. The result is
given by Eq.\ \eqref{nren} with $C \to C_d$ where $C_d$ denotes the
correlation matrix in the diagonal basis and has contribution from
both $C$ and $F$.

Next, we derive the results concerning the correlation functions $C$
and $F$ for fermions obtained earlier in Refs.
\onlinecite{peschelref,henleyref}. To this end, we first note that
in the diagonal basis the correlation matrix, as obtained from Eq.\
\eqref{quadres2}, is given by
\begin{eqnarray}
C_d = \left(\begin{array}{cc} 1- \eta n & 0 \\
0 & n \end{array} \right) ,
\label{cmdiag}
\end{eqnarray}
where $n$ is a diagonal matrix whose eigenvalues are obtained by
diagonalizing the correlation matrix of bosons or fermions. In
contrast, in the off-diagonal basis, the correlation matrix $C$, as
obtained in Refs.\ \onlinecite{peschelref,henleyref}, is given by
\begin{equation}
C = \left(\begin{array}{cc} I-\eta C & F \\
\eta F^{\ast} & C \end{array} \right) . \label{cmph}
\end{equation}
We note that $C_d$ and $C$ must be related by a diagonalizing matrix
$U$: $C_d = U C U^{\dagger}$. Moreover, the elements of $U$ can be
constructed out of $2L$ component eigenvector $\psi_0=(u,v)$ of $C$,
one can write $C \psi_0 = C_d \psi_0$. This leads to
\begin{align}
(1-\eta C) u +F v&=(1-\eta n) u , \\
\eta F^{\ast} u +C v&=n v .
\end{align}
Assuming $F$ to be real, this gives, for fermions,
\begin{align}
\left(C-\frac{1}{2} \right)u+ F v&=-\left(n-\frac{1}{2} \right)u , \\
-F u+\left(C-\frac{1}{2} \right)v&=\left(n-\frac{1}{2} \right)v .
\end{align}
Introducing $\phi=u+v, \psi=u-v$, we finally obtain
\begin{align}
\left(C-\frac{1}{2}-F \right)\phi &=-\left(n-\frac{1}{2} \right) \psi  ,\\
\left(C-\frac{1}{2}+F \right)\psi &= -\left(n-\frac{1}{2} \right) \phi .
\end{align}
This set of equations can be combined into one, as,
\begin{equation}
\left(C-\frac{1}{2}-F \right)\left(C-\frac{1}{2}+F \right)\phi
= \left(n-\frac{1}{2} \right)^2 \phi .
\end{equation}
This is exactly Peschel's equation with the
identification $n= {\rm Diag}(\epsilon_{\ell})$ where $(n_{\ell}
-\frac{1}{2})^2 =\frac{1}{4}\tanh^2(\epsilon_{\ell} /2)$ and
$\epsilon_{\ell}$ are the eigenvalues of the entanglement spectrum.
This can be easily seen from the fact that $\nu_{\ell} $ are the
eigenvalues of the covariance matrix and they are related to
$\epsilon_{\ell}$ by $n_{\ell}=1/(e^{\epsilon_{\ell}}+1)$ for
fermions. The choice of the negative root of the above equation leads
to the results of Sec.~II.B.

\end{document}